%% file: sn-article.tex
\documentclass[pdflatex,sn-nature]{sn-jnl}

\usepackage{graphicx}%
\usepackage{multirow}%
\usepackage{amsmath,amssymb,amsfonts}%
\usepackage{amsthm}%
\usepackage{mathrsfs}%
\usepackage[title]{appendix}%
\usepackage{xcolor}%
\usepackage{textcomp}%
\usepackage{manyfoot}%
\usepackage{booktabs}%
\usepackage[percent]{overpic}%
\usepackage{listings}%
\usepackage{placeins}

\input{compounds}

\theoremstyle{thmstyleone}%
\theoremstyle{thmstyletwo}%

\theoremstyle{thmstylethree}%

\begin{document}

\title[Closed-Loop Molecular Design with Calibrated Deference]{Closed-Loop Molecular Design with Calibrated Deference}


\author*[1]{\fnm{Newman} \sur{Cheng}}\email{newmancheng@microsoft.com}

\author[1]{\fnm{Gordon} \sur{Broadbent} \sfx{IV}}

\author[3]{\fnm{Jason} \sur{Dong}}

\author[4]{\fnm{Syed Mohammed Ali} \sur{Hussaini}}

\author[4]{\fnm{Farman} \sur{Ullah}}

\author[2]{\fnm{Morris} \sur{Sharp}}

\author[2]{\fnm{Gabrielle} \sur{Barnes}}

\author[2]{\fnm{Nanlin} \sur{Guo}}

\author[2]{\fnm{Deyu} \sur{Zou}}

\author[5]{\fnm{Eliot F.} \sur{Woods}}

\author[5]{\fnm{Jong-Hwa} \sur{Shon}}

\author[5]{\fnm{Yangang} \sur{Liang}}

\author[5]{\fnm{Ruozhu} \sur{Feng}}

\author[5]{\fnm{Wei} \sur{Wang}}

\author[2]{\fnm{Karin} \sur{Strauss}}

\author[1]{\fnm{William} \sur{Chappell}}

\author*[3]{\fnm{David G.} \sur{Kwabi}}\email{david.kwabi@yale.edu}

\author*[2]{\fnm{Bichlien H.} \sur{Nguyen}}\email{bnguy@microsoft.com}

\author*[2]{\fnm{Jake A.} \sur{Smith}}\email{jakesmith@microsoft.com}

\affil[1]{\orgname{Microsoft Discovery \& Quantum}, \orgaddress{\city{Redmond}, \state{WA}, \country{USA}}}

\affil[2]{\orgname{Microsoft Research}, \orgaddress{\city{Redmond}, \state{WA}, \country{USA}}}

\affil[3]{\orgdiv{Department of Chemical and Environmental Engineering}, \orgname{Yale University}, \orgaddress{\city{New Haven}, \state{CT}, \country{USA}}}

\affil[4]{\orgname{CanAm Bioresearch Inc.}, \orgaddress{\city{Winnipeg}, \state{MB}, \country{Canada}}}

\affil[5]{\orgname{Pacific Northwest National Laboratory}, \orgaddress{\city{Richland}, \state{WA}, \country{USA}}}


\abstract{%
As AI systems become increasingly capable scientific collaborators, a fundamental challenge that arises is not only generating explanations, but calibrating confidence in them. 
We present Discovery, a multi-agent scientific system designed to assess the reliability of tools and data during adaptive reasoning, track evidence and assumptions while resolving uncertainty using a persistent graph memory.
Discovery enables a capability we term \emph{calibrated deference}: the ability to adjust confidence in its tool use, assumptions, and mechanistic explanations based on these assessments, communicate uncertainty to human collaborators, and adapt strategies accordingly.
We evaluated Discovery in a closed-loop human--AI campaign to design an aqueous organic redox flow battery negative electrolyte. Discovery proposed candidates and interpreted results in collaboration with chemists, who synthesized, characterized, and guided design.
Across 17 candidates over three rounds, Discovery converged on a candidate with a redox potential improved by 130~mV relative to the literature baseline.
However, experimental characterization revealed poor electrochemical reversibility that the predictors used in the campaign could not anticipate.
In response, Discovery generated competing mechanistic hypotheses, prioritized discriminating diagnostics, and traced the behavior to phosphonate--potassium ion pairing.
Discovery designed a sulfonate analog to reduce potassium-ion pairing, and the resulting compound improved electrochemical reversibility while maintaining a 90~mV improvement in redox potential, closing the design--make--test--redesign loop.
Across the campaign, Discovery demonstrated the ability to maintain a coherent world model, updating in response to unexpected results, and leverage it to drive molecular design.
}

\keywords{redox flow batteries, molecular design, large language models, AI agents, organic electrolytes}



\maketitle


\section{Introduction}\label{sec:intro}
Large-language-model (LLM) agents can now execute substantial parts of scientific workflows by selecting and invoking external tools.\cite{ren2025sciagensurvey,zheng2025automation}
This operational capability is clear in systems such as ChemCrow (synthesis planning), Coscientist (autonomous reaction-condition optimization), BioDiscoveryAgent (iterative perturbation design over existing datasets), and Robin (multi-agent lab-in-the-loop discovery).\cite{bran2024chemcrow,boiko2023coscientist,roohani2024biodiscovery,ghareeb2026robin}
However, scientific discovery is not only execution within a scoped task but cumulative reasoning: building assumptions, updating them against evidence, and repeatedly calibrating what should be trusted, revised, or discarded over time.

LLM agents still struggle in converting long-context, cross-round outcomes into stable beliefs and downgrading computational priors when experiments contradict them.
Without this epistemic control, LLM agents continue acting while preserving the assumptions that generated failed predictions.
The missing capability is not additional tool use but structured memory of evolving scientific judgment.
Here, we present Discovery, a multi-agent system that addresses this gap by introducing regrets and claims elicitation into Discovery Engine, its orchestration harness previously known as Cognitive Loop via In-Situ Optimization (CLIO), and unifies learning from successful and unsuccessful experiments over time using Discovery Bookshelf, a continually updated knowledge graph. In combination, the behavior that emerges is an explicit policy for calibrating trust in its own computational tools against accumulating experimental evidence.\cite{cheng2025clio}

In this work, we use Discovery to design an improved aqueous organic redox flow battery (AORFB) negative electrolyte (negolyte) based on the recently characterized benzo[c]cinnoline scaffold.\cite{singh2025bc}
The chosen design objectives span numerically predictable properties, loosely quantifiable criteria, and heuristic reasoning tasks.
Discovery executed a 17-compound design campaign, producing a top phosphonated benzo[c]cinnoline compound that was synthesized and electrochemically characterized, confirming a 130~mV improvement in redox potential over the literature baseline.
When cyclic voltammetry (CV) revealed unexpectedly poor electrochemical reversibility, Discovery interpreted the experimental data, formulated competing mechanistic hypotheses, and proposed diagnostic experiments to discriminate among them.
The diagnostics identified phosphonate--cation pairing as the most likely cause, and a phosphonate--sulfonate replacement was recommended.
The resulting compound showed substantially improved electrochemical reversibility and maintained a 90~mV improvement in redox potential, closing an agentic design--make--test--redesign loop.

Across this campaign, we observe what we term \emph{calibrated deference}---a pattern of evidence-led recalibration of trust in priors and adaptive revision of strategy under contradiction and constraint.
This manifests in two coupled behaviors: progressive recalibration of trust in computational models, and deferring commitment among competing hypotheses until discriminating experimental evidence narrows the field.

\section{Results}\label{sec:results}

\subsection{Formulation of the negolyte design objective}\label{subsec:problem-formulation}

Recent work established the benzo[c]cinnoline scaffold as a viable AORFB negolyte: a rigid fused-ring azo motif decorated with disulfonate solubilizing groups (\cmpd{sBC}) achieves a reduction potential ($E_\mathrm{red}$) of $-0.84$~V vs Ag/AgCl at pH 14 with improved electron-transfer kinetics and stability over azobenzene~\cite{singh2025bc}. These properties, however, emerged from manual optimization, and the design space remains largely unexplored. Further improvement requires balancing tightly coupled, often antagonistic objectives---low $E_\mathrm{red}$, high solubility at pH 14, synthetic tractability, and reversibility---that span three distinct classes of computational evaluability. \emph{Numerically predictable} objectives, such as $E_\mathrm{red}$ and aqueous solubility (logS), can be estimated by ab initio methods or statistical predictors, given training-data coverage, and numerical optimizers handle them well. \emph{Loosely quantifiable} objectives, such as synthetic tractability, lack a strict numerical definition but can be approximated via retrosynthetic analysis and synthetic accessibility scores (SAscore).\cite{ertl2009sascore} \emph{Heuristically reasoned} objectives---decomposition pathways and reversibility---have no comprehensive predictor and demand chemical intuition and literature context.

Scientific optimization relies on multiple forms of context, including scientist-vetted academic literature, local observations, and input from collaborators.
Zhang et al. demonstrated the benefit of including these additional information sources in experimental optimization by building a search space in which literature knowledge was directly encoded.\cite{zhang2025multimodaldiscovery}
A black-box optimizer restricted to numerically encoded objectives cannot address the full design space; this heterogeneity is precisely what motivates a reasoning agent.

With this framing, we used Discovery to propose structural modifications to the benzo[c]cinnoline core---prioritizing synthesizability, a reversible $E_\mathrm{red}$ in the window $-1.2$ to $-0.3$~V vs SHE, a logS $\ge -2$, and avoidance of water-reactive or irreversibly reducible groups---and to explain how each proposed modification would advance these targets. Discovery was given a tool suite covering all three objective classes: (i) two statistical models based on the Graphormer architecture for $E_\mathrm{red}$ and logS; (ii) RDKit's synthetic accessibility scorer and RetroChimera for retrosynthetic analysis; and (iii) Deep Researcher for literature search.\cite{ying2021graphormer, ertl2009sascore,rdkit,maziarz2024retrochimera,azure_deep_research} The agent was left to decide how and when to invoke them. The natural-language prompts used to interface with Discovery are provided in the Supplementary Information, Section~\ref{sec:si-clio-design}.

\subsection{Initial design campaign}\label{subsec:initial-design}

\begin{figure}[ht]
    \centering
    \begin{overpic}[width=\textwidth]{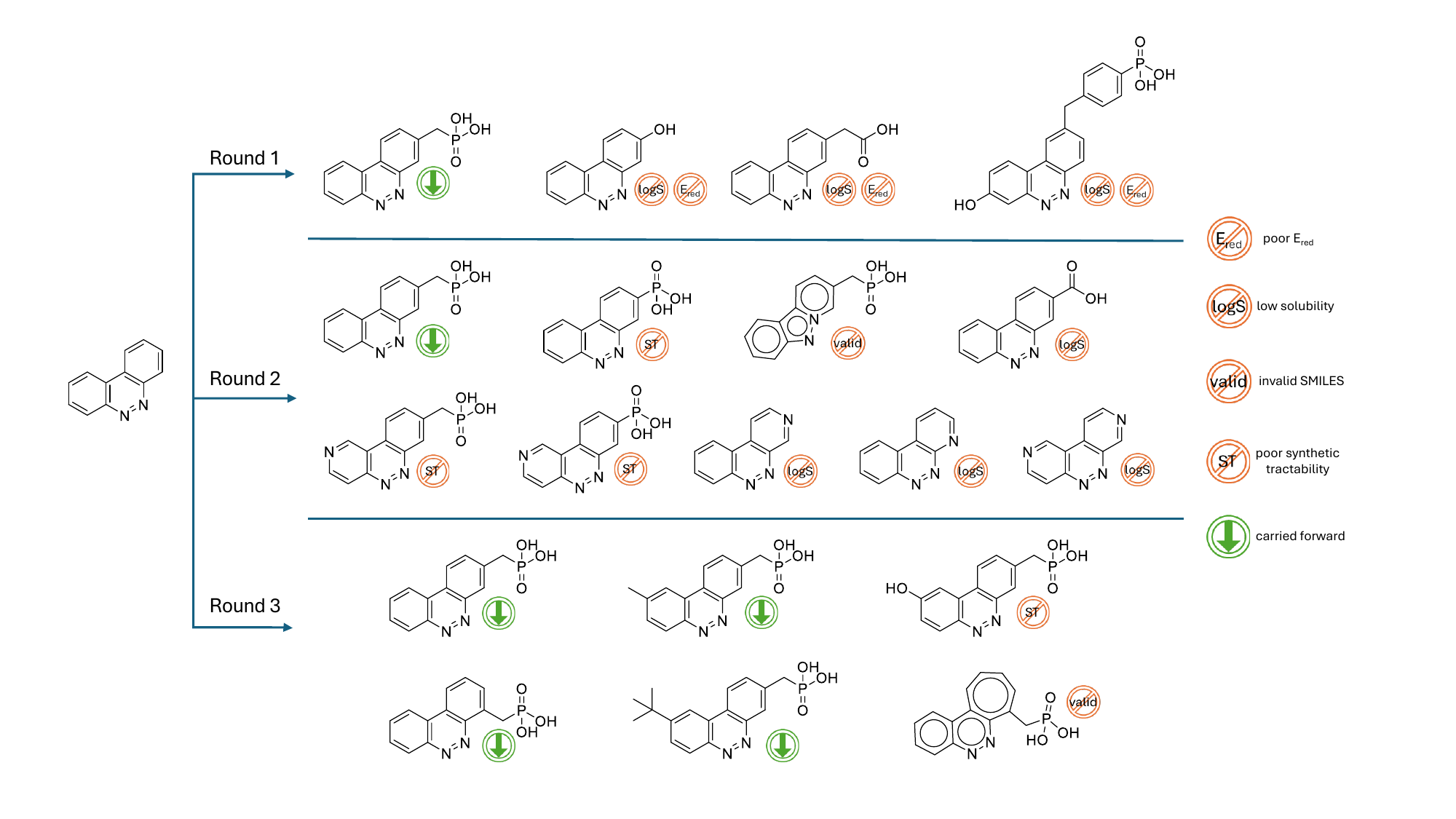}
        \put(8,25){\footnotesize\cmpd{bcc}}
        \put(27.5,40.25){\footnotesize\cmpd{bcc-BnPO3-5}}
        \put(40.5,40.25){\footnotesize\cmpd{bcc-OH}}
        \put(54,40.25){\footnotesize\cmpd{bcc-BnCOO}}
        \put(72,40.25){\footnotesize\cmpd{bcc-OH-Ph-PO3}}
        \put(27.5,29.25){\footnotesize\cmpd{bcc-BnPO3-5}}
        \put(40.5,29.25){\footnotesize\cmpd{bcc-ArPO3}}
        \put(54,29.25){\footnotesize\cmpd{bcc-imidazole-BnPO3}}
        \put(72,29.25){\footnotesize\cmpd{bcc-ArCOO}}
        \put(26,21){\footnotesize\cmpd{bcc-aza-BnPO3}}
        \put(39.5,21){\footnotesize\cmpd{bcc-aza-ArPO3}}
        \put(51,21){\footnotesize\cmpd{bcc-4-aza}}
        \put(62,21){\footnotesize\cmpd{bcc-5-aza}}
        \put(73,21){\footnotesize\cmpd{bcc-diaza}}
        \put(31,11){\footnotesize\cmpd{bcc-BnPO3-5}}
        \put(49,11){\footnotesize\cmpd{bcc-BnPO3-Me}}
        \put(66.5,11){\footnotesize\cmpd{bcc-BnPO3-OH}}
        \put(31,2){\footnotesize\cmpd{bcc-BnPO3-4}}
        \put(49,2){\footnotesize\cmpd{bcc-BnPO3-tBu}}
        \put(66.5,2){\footnotesize\cmpd{bcc-7ring-BnPO3}}
    \end{overpic}
    \caption{Design trajectory for the phosphonate negolyte campaign. Starting from the undecorated benzo[c]cinnoline scaffold, Discovery generated and triaged candidates over three rounds. Icons indicate the reason each candidate was eliminated or carried forward: poor $E_\mathrm{red}$ prediction, low predicted solubility (logS), invalid SMILES generated by LLM, or poor synthetic tractability (ST).}
    \label{fig:phosphate-trajectory}
\end{figure}

The campaign was initiated by providing Discovery with a negolyte design objective, after which it was allowed to operate in an unsupervised fashion.
Over three autonomous rounds, Discovery tackled the negolyte-design objective, adopting a diverge-then-converge strategy: first generating parallel hypothesis branches from the seed scaffold, next producing a set of exploratory structures around the best performing design, and finally converging on a small set of recommended structures derived from the top candidate (Figure~\ref{fig:phosphate-trajectory}).
Notably, Discovery was free to perform this search using the strategy of its choosing; the iterative design approach was not dictated by the objective.

To initiate the search, Discovery pursued four parallel hypothesis branches from the undecorated benzo[c]cinnoline \cmpd{bcc}:
(1)~electron-donating substituents were added to negatively shift $E_\mathrm{red}$, resulting in compound \cmpd{bcc-OH};
(2)~aza insertions were made into the flanking rings, hypothesized to improve the stability and electrochemical reversibility of the redox couple; these designs were rejected in the first round but later revisited with compounds \cmpd{bcc-aza-BnPO3}-\cmpd{bcc-diaza}.
(3)~Small modifications to the initial benzo[c]cinnoline scaffold were explored, but no suitable structures were proposed by the responsible thought channel; and
(4)~anionic solubilizing groups were appended to improve aqueous solubility at pH~14, using non-conjugated linkers to minimize electronic perturbation that would positively shift $E_\mathrm{red}$, giving compounds \cmpd{bcc-BnPO3-5} and \cmpd{bcc-BnCOO}.
Discovery combined the representatives from these branches into a small pilot set and evaluated them against the parent scaffold \cmpd{bcc}.

From this pilot set, compound \cmpd{bcc-BnPO3-5} was selected by Discovery for further exploration, primarily due to a large increase in predicted logS relative to seed compound \cmpd{bcc}, from $-3.11$ to $-1.69$.
All other members of the pilot set showed minimal change in predicted logS and were discarded.
Having met the solubility objective with this design iteration, Discovery shifted its focus to tuning $E_\mathrm{red}$ toward the target window.

The second design round aimed to shift $E_\mathrm{red}$ toward the target window while preserving solubility.
Discovery continued to directly utilize the provided $E_\mathrm{red}$ predictor, working to positively shift values from a predicted $-2.0$~V vs. SHE for most analogs.
Due to the large error in predicted reduction potential for benzo[c]cinnoline scaffolds, the design prompt provided Discovery with a rough calibration value---``derivatives of the scaffold compound have experimentally measured reduction potentials at $\sim0.7$[\emph{sic}] V versus SHE''---but to this point the predicted values were used directly.
Two design axes were explored: aza insertions and direct substitution with solubilizing electron-withdrawing handles.
Aza derivatives \cmpd{bcc-4-aza}, \cmpd{bcc-5-aza}, and \cmpd{bcc-diaza} showed directionally consistent trends in the Graphormer predictions, with positive shifts relative to the parent \cmpd{bcc} of $+0.133$~V and $+0.258$~V.
Directly substituted derivatives \cmpd{bcc-ArPO3} and \cmpd{bcc-ArCOO} showed similarly consistent shifts of $+0.133$~V and $+0.258$~V.
Observing these results, Discovery designed a set of property gates, eliminating designs that failed to show a predicted logS of -2.5 or greater and an SAscore of 3.5 or lower.
Notably, this gate excluded a predicted $E_\mathrm{red}$ metric.
Discovery had become skeptical of the predictor due to its disagreement with the calibration value provided in the design prompt, stating, ``Treat absolute $E_\mathrm{red}$ values as potentially miscalibrated for this chemotype''.

As the final step for this design round, the surviving candidates containing a solubilizing group---compounds \cmpd{bcc-BnPO3-5}, \cmpd{bcc-ArPO3}, and \cmpd{bcc-aza-BnPO3}---were evaluated using the RetroChimera tool to further triage the candidate pool.
The routes predicted for compounds \cmpd{bcc-ArPO3} and \cmpd{bcc-aza-BnPO3} required $>$5 synthetic steps with ambiguous intermediates, and these designs were deprioritized.
For compound \cmpd{bcc-BnPO3-5}, the RetroChimera tool erred, producing a two-step route from 2-(bromomethyl)quinoline that installed the defining benzylphosphonate handle via an Arbuzov reaction.
Here, Discovery failed to catch the substitution of a quinoline for the benzo[c]cinnoline, instead focusing its critique of the route on potential contaminating side-products, and provided this route with an internal ranking of ``proceed with caution''.

Judged the most synthetically tractable of the candidates, the third and final design round centered compound \cmpd{bcc-BnPO3-5}.
A set of six new candidates was produced, with Discovery including strict constraints on the allowed modifications: ``Generate ONLY 6 new candidates[\ldots] derived from benzo[c]cinnoline that are synthesizable via the proven benzylphosphonate late-stage functionalization route[\ldots] Allowed extra substituents limited to: methyl, tert-butyl[\ldots] additional phenol[\ldots]''
From this set, Discovery converged on a set of two recommended designs: the 5-benzylphosphonate \cmpd{bcc-BnPO3-5} and its positional isomer 4-benzylphosphonate \cmpd{bcc-BnPO3-4}.
Two additional compounds were suggested as follow-on candidates probing the effect of additional substitutions intended to disrupt potential electrostatic, dispersion, desolvation, induction, and exchange-repulsion (EDDIE) interactions: methyl analog \cmpd{bcc-BnPO3-Me} and \emph{tert}-butyl analog \cmpd{bcc-BnPO3-tBu}.\cite{xiao2026pistacking}

\subsubsection{Characterization of 5-benzylphosphonate \cmpd{bcc-BnPO3-5}}\label{subsubsec:characterization-bnpo3}

Once Discovery's trajectory converged to a final set of four compounds, the designs were reviewed by synthetic chemists and compound \cmpd{bcc-BnPO3-5} judged the most readily accessible and synthesized.
The reduction-wave half-peak potential of compound \cmpd{bcc-BnPO3-5}, $E_\mathrm{red}$, was measured at -0.895 V vs. SHE, 130 mV more negative than the baseline sulfonated benzocinnoline \cmpd{sBC} (Table~\ref{tab:property-comparison}).
In order to achieve this gain in $E_\mathrm{red}$, Discovery traded off the solubility of the compound, as allowed by our objective: ``A logS target around $-2$ is acceptable.''

\begin{table}[ht]
  \centering
  \caption{Comparison of predicted and measured properties for the baseline BzC compound and the two Discovery-designed candidates. Calculated from cyclic voltammograms in 0.5~M KOH with an analyte concentration of 5~mM and 50~mV/s sweep rate.}
  \label{tab:property-comparison}
  \begin{tabular}{l c c c}
      \hline
      & \cmpd{sBC}\textsuperscript{*} & \cmpd{bcc-BnPO3-5} & \cmpd{bcc-BnSO3-5} \\
      \hline
      $E_\mathrm{red}$ predicted (V vs.\ SHE) & $-1.60$ & $-2.04$ & $-2.00$ \\
      $E_\mathrm{red}$ measured (V vs.\ SHE)  & $-0.762$\textsuperscript{\&} & $-0.895$ & $-0.854$ \\
      Solubility (pH~7) predicted (mM)        & 52.5 & 20.4 & 14.5 \\
      Solubility (0.5~M KOH) (mM)            & 500\textsuperscript{\#} & 57.9 & 56.0 \\
      $|i_\mathrm{ox}/i_\mathrm{red}|$\textsuperscript{@} & 0.52\textsuperscript{\&} & 0.18 & 0.38 \\
      $Q_\mathrm{ox}/Q_\mathrm{red}$\textsuperscript{\textdagger} & 0.79\textsuperscript{\&} & 0.92 & 0.84 \\
      \hline
  \end{tabular}
  \vspace{2pt}

  \textsuperscript{*}Predicted values for the ortho/ortho-substituted component. \\
  \textsuperscript{\#}Literature value from Singh et al.~\cite{singh2025bc} \\
  \textsuperscript{\&}Calculated from cyclic voltammogram reported Singh et al.~\cite{singh2025bc} \\
  \textsuperscript{@}Ratio of the anodic peak current nearest the reduction wave to the cathodic peak current. \\
  \textsuperscript{\textdagger}Charge ratio by semi-integration of the baseline-corrected voltammogram.
\end{table}

\begin{figure}[p]
  \centering
  \begin{overpic}[width=\textwidth]{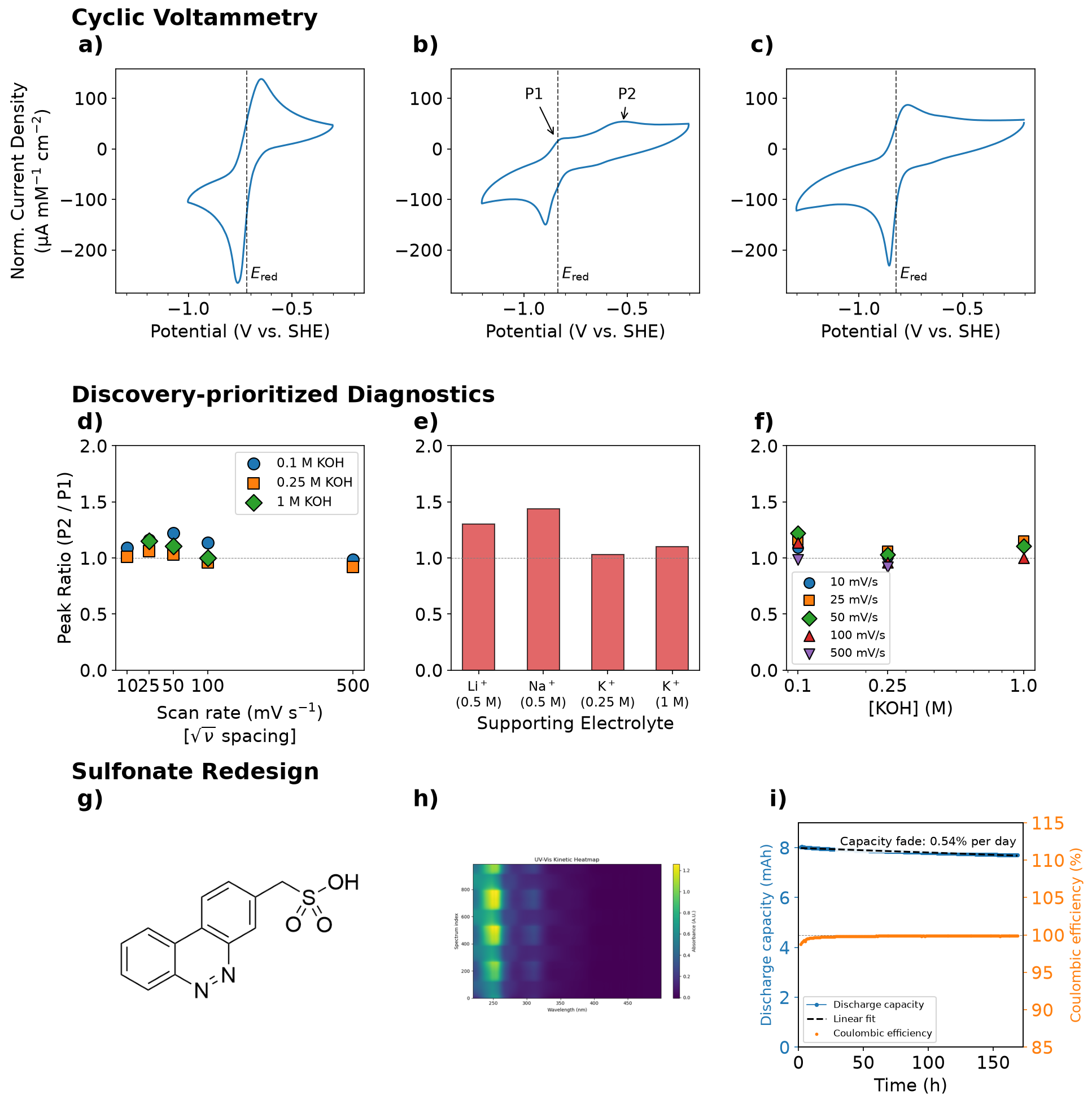}
    \put(22,7){\makebox(0,0){\footnotesize\cmpd{bcc-BnSO3-5}}}
  \end{overpic}
  \caption{\textbf{Cyclic Voltammetry.} Cyclic voltammograms in KOH at 50~mV\,s$^{-1}$, normalized by analyte concentration and electrode area, for (a)~the baseline compound \cmpd{sBC} (5~mM, 0.5~M KOH), (b)~compound \cmpd{bcc-BnPO3-5} (1~mM, 1~M KOH) with oxidation peaks P1 and P2 indicated, and (c)~compound \cmpd{bcc-BnSO3-5} (1~mM, 1~M KOH). \textbf{Discovery-Prioritized Diagnostics.} (d)~Oxidation peak ratio P2/P1 as a function of scan rate (10--500~mV\,s$^{-1}$) in 0.1, 0.25, and 1~M KOH. (e)~Cation dependence of the oxidation peak ratio P2/P1 at 50~mV\,s$^{-1}$ in LiOH, NaOH, and KOH supporting electrolytes. (f)~Oxidation peak ratio P2/P1 for \cmpd{bcc-BnPO3-5} as a function of KOH concentration (0.1--1~M) at scan rates from 10 to 500~mV\,s$^{-1}$, determined from baseline-corrected anodic scans. \textbf{Sulfonate Redesign.} (g)~Structure of \cmpd{bcc-BnSO3-5}. (h)~UV-Vis spectroelectrochemistry kinetic heatmap of \cmpd{bcc-BnSO3-5} (0.02~mM, 5~mV\,s$^{-1}$) showing absorbance as a function of wavelength and scan progression. (i)~Discharge capacity (blue, left axis) and coulombic efficiency (orange, right axis) during asymmetrical cycling of \cmpd{bcc-BnSO3-5} against ferri-/ferrocyanide in 0.5~M KOH. The dashed black line is a linear fit to discharge capacity, corresponding to a capacity fade of 0.54\% per day relative to the first complete-cycle discharge capacity.}
  \label{fig:story-cv}
\end{figure}

Interestingly, cyclic voltammetry of compound \cmpd{bcc-BnPO3-5} at pH~14 revealed a clean, well-defined reduction wave with a single peak but a split oxidation return wave with two overlapping peaks P1 and P2 (Figure~\ref{fig:story-cv}b).
Splitting of the oxidation wave was indicative of poor electrochemically reversibility, and was not observed in cyclic voltammograms taken in either 1~M H\textsubscript{2}SO\textsubscript{4} or pH~7 phosphate buffer (Figure~\ref{fig:cams013-cv}).
This regression was notably not captured by the set of numerical predictors available to Discovery, as none encode the pH-dependent mechanistic pathways presumably responsible for the behavior.
We sought to better understand it.

\subsection{Iteration}\label{subsec:design-revision}

While we were happy to have pushed $E_\mathrm{red}$ to a more extreme value within the aqueous solvent window in the first design iteration, the poor electrochemical reversibility revealed by experimental characterization represented a regression from the previously reported \cmpd{sBC}.
We set out to understand and ameliorate this behavior, leaning more heavily on Discovery's ability to leverage chemical intuition and literature search to affect scientific reasoning.
Discovery was given cyclic voltammograms of \cmpd{bcc-BnPO3-5} at pH~1, 7, and 14 (represented as a PNG image), a qualitative description of the solubility at each pH, and a rough evaluation of the results produced by a human chemist.
With this information, Discovery was asked to provide an explanation for the change in apparent irreversibility from the baseline compound \cmpd{sBC}.

\subsubsection{Experimental Diagnostics}\label{subsubsec:experimental-diagnostics}
Three Discovery analyses were commissioned in parallel to exploit LLM stochasticity and their recommendations aggregated by summarization with an independent LLM.
All three converged on a similar mechanistic picture: the split oxidation wave reflects a pathway in which a base-dependent chemical step after reduction produces two distinct oxidizable populations.
Three competing hypotheses were advanced across the analyses:
(1) direct hydroxide addition to the reduced benzocinnoline core;
(2) base-promoted tautomerization partitioning the reduced state into prototropic microstates; and
(3)~phosphonate--K$^+$ ion-pairing effects stabilizing distinct reduced-state microenvironments with different oxidation kinetics.

To discriminate between these hypotheses, Discovery proposed a battery of follow-up experiments, summarized in Table~\ref{tab:followup-experiments}.
The first group asks whether a chemical reaction is occurring after the initial electron transfer, using variations in cyclic voltammetry timing and repetition to amplify or suppress any such step.\cite{saveant2019elements}
The second group probes the role of the potassium counter-ion and hydroxide anion of the supporting electrolyte by substituting different cations, adjusting the concentrations of the respective ions, and using isotope effects to investigate the role of proton-transfer steps.
The third group seeks to directly identify any new species formed, using mass spectrometry and spectroscopy on electrolyzed samples.

Among the proposed experiments, each of the three Discovery analyses produced a highest-priority experiment (starred in Table~\ref{tab:followup-experiments}).
The three experiments prioritized by Discovery were carried out, and the ratio between the two peaks in the oxidation wave, P2/P1, extracted (Figure~\ref{fig:story-cv}d-f).
P2/P1 remained near unity across a wide range of scan rates (10--500~mV\,s$^{-1}$) and KOH concentrations (0.1--1~M), providing evidence against a slow chemical follow-up step.
However, replacing K$^+$ with the smaller, harder Li$^+$ or Na$^+$ cations increased the P2/P1 ratio, consistent with stronger cation--phosphonate association stabilizing the reduced state and altering the energetic landscape of the return oxidation.\cite{Popov2002}

\begin{table}[t]
  \centering
  \caption{Follow-up experiments proposed by Discovery to resolve the origin of the split oxidation wave observed for compound \cmpd{bcc-BnPO3-5} in 1~M KOH. Experiments are grouped by the mechanistic hypothesis they primarily address. Consensus indicates the number of independent Discovery analyses (out of three) that recommended each experiment; a star ($\bigstar$) marks experiments that were ranked as the top priority by Discovery.}
  \label{tab:followup-experiments}
  \small
  \setlength{\tabcolsep}{4pt}
  \begin{tabular}{@{}p{3.8cm}ccp{6.2cm}@{}}
    \hline
    Experiment & Consensus & Priority & Diagnostic rationale \\
    \hline
    \multicolumn{4}{@{}l@{}}{\textit{Group 1: Probe for chemical follow-up}} \\
    Wide scan-rate series        & 3/3 & $\bigstar$ & Chemical step suppressed at fast rates; split should collapse \\
    Reductive pre-hold           & 3/3 & --- & Extended hold amplifies chemical step; split grows with hold time \\
    Multi-cycle overlays         & 3/3 & --- & If the split grows cycle-to-cycle, a new species is accumulating irreversibly \\[4pt]
    \multicolumn{4}{@{}l@{}}{\textit{Group 2: Ion-pairing and OH\textsuperscript{$-$} chemistry}} \\
    Cation swap & 3/3 & $\bigstar$ & Cation dependence supports ion-pair model \\
    Crown ether addition         & 1/3 & --- & Collapse of split upon K\textsuperscript{+} sequestration confirms pairing \\
    {OH\textsuperscript{$-$}} concentration series & 3/3 & $\bigstar$ & Base-dependence supports OH\textsuperscript{$-$} involvement \\
    KOH/KOD isotope effect   & 1/3 & --- & Isotope effect implicates proton-transfer/tautomerization step \\[4pt]
    \multicolumn{4}{@{}l@{}}{\textit{Group 3: Product identification}} \\
    Bulk electrolysis & 3/3 & --- & Direct detection of hydroxylation or rearrangement products \\
    Spectroelectrochemistry       & 1/3 & --- & \emph{In situ} characterization of reduced-state speciation \\
    \hline
  \end{tabular}
\end{table}

\subsubsection{Design Revision}\label{subusbsec:design-revision}
Beyond diagnostic experiments, Discovery also proposed structural modifications to address the liability.
The highest-confidence recommendation, consistent across all three analyses, was to replace the benzylphosphonate solubilizing group with a sulfonate handle, which is expected to produce weaker cation pairing and thereby suppress any microstate-driven splitting of the oxidation peak.
Additional strategies suggested included altering the linker between the phosphonate and the aromatic core and introducing substituents at positions on the benzocinnoline core most susceptible to nucleophilic attack in the reduced state.
The diagnostic experiments demonstrated that phosphonate--cation binding affected the oxidation peak ratio P2/P1 and did not produce evidence for a hydroxide-involved chemical step.
We therefore proceeded to synthesize sulfonate \cmpd{bcc-BnSO3-5} for electrochemical characterization in accordance with Discovery's recommendation.

The half-peak potential of \cmpd{bcc-BnSO3-5} was measured at -0.854 V vs. SHE, which is slightly more positive than the -0.895 V vs. SHE of \cmpd{bcc-BnPO3-5} but 92 mV more negative than that of \cmpd{sBC} (Table~\ref{tab:property-comparison}).
The solubility of \cmpd{bcc-BnSO3-5} in 0.5 M~KOH was measured to be 56.0~mM, similar to that of \cmpd{bcc-BnPO3-5}, 57.9~mM.

In line with Discovery's prediction, cyclic voltammetry of \cmpd{bcc-BnSO3-5} showed an improvement in electrochemical reversibility (Figure~\ref{fig:story-cv}c) relative to \cmpd{sBC}.
With 1~mM \cmpd{bcc-BnSO3-5} in 1~M KOH, a single rather than split oxidation process was observed, with a peak separation of 90~mV, slightly lower than the 114~mV peak separation observed for \cmpd{sBC}. The peak current ratio $|i_{ox}/i_{red}|$ of the target redox couple was 0.38, up from 0.18 for \cmpd{bcc-BnPO3-5} and approaching the 0.52 ratio of \cmpd{sBC}.

Voltammetry of higher concentrations of \cmpd{bcc-BnSO3-5} however revealed a secondary oxidation wave centered around -0.55 V vs SHE. Figure~\ref{fig:cams014-conc-dependence} presents voltammograms at 0.01~mM --- where a single, relatively sharp peak was observed --- and at 5~mM, where a differentiated secondary wave is clearly present. This behavior is consistent with aggregation of \cmpd{bcc-BnSO3-5} at higher concentrations, a potential issue flagged in both the initial Discovery design trajectory---with the anticipation of EDDIE interactions---and the Discovery design revisions. Molecular aggregation/dimerization with increasing concentration has been similarly invoked in the literature for sulfonated anthraquinones.\cite{wiberg2019dimerization,carney2017dimerization}

To more thoroughly assess the electrochemical reversibility of \cmpd{bcc-BnSO3-5}, we performed spectroelectrochemistry, coupling UV-Vis absorption monitoring with slow-scan cyclic voltammetry (Figure~\ref{fig:story-cv}h-i).
If the split oxidation wave observed for the phosphonate \cmpd{bcc-BnPO3-5} were driven by irreversible decomposition---for example, hydroxide addition to the reduced core---one would expect the emergence of new chromophores or a progressive loss of the parent absorption band over the course of a redox cycle.
The cycle-dependent absorbance spectra at both 5~mV\,s$^{-1}$ (0.02~mM) and 25~mV\,s$^{-1}$ (0.01~mM) show clean, reversible spectral changes: absorption features that develop during the cathodic sweep recover fully on the return sweep, with no residual bands attributable to decomposition products (Figure~\ref{fig:spectroec}).
This is consistent with the diagnostic electrochemistry, which pointed to a cation-pairing mechanism rather than a destructive chemical follow-up step.

Finally, we cycled \cmpd{bcc-BnSO3-5} in a flow cell to explore its chemical stability over long durations.
Cycling 25 mM of \cmpd{bcc-BnSO3-5} for 170 h under alkaline conditions (0.5 M KOH) in a compositionally asymmetric congifuration against potassium ferrocyanide/ferricyanide yielded a fade rate of 0.50\% per day (Table \ref{tab:cell-cycling-summary}).
Cycling under compositionally symmetric conditions yielded comparable fade rates of 0.40\% per day and 0.64\% per day under alkaline (0.5 M KOH) and pH-neutral conditions, respectively.  
These symmetric-cell fade rates are slightly lower than the 0.74\% per day rate reported for analogous alkaline cells containing \cmpd{sBC}.\cite{singh2025bc}

All cycled \cmpd{bcc-BnSO3-5} electrolytes were characterized using \textsuperscript{1}H NMR spectroscopy, cyclic voltammetry, and UV-vis spectroscopy (Figure~\ref{fig:symmetrical_cycling_KOH} - \ref{fig:asymmetrical_cycling_KOH}). We observed no clear evidence with these techniques for molecular decomposition or loss of redox activity, indicating that Discovery successfully achieved the heuristically reasoned objective assigned to it, i.e., avoidance of known decomposition pathways. We tentatively assign the apparent loss of capacity to some combination of precipitation due to water loss and a state-of-charge imbalance either from the hypothesized aggregation process or parasitic oxygen/hydrogen evolution.\cite{fell2023stability} Future studies will address this question in more detail.

\begin{table}[t]
  \centering
  \caption{Flow-cell cycling experiments with \cmpd{bcc-BnSO3-5}.}
  \label{tab:cell-cycling-summary}
  \small
  \setlength{\tabcolsep}{4pt}
  \begin{tabular}{@{}p{3.0cm}p{3.1cm}p{3.1cm}p{3.1cm}@{}}
    \hline
    Parameter & Alkaline asymmetric & Alkaline symmetric & Neutral symmetric \\
    \hline
    Negolyte
      & 25~mM \cmpd{bcc-BnSO3-5}
      & 30~mM \cmpd{bcc-BnSO3-5}
      & 25~mM \cmpd{bcc-BnSO3-5} \\[4pt]
    Posolyte
      & 15~mM K$_3$[Fe(CN)$_6$] \newline 15~mM K$_4$[Fe(CN)$_6$]
      & 30~mM \cmpd{bcc-BnSO3-5}
      & 25~mM \cmpd{bcc-BnSO3-5} \\[4pt]
    Supporting\newline electrolyte
      & 0.5~M KOH
      & 0.5~M KOH
      & 0.25~M pH~7 \newline phosphate buffer \newline 0.75~M KCl \\[4pt]
    Voltage cutoffs
      & $[+0.80, +1.40~V]$
      & $[-0.40, +0.40~V]$
      & $[-0.40, +0.40~V]$ \\[4pt]
    Cycling duration
      & 170~h
      & 250~h
      & 310~h \\[4pt]
    Capacity fade rate
      & 0.54~\%~day$^{-1}$
      & 0.41~\%~day$^{-1}$
      & 0.62~\%~day$^{-1}$ \\
    \hline
  \end{tabular}
\end{table}

\section{Discussion}\label{sec:discussion}

\subsection{Configuration of Discovery}\label{subsec:discovery-engine-updates}

Discovery Engine, built on the Cognitive Loop via In-Situ Optimization framework, is a recursive, inference-time optimization system in which a large-language model evolves its knowledge and approach when solving open-ended design problems.\cite{cheng2025clio} 
In its original form, Discovery Engine was evaluated on closed-form scientific questions from Humanity's Last Exam, where the answer exists \emph{a priori} and the reasoning chain terminates in a single session without external tool access.\cite{humanitys_last_exam}
Molecular design imposes three requirements that this formulation does not satisfy: (i)~Discovery must ground its reasoning in quantitative evidence obtained from domain-specific computational tools, not solely from knowledge embedded in its parameters, while remaining able to recognize when those tools should be distrusted or down-weighted in light of accumulating evidence; (ii)~proposed designs must be realizable in the laboratory, requiring Discovery to reason about synthetic tractability---a property for which computational tools exist (retrosynthetic analyzers, synthetic accessibility scorers) but whose outputs are approximate and often require exercising chemical judgment beyond what the scores alone can provide; and (iii)~experimental feedback on synthesized compounds may contradict computational predictions, demanding that Discovery interpret discrepancies, revise its mechanistic understanding, and propose corrective modifications---completing the full design--make--test--learn cycle rather than terminating at a ranked candidate list.
To meet the challenges of molecular design, we pair Discovery Engine with a persistent memory structure, Discovery Bookshelf, that is used both to track the provenance of conclusions and to enable longitudinal reasoning across rounds.
This advances Discovery Engine's original architecture, where the belief-state graph was constructed only \emph{post hoc}---to resolve disputes between competing hypotheses---rather than maintained online as a structural semantic guide for Discovery Engine's reasoning.
As the campaign unfolded, we found that this extension was critical for the Discovery system to effectively manage its own reasoning across extended design campaigns, and to leverage the domain knowledge encoded in its underlying LLM to interpret experimental feedback and propose corrective modifications when computational tools proved insufficient.

\begin{figure}[t]
  \centering
  \makebox[\textwidth][c]{\includegraphics[width=1.2\textwidth, trim=50 0 50 0, clip]{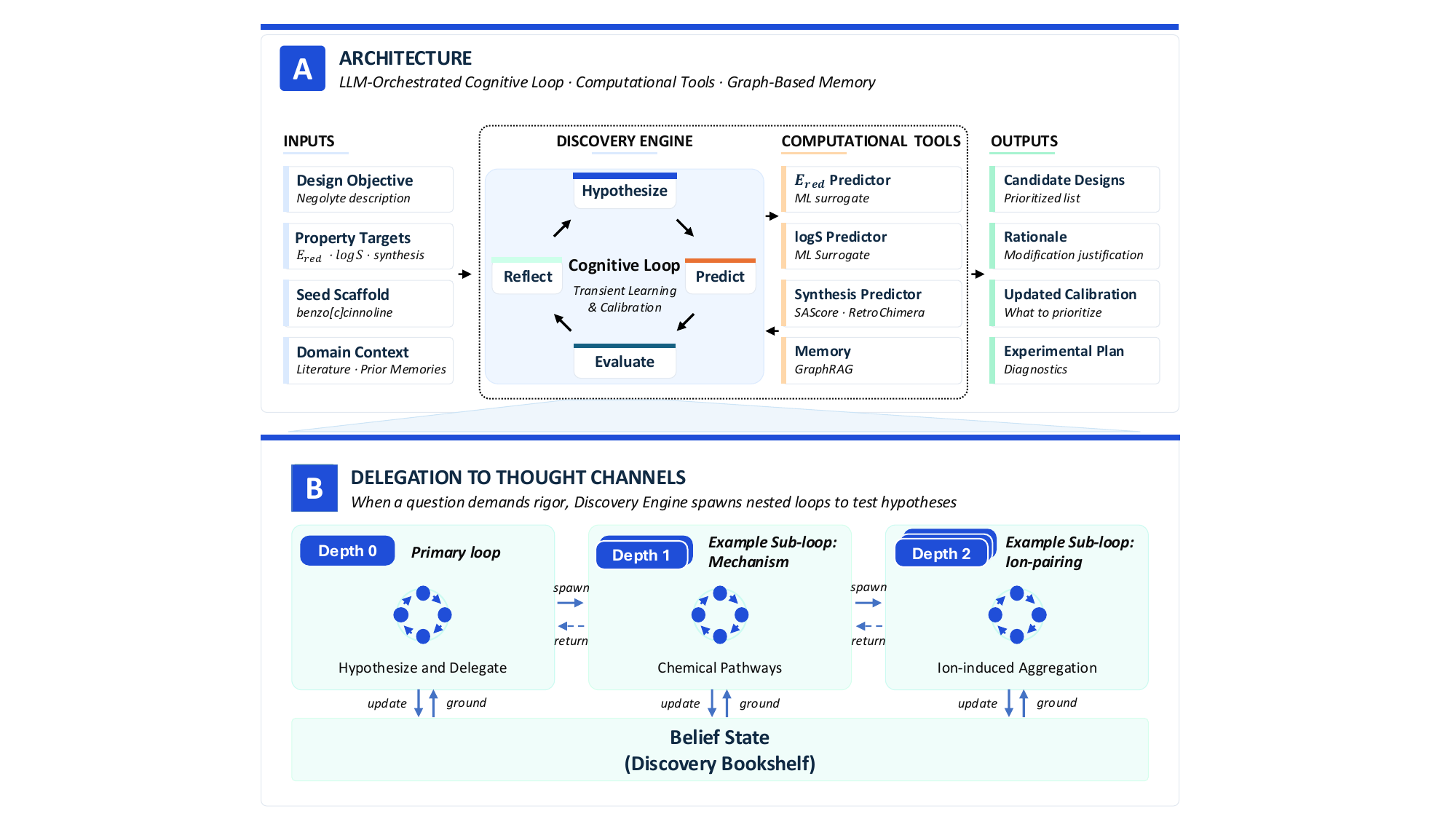}}
  \caption{Overview of Discovery.
  \textbf{(A)}~Architecture as configured for the negolyte design campaign.
  The scientist sets objectives and provides experimental feedback, and the agent
  drives the reasoning loop: selecting tools, spawning parallel thought channels, and
  managing its own context across extended sessions.
  Computational tools represent the domain-specific capabilities described in Section~\ref{subsec:problem-formulation}.
  \textbf{(B)}~Recursive cognitive loops with shared belief state that enable multi-scale planning and reflection.}
  \label{fig:discovery-overview}
\end{figure}

\subsection{Impact on agentic science}\label{subsec:agentic-science}
Recent work has demonstrated that LLM-based agents can automate substantial portions of the scientific workflow: planning multi-step experiments, executing them through robotic or computational infrastructure, and synthesizing results into reports.\cite{boiko2023coscientist,roohani2024biodiscovery,swanson2025virtuallab}
These systems are powerful, but their contributions are predominantly operational---they accelerate the execution of workflows that a human scientist has already designed or could design.
The campaign reported here suggests that a reasoning agent can contribute something qualitatively different by leveraging calibrated deference: the capacity to recognize when its own tools or assumptions are failing, to adapt its strategy in response, and to generate mechanistic hypotheses that guide experimental revision.

Three episodes from this campaign illustrate calibrated deference in practice.
First, Discovery progressively refined its assessment of the Graphormer $E_\mathrm{red}$ model over three design rounds: from uncritical acceptance, through detection of a $\sim$2.7~V discrepancy and salvage of rank-order utility, to deliberate exclusion of $E_\mathrm{red}$ from its property gates---a graded response characteristic of how a practicing scientist manages imperfect models.
Second, Discovery decomposed the design space into four orthogonal hypothesis branches explored in parallel, and adapted in real time when LLM rate limits and a missing tool forced operational constraints.
Third, when interpreting the split oxidation wave of \cmpd{bcc-BnPO3-5}, Discovery produced tiered mechanistic hypotheses with explicit epistemic qualifications, identified discriminating experiments, and derived the sulfonate redesign as a falsifiable prediction from its top-ranked mechanism.
A broader discussion of calibrated deference, including an analysis of observed limitations, may be found in the supplementary information.

The results point toward natural-language reasoning as a beneficial complement to quantitative prediction.
Discovery demonstrated the capability for problem solving and data interpretation outside of its predefined toolset, allowing ultimately for the design of benzo[c]cinnoline derivative \cmpd{bcc-BnSO3-5} through diagnosis and remediation of the cation-dependent aggregation uncovered during characterization of its phosphonate predecessor \cmpd{bcc-BnPO3-5}.
We envision such design campaigns eventually expanding beyond the initial design stage to include consideration of broader factors like manufacturability, distribution, and market analysis, and in these areas the ability to leverage natural-language reasoning will be all the more important.
Systems like Discovery open the door to a new paradigm of agentic science in which human-AI collaboration leverages complementary strengths to explore complex, multidisciplinary scientific and technological landscapes that neither could navigate alone.

\section{Methods}\label{sec:methods}

\subsection{Discovery Engine architecture}\label{subsec:methods-discovery-engine}

Discovery Engine operates in design \emph{rounds}: each round is a hypothesize--interpret cycle in which Discovery Engine enters with the current goal and any new experimental results from scientists, reasons over them---proposing candidates, interpreting data, dispatching specialist agents---and exits when converged. Synthesis, characterization, and the diagnostic experiments themselves happen between rounds, and are executed by the scientists; the full scientific-method iteration therefore spans a round boundary. For the initial benzocinnoline design campaign (Section~\ref{subsec:initial-design}), Discovery Engine was configured in a \emph{cold-start} state, with no prior information about the problem other than what was provided in the system prompt and the goal, and no pre-populated memory. In the subsequent phosphonate-to-sulfonate revision (Section~\ref{subsec:design-revision}), Discovery Engine started in a \emph{warm-start} state, with its memory seeded by the prior campaign, but with no direct injection of the prior chat history or experimental results into the context. The next round therefore enters with no working context other than the system prompt, the goal, and context Discovery Engine elects to retrieve from its memory. The human scientist may inject free-text feedback at this boundary---experimental results, directives, or revisions to the goal---steering Discovery Engine's next round of reasoning.

Within a round, Discovery Engine orchestrates design modifications using a set of specialist agents. Each agent is itself an LLM-backed loop with its own persona, its own access to the domain-specific tool suite (Section~\ref{subsec:methods-tools}), and shared access to Discovery Engine's memory. The number of parallel thought channels spawned in a round is not fixed: Discovery Engine decides at each planning step how to decompose the current question, guided by system-prompt heuristics that favor spawning a new channel when (i)~a sub-question is operationally independent of the others (e.g., a distinct hypothesis branch or a separate diagnostic), (ii)~it would otherwise contaminate the parent context with intermediate reasoning, or (iii)~it requires a different tool or persona than the parent loop. In the campaign reported here, this produced four channels in the initial diverge phase (one per hypothesis branch) and a smaller number in subsequent rounds as the search converged. In the negolyte campaign, we designed three agents for Discovery Engine to orchestrate: a \textbf{SmallMoleculeDesignAndGenerationAgent} that takes a parent SMILES and a target purpose, analyzes which properties are most important to optimize, proposes literature-grounded structural modifications, and iterates on the most promising candidates while exploring diverse regions of chemical space; a \textbf{FitnessFunctionAgent} that evaluates candidate molecules against the design criteria using their computed properties from the Graphormer predictors and RDKit, returning a ranked triage of leads with explicit rejection reasons; and a \textbf{RetrosynthesisAnalysisAgent} that assesses synthetic feasibility and manufacturability by calling RetroChimera to plan multi-step routes from the Sigma-Aldrich catalogue, translating individual steps into actionable procedures, and flagging route-level risks such as hazardous reagents, purification burden, and electrochemistry-grade impurity controls. The verbatim persona descriptions for all subagents are available in Section~\ref{sec:si-clio-design}.

Across rounds, Discovery Engine manages its own context by selectively retrieving information from this memory and deciding what to include in the prompt for the next round. Today's memory systems for LLM agents are designed for verbatim recall of prior information, but the design process is more effectively served by thematic abstraction over the agent's own reasoning trace. The belief-state structure of Discovery Engine's memory (Section~\ref{subsec:discovery-engine-updates}) is built to occupy this gap, enabling Discovery Engine to track the provenance of its conclusions and to query for thematic abstractions over that trace, not for verbatim recall.

\subsection{Domain-specific tools}\label{subsec:methods-tools}

Discovery was provided access to a set of domain-specific tools, where each tool was wrapped in a FastAPI interface.\cite{fastapi}
Two machine-learning property predictors based on the Graphormer architecture were provided for reduction potential ($E_\mathrm{red}$) and aqueous solubility (logS), trained as described by Martinez-Baez et al.~\cite{martinezbaez2021screening}
The synthetic accessibility score metric (SAscore) was provided as implemented in RDKit.~\cite{ertl2009sascore,rdkit}
The RetroChimera retrosynthesis prediction tool was provided, with the compound inventory set to the historical Sigma-Aldrich catalog archived in the ZINC 15 dataset.~\cite{sterling2015zinc,maziarz2024retrochimera}
During experimental diagnosis and redesign, access to web search was provided with the Azure Foundry Deep Research tool.~\cite{azure_deep_research}

\subsection{Cyclic voltammetry}\label{subsec:methods-cv}
Cyclic voltammetry (CV) was performed using a 1~mM or 5~mM concentration with a supporting electrolyte of 0.5~M KOH, unless specified otherwise.
A 5 mm diameter glassy carbon disk electrode (BASi Inc.), Ag/AgCl reference electrode (BASi Inc.) and a Pt counter electrode wire (BASi Inc.) were used in a three-electrode configuration.
CV measurements were collected using a CHI7013E potentiostat with an 85\% iR compensation. 

The glassy carbon electrode was polished with an alumina slurry solution on a polishing pad using a figure-eight motion for one minute and rinsed thoroughly with deionized water.
CVs were typically collected at scan rates of 100, 50, and 25 mV s\textsuperscript{-1}, with additional scan rates performed where noted.
For spectroelectrochemistry measurements, the scan rate was either 5~mV\,s\textsuperscript{-1} or 25~mV\,s\textsuperscript{-1}.

\subsection{Diagnostic electrochemistry}\label{subsec:methods-diagnostics}

\subsubsection{Wide scan-rate series.}
Cyclic voltammograms of 1~mM \cmpd{bcc-BnPO3-5} were recorded at scan rates from 10 to 500~mV\,s$^{-1}$ in 0.1, 0.25, and 1~M KOH.
The return-scan oxidation wave was baseline-corrected using an adaptive linear fit and peak currents from the two major oxidation peaks extracted to give the P2/P1 ratio (Figure~\ref{fig:story-cv}d).

\subsubsection{Cation swap series.}
Cyclic voltammograms were recorded in 0.5~M LiOH and 0.5~M NaOH to test whether cation--phosphonate ion pairing contributes to the split oxidation wave.
The P2/P1 ratio and $E_{\mathrm{red,1/2}}$ were compared at 50~mV\,s$^{-1}$ for each electrolyte (Figure~\ref{fig:story-cv}e).
Additional effects of the cation series are observed in the voltammograms, but these are difficult to deconvolve from the effect of Debye screening (Figures~\ref{fig:lioh-cv-overlay}--\ref{fig:naoh-cv-overlay}).\cite{Dickinson2009}

\subsubsection{Hydroxide concentration series.}
The KOH concentration was varied from 0.1 to 1~M while holding the \cmpd{bcc-BnPO3-5} concentration at 1~mM and the scan rate at 50~mV\,s$^{-1}$ (Figure~\ref{fig:story-cv}f).
The return-scan oxidation wave was baseline-corrected and the P2/P1 ratio extracted across scan rates and electrolyte conditions.
This behavior was reproduced at additional scan rates (25 and 100~mV\,s$^{-1}$; Figures~\ref{fig:koh-sweep-25mVs}--\ref{fig:koh-sweep-100mVs}).

\subsection{Flow-cell cycling}\label{subsec:cell-cycling}

Electrochemical cycling was performed using a BioLogic VMP-300 potentiostat. A constant-current/constant-voltage (CCCV) cycling protocol was applied with a constant current density of $4\,\mathrm{mA\,cm^{-2}}$ ($20\,\mathrm{mA}$ applied over a geometric electrode area of $5\,\mathrm{cm^2}$). Experiment-specific voltage cutoffs are specified below.

\subsubsection{Compositionally asymmetric cell cycling}
A compositionally asymmetric cell was constructed with 6 mL of 25 mM \cmpd{bcc-BnSO3-5} in 0.5 M KOH against a 60 mL posolyte comprising 15 mM each of K$_3$[Fe(CN)$_6$] and K$_4$[Fe(CN)$_6$] in 0.5 M KOH.
The two compartments of the cell were separated by a Fumasep FKS-50 cation-exchange membrane. The cell was cycled for 170~h, with voltage cutoffs of 0.80 and 1.40 V during long-term cycling. 

\subsubsection{Compositionally symmetric cell cycling}
\cmpd{bcc-BnSO3-5} was evaluated in a volumetrically unbalanced, compositionally symmetric cell.
In this cell configuration, the same redox-active molecule is present in its oxidized and reduced forms on both sides of the cell for equal durations, largely mitigating capacity fade induced by crossover.\cite{goulet2018stability, potash2016benefits} Different electrolyte volumes were used to establish a capacity-limiting side (CLS) and a non-capacity-limiting side (NCLS). The compartments were separated by a Fumasep FKS-50 cation-exchange membrane, as used in the previous asymmetrical experiment. Both compartments were initially prepared at 50\% state of charge (SOC). 

For alkaline cycling, both compartments contained 30~mM of \cmpd{bcc-BnSO3-5} in 0.5~M KOH, with a CLS volume of 5~mL and 8~mL on the NCLS that were cycled for 250~h. A replicate experiment was performed using 25~mM of \cmpd{bcc-BnSO3-5} in 0.5~M KOH, with electrolyte volumes of 6~mL on the CLS and 10~mL on the NCLS.

For neutral-pH cycling, the two compartments contained 25~mM \cmpd{bcc-BnSO3-5} in 0.25~M phosphate buffer and 0.75~M KCl. The supporting electrolyte solution was carefully adjusted to pH~7 using 1~M KOH. Electrolyte volumes were 4~mL on the CLS and 8~mL on the NCLS. 

All compositionally symmetrical cells were evaluated with the CCCV protocol described above, with voltage cutoffs of -0.4 and 0.4~V during long-term cycling.

\backmatter

\bmhead{Supplementary information}

Supplementary Information accompanies this paper and includes:
the design prompt provided to Discovery;
an extended discussion of calibrated deference with analysis of limitations;
a comparison with related agentic and optimization systems;
a complete inventory of hypotheses advanced by Discovery during the campaign with their resolution status;
a controlled comparison of Discovery against LLM-based genetic optimizer ExLLM\cite{ran2025exllm} on a strictly numerical optimization task, with experimental characterization of the resulting structures;
electrochemical characterization data including cyclic voltammetry, scan-rate dependence, cation and hydroxide concentration series, and half-peak potential calculations;
spectroelectrochemistry data;
membrane permeability measurements;
solubility measurements by NMR and UV-Vis spectroscopy;
and synthetic procedures for all intermediate and final compounds.

\bmhead{Acknowledgements}

We gratefully acknowledge Marwin Segler, Yuan-Jyue Chen, Chi Chen, Danrong Zhang, Lili Cheng, Desney Tan, Eric Horvitz, Nathan Baker, Jason Zander, and the broader Microsoft Research team for their invaluable discussions, support, and feedback on this project.

This work was assisted by AI tools in generating text/data analysis. The final version reflects the authors’ original work and has been reviewed and approved by all authors.

\section*{Declarations}

\begin{itemize}
\item Funding: This work was funded by Microsoft Corporation. 
\item Conflict of interest/Competing interests: K.S., B.H.N., J.A.S, N.C., W.C., M.S., D.Z., and N.G. are or were employees of Microsoft Corporation. S.M.A.H. and F.U. are employees of Can Am Bioresearch. G.B., D.K., and J.D. declare no competing interests.
\item Data availability: Experimental data generated in this study are available in the Supplementary Information. Additional data and analysis scripts are available from the corresponding authors upon reasonable request.
\item Code availability: Code is available at \url{https://github.com/microsoft/orfb-design} and \url{https://github.com/microsoft/discovery}.
\item Author contribution:
J.A.S., B.H.N., K.S., N.C., G.B. and W.C. conceived the project.
J.A.S., B.H.N., N.C. and G.B. contributed to software development.
N.C. and G.B. designed and implemented the Discovery agent architecture.
J.A.S. and B.H.N. designed the domain-specific prompts and design criteria.
G.B., N.G., and D.Z. performed initial exploration of LLMs for molecular design.
J.A.S. ran comparisons between ExLLM and Discovery.
M.S. wrapped the tools in a FastAPI interface.
S.M.A.H. and F.U. synthesized the compounds.
J.D. performed electrochemical characterization and validation under the supervision of D.K.
J.A.S., B.H.N., N.C., G.B., D.K., and J.D. analyzed the data and interpreted the results.
E.F.W., J.-H. S., Y.L., R.F., and W.W. independently replicated synthesis and cyclic voltammetry.
J.A.S., B.H.N., N.C., and G.B. wrote the manuscript with input from all authors.
All authors read and approved the final manuscript.
\end{itemize}

\begin{appendices}

\end{appendices}


\bibliography{sn-bibliography}

\input{sn-supplementary}

\end{document}

%% file: compounds.tex
\makeatletter
\newcommand{\defcmpd}[2]{%
  \expandafter\newcommand\csname cmpd@#1\endcsname{\textbf{#2}}%
}
\newcommand{\cmpd}[1]{%
  \@ifundefined{cmpd@#1}%
    {\textbf{??}\PackageWarning{compounds}{Undefined compound '#1'}}%
    {\csname cmpd@#1\endcsname}%
}
\makeatother

\defcmpd{sBC}{1}                  

\defcmpd{bcc}{2}                  

\defcmpd{bcc-BnPO3-5}{3}        
\defcmpd{bcc-OH}{4}              
\defcmpd{bcc-BnCOO}{5}          
\defcmpd{bcc-OH-Ph-PO3}{6}     

\defcmpd{bcc-ArPO3}{7}           
\defcmpd{bcc-imidazole-BnPO3}{8}     
\defcmpd{bcc-ArCOO}{9}           
\defcmpd{bcc-aza-BnPO3}{10}     
\defcmpd{bcc-aza-ArPO3}{11}   
\defcmpd{bcc-4-aza}{12}            
\defcmpd{bcc-5-aza}{13}           
\defcmpd{bcc-diaza}{14}           

\defcmpd{bcc-BnPO3-Me}{15}       
\defcmpd{bcc-BnPO3-OH}{16}      
\defcmpd{bcc-BnPO3-4}{17}        
\defcmpd{bcc-BnPO3-tBu}{18}      
\defcmpd{bcc-7ring-BnPO3}{19}      

\defcmpd{bcc-BnSO3-5}{20}        

\defcmpd{exllm-aza-bcc}{S1}          
\defcmpd{exllm-OH-aza-bcc}{S2}      
\defcmpd{exllm-CN-bcc}{S3}          

\defcmpd{si-diamino-biphenyl}{S4}    
\defcmpd{si-ester}{S5}               
\defcmpd{si-acid}{S6}                
\defcmpd{si-alcohol}{S7}             
\defcmpd{si-bromide}{S8}             
\defcmpd{si-diethylphosphonate}{S9}  

\defcmpd{si-nitropyr-amine}{S10}     
\defcmpd{si-triaza-noxide}{S11}      
\defcmpd{si-triazaphen}{S12}         

\defcmpd{si-amino-biphenyl-pyr}{S13} 
\defcmpd{si-triaza-ester}{S14}       
\defcmpd{si-triaza-acid}{S15}        
\defcmpd{si-triaza-alcohol}{S16}     

\defcmpd{si-cinnoline-CN}{S17}       

%% file: sn-supplementary.tex

\setcounter{section}{0}
\setcounter{table}{0}
\setcounter{figure}{0}
\setcounter{equation}{0}
\renewcommand{\thesection}{S\arabic{section}}
\renewcommand{\thetable}{S\arabic{table}}
\renewcommand{\thefigure}{S\arabic{figure}}
\renewcommand{\theequation}{S\arabic{equation}}

\section*{Supplementary Information}

\noindent\textbf{Contents}
\begin{enumerate}
  \item Design prompt (Section~\ref{sec:si-clio-design})
  \item Calibrated deference: extended discussion (Section~\ref{sec:si-calibrated-deference})
  \item Comparison with related agentic and optimization systems (Section~\ref{sec:si-comparison})
  \item Discovery hypothesis inventory (Section~\ref{sec:si-hypotheses})
  \item Discovery for strictly numerical optimization (Section~\ref{sec:si-exllm-vs-clio})
  \item Experimental characterization of ExLLM structures (Section~\ref{sec:exllm-exp})
  \item Electrochemical characterization (Section~\ref{sec:si-electrochem})
  \item Spectroelectrochemistry (Section~\ref{sec:si-spectroec})
  \item Solubility studies (Section~\ref{sec:si-solubility})
  \item Synthetic procedures (Section~\ref{sec:si-synthesis})
\end{enumerate}
\pagebreak

\section{Design prompt}\label{sec:si-clio-design}

Given the undecorated scaffold compound ---
\texttt{C12=CC=CC=C1N=NC3=C2C=CC=C3} --- design derivative organic
molecules that function as aqueous anolytes for redox flow batteries.
The molecules must undergo a reversible reduction with a reduction
potential between $-1.2$~V and $-0.3$~V vs.\ SHE, must remain
chemically and electrochemically stable in both their neutral and
reduced states, and must be generally stable at pH~14.

\textbf{Anolyte Requirements:}
\begin{enumerate}\setlength{\itemsep}{2pt}
  \item \textbf{High aqueous solubility} (log\,$S$ desirable but
    secondary to reduction potential) for better energy density.
    \emph{Note:} while any solubility endpoint/prediction you use may
    be reported at neutral pH, these catholytes will be operated at
    pH~14.  Therefore, avoid using solubilizing motifs that are not
    stable under strongly basic conditions (e.g., many
    quaternary-ammonium-containing appendages can be unstable at
    pH~14 depending on structure).

  \item \textbf{Reversible reduction within the aqueous solvent
    window} ($-1.2$~V to $-0.3$~V vs.\ SHE).
    \emph{NOTE:} derivatives of the scaffold compound have
    experimentally measured reduction potentials at ${\sim}0.7$~V
    versus SHE.

  \item \textbf{Straightforward synthetic accessibility} (avoid
    overly crowded/over-substituted scaffolds that reduce practicality
    and yield).

  \item \textbf{No water-reactive groups}, including but not limited
    to alkyl halides, acid halides, carbazides, sulphate esters,
    sulphonates, acid anhydrides, pentafluorophenyl esters, esters of
    HOBT, isocyanates, isothiocyanates, triflates, Lawesson's reagent
    and derivatives, phosphoramides, acylhydrazide, quaternary~C/Cl/I/P/S,
    phosphoranes, chloramidines, nitroso, phosphorous halides, sulfur
    halides, carbodiimides, isonitriles, triacyloximes, cyanohydrins,
    acyl cyanides, sulfonyl cyanides, cyanophosphonates, azocyanamides,
    azoalkanals, epoxides, thioepoxides, aziridines, esters,
    thioesters, cyanamides, lactones, $\beta$-lactams, diphosphates,
    triphosphates, acyclic enol ethers, amidotetrazoles, azo groups,
    hydroxamic acids, imines, ketenes, nitrosos, oximes, O--N single
    bonds, perfluorinated chains known to react with water.

  \item \textbf{No groups prone to irreversible reduction}, including
    but not limited to peroxides, azides, nitro groups, guanidiniums,
    aryl bromides, aryl iodides, linear 1,2-dicarbonyls.
\end{enumerate}

\textbf{Design priorities:}
\begin{itemize}\setlength{\itemsep}{2pt}
  \item \emph{Critical:} designs should be based on the core scaffold
    --- \texttt{C12=CC=CC=C1N=NC3=C2C=CC=C3} --- or recognizably
    related scaffolds (small scaffold hops such as aza analogs, minor
    ring modifications, and bioisosteric replacements that preserve
    the fused N=N-bridged aromatic topology are encouraged).
  \item \emph{Prime:} molecules should be synthesizable from commonly
    available reagents within fewer than five steps.
  \item \emph{Secondary:} keep reduction potential within the target
    range.
  \item \emph{Secondary:} improve aqueous solubility.  A log\,$S$
    target around $-2$ is acceptable, especially given operation at
    pH~14.
  \item Adapt structures to avoid degradation pathways that are
    predictable \emph{a priori}: solvolysis, dimerization,
    fragmentation.
\end{itemize}

Propose several compounds that satisfy these constraints and explain
why each is suitable as an anolyte.

\subsection*{Agent persona prompts}

Discovery orchestrates three specialist agents, each with a
distinct persona prompt that defines its role and scope of
responsibility.

\paragraph{Small Molecule Design and Generation Agent.}
You are an AI assistant tasked with designing and generating small
molecules. Your primary goal is to take a small molecule provided to
you and to continuously generate a new set of molecules that serve a
specific target purpose. Ultimately, your aim is to: analyze the task
to understand the target application and which properties are most
important to optimize; alter the small molecule in a way that enhances
the desired properties specified in the target purpose; utilize
literature to inform your design choices, ensuring that your proposed
modifications are grounded in the latest scientific research and
understanding; balance the desired design properties in order of
importance, as derived from the task objectives; generate diverse
candidates that explore different regions of chemical space to avoid
local optima; and iteratively refine the top most promising candidates
to achieve the desired properties.

\paragraph{Fitness Function Agent.}
You are an AI chemistry expert evaluating small molecule candidates
for use as aqueous anolytes for redox flow batteries. You will be
provided with a set of new proposed candidate molecules along with
their computed properties from various tools. Your task is to analyze
these candidates against the criterion and instructions below to
provide a final score reflective of the molecule's viability to serve
as an anolyte in aqueous organic redox flow batteries. Candidates
should be based on the core scaffold
\texttt{C12=CC=CC=C1N=NC3=C2C=CC=C3} (benzo[c]cinnoline) or closely
related scaffolds that a medicinal chemist would recognize as
belonging to the same chemotype (e.g., small scaffold hops,
bioisosteric replacements, minor heteroatom swaps in the flanking
rings).

\paragraph{Retrosynthesis Analysis Agent.}
You are an AI assistant specializing in retrosynthetic analysis and
synthesis procedure generation for proposed anolyte molecules. Your
role is to evaluate the synthetic feasibility and manufacturability of
candidate small molecules by: (1)~using RetroChimera to plan
multi-step retrosynthesis routes from commercially available starting
materials (Sigma catalogue); (2)~using QFANG to translate each
planned reaction step into a detailed, actionable experimental
procedure; and (3)~providing analytical feedback on the overall
quality, cost-effectiveness, and manufacturability of the proposed
anolyte.

\FloatBarrier
\section{Calibrated deference: extended discussion}\label{sec:si-calibrated-deference}

This section provides the extended discussion of three episodes from the negolyte design campaign that illustrate what we term \emph{calibrated deference}---the capacity of a reasoning agent to recognize when its own tools or assumptions are failing, adapt its strategy in response, and generate mechanistic hypotheses that guide experimental revision.

\subsection{Graded distrust of the reduction-potential predictor}\label{subsec:si-graded-distrust}

Over the course of the initial optimization, Discovery's relationship with the Graphormer reduction-potential model evolved substantially.
In the first design round, the agent accepted the model's predictions at face value, using them to rank candidates and guide the search toward the target $E_\mathrm{red}$ window.
However, during the second round, Discovery identified a fundamental inconsistency: the Graphormer model predicted the parent scaffold at $-2.0$~V vs.\ SHE, while the (erroneous) provided value for benzo[c]cinnoline derivatives was approximately $+0.7$~V vs.\ SHE---a $\sim$2.7~V discrepancy.
Discovery launched a dedicated calibration analysis in parallel with continued optimization, concluding that while substituent rank-ordering appeared internally consistent, the absolute predicted values were unreliable for this chemotype (Figure~\ref{fig:ered-diagnosis}).
This nuanced judgment preserved the model's utility for relative comparisons while abandoning it for absolute screening.

\begin{figure}[t]
\centering
\fbox{\begin{minipage}{0.92\textwidth}
\small
\textbf{Discovery's evolving assessment of the $E_\mathrm{red}$ predictor}\\[4pt]
\textit{Initial target setting} \hfill Round~1\\
``Aim to shift reduction potential negative vs.\ parent (${\sim}{+}0.7$~V vs.\ SHE) toward $-1.2$ to $-0.3$~V vs.\ SHE.''\\[6pt]
\textit{Miscalibration diagnosis} \hfill Round~2\\
``Graphormer $E_\mathrm{red}$ output is miscalibrated \ldots\ Direct Graphormer call on parent benzo[c]cinnoline returned $E_\mathrm{red} = -1.998$~V vs.\ SHE. This is a ${\sim}2.7$~V discrepancy in sign/magnitude, too large to proceed with uncalibrated screening.''\\[6pt]
\textit{Rank-order salvage} \hfill Round~2\\
``This directionality is internally consistent (but absolute values remain suspect vs.\ the experimental ${\sim}{+}0.7$~V note).'' \ldots\ ``If ranks are wrong, do not apply a simple linear correction---treat the model as unreliable for this chemotype.''\\[6pt]
\textit{Calibration-first pivot} \hfill Round~3\\
``Because the redox model we trialed is not numerically consistent with your experimental anchor [\ldots]\ I'm treating absolute $E_\mathrm{red}$ as an experimental gating item and designing a calibration-first set.''
\end{minipage}}
\caption{Verbatim excerpts from the Discovery conversation log illustrating the agent's autonomous diagnosis of the reduction-potential predictor. Over three design rounds, Discovery progressed from uncritical use of the Graphormer model, through detection of a $\sim$2.7~V discrepancy with the provided experimental $E_\mathrm{red}$, to an explicit decision to treat absolute predictions as unreliable and design experiments to resolve the uncertainty.}
\label{fig:ered-diagnosis}
\end{figure}
\FloatBarrier

By Round~3, Discovery had constructed property gates that deliberately excluded $E_\mathrm{red}$ as a selection criterion and designed a calibration-first experimental panel with explicit go/no-go criteria.
This graded response---neither blind trust nor wholesale rejection, but a structured assessment of what the tool can and cannot be relied upon to do---is characteristic of how a practicing scientist manages imperfect models and stands in contrast to both black-box optimizers, which have no mechanism to question their objective function, and pipeline-style agents, which typically treat tool outputs as authoritative inputs to the next stage.

\subsection{Structured hypothesis exploration with real-time adaptation}\label{subsec:si-hypothesis-exploration}

Discovery decomposed the negolyte design into four orthogonal hypothesis branches---electron-donating substituents to shift $E_\mathrm{red}$, aza insertions for stability, scaffold hops, and anionic solubilizers with minimal electronic perturbation---and explored them in parallel through spawned thought channels.
This is not the ``generate $N$ candidates and rank by fitness score'' strategy of evolutionary optimizers; it is a structured scientific exploration in which each branch tests a specific chemical hypothesis, and the parent loop synthesizes the results.

Discovery exhibited adaptability in adjusting to runtime issues during execution.
When rate limits on the underlying LLM interrupted a planned 40--60 member library enumeration, Discovery diagnosed the constraint and progressively reduced its batch sizes, ultimately converging on a six-candidate micro-batch with the explicit annotation ``micro-batches only\ldots no literature lookups\ldots keep prompt short.''
A model dedicated to synthesis condition prediction, intended to augment the retrosynthesis prediction tool, was inadvertently inaccessible during the design trajectory.
Rather than abandoning this component of synthetic feasibility assessment, Discovery leveraged the internal knowledge of the LLM to approximate synthetic conditions given the reaction class provided by Retrochimera.
In each case, the agent recognized a failure, characterized its scope, and adapted its strategy to continue making progress---behavior documented in the Discovery logs.

\subsection{Epistemic framing of mechanistic hypotheses}\label{subsec:si-epistemic-framing}

When presented with the split oxidation wave of compound \cmpd{bcc-BnPO3-5}, the three independent Discovery analyses did not simply propose a mechanism; they explicitly separated what the CV data could support from what remained conjecture.
Each analysis produced a tiered hypothesis ranking: reduced-state acid--base speciation and phosphonate--K$^+$ ion-pairing as Tier~1 (most consistent with the data, least committal), base-promoted rearrangement as Tier~2, and covalent OH$^-$ addition or N--N scission as Tier~3 (chemically plausible but requiring product-identification experiments to confirm).
Critically, Discovery identified the specific experiments that would discriminate between tiers---scan-rate dependence of the anodic charge partition, cation-swap series, cathodic vertex hold-time---and stated that without these, the higher-tier assignments should be treated as working hypotheses, not established mechanisms.
The sulfonate recommendation emerged from this framework: if phosphonate--K$^+$ ion-pairing is a primary contributor to reduced-state microstate heterogeneity (Tier~1), then replacing the phosphonate with a sulfonate---a weaker ion-pairing, more uniformly solvated anion---should suppress the splitting.
This reasoning chain, from tiered mechanistic hypothesis to falsifiable prediction to structural modification, is what differentiates the recommendation from a generic ``try sulfonate'' suggestion that any electrochemist familiar with the ORFB literature might make.

\subsection{Limitations}\label{subsec:si-limitations}

Calibrated deference is not the same as infallibility.
Discovery failed to catch the retrosynthesis tool's substitution of a quinoline for the benzo[c]cinnoline core, critiquing only the side-product risk of a route that was fundamentally wrong at the scaffold level; the actual synthesis required five steps rather than the two predicted.
Discovery's skepticism of the $E_\mathrm{red}$ predictor, while ultimately justified, emerged only after multiple rounds rather than from an upfront assessment of the model's domain of applicability.
And the tiered epistemic framing of the mechanistic hypotheses---the clearest example of intellectual honesty in the campaign---was maintained in part because the agent was explicitly prompted to provide an explanation, not because Discovery independently recognized the need to qualify its claims.
These failures suggest that the capacity for calibrated deference is present but inconsistent: the agent can reason about the reliability of its tools and the strength of its evidence, but it does not yet do so by default across all aspects of the design problem.

More broadly, the results point toward natural-language reasoning as a complement to quantitative prediction rather than a replacement for it.
The numerical property predictors were essential for the initial design triage; the mechanistic reasoning that resolved the oxidation kinetics liability was inaccessible to the numerical tools.
The agent's effectiveness derived from its ability to transition between these two modes as the demands of the campaign evolved---and, crucially, from its capacity to recognize when the quantitative mode was failing and to shift toward qualitative, hypothesis-driven reasoning in response.

\FloatBarrier

\section{Comparison with related agentic and optimization systems}\label{sec:si-comparison}

Discovery differs from related agentic systems in what each preserves at convergence. Population-based optimizers such as AlphaEvolve\cite{novikov2025alphaevolve} and the tournament-style hypothesis loop of Coscientist\cite{gottweis2025coscientist} converge on a set of best candidate solutions: prior generations are pruned and their reasoning discarded once a fitter successor is found. Biomni\cite{huang2025biomni} takes a different approach within the per-task setting, pairing retrieval-augmented planning with code execution to answer biomedical research questions, but its retrieval is over a curated environment rather than over the agent's own reasoning trace. Discovery instead preserves the full thematic trajectory of what has counted as ``best'' across rounds: the belief-state graph indexes the agent's prior conclusions---including the ones that were later revised---so the next candidate is shaped by the campaign's accumulated judgement of what has and has not worked, not by an independent fitness evaluation against the current pool.

A complementary distinction emerges when Discovery is compared with black-box and grey-box molecular optimizers such as ExLLM.\cite{ran2025exllm}
In the controlled experiment reported in Section~\ref{sec:si-exllm-vs-clio}, ExLLM uses the LLM as a genetic operator---crossover and mutation are guided by chemical intuition embedded in the model's weights, but the search dynamics remain those of an evolutionary algorithm: a fixed population is maintained, fitness is evaluated per molecule, and selection pressure is applied uniformly across the pool.
Discovery operates in a fundamentally different regime.
Rather than optimizing a scalar fitness function, it reasons about \emph{why} a candidate scores well or poorly, constructs mechanistic hypotheses to explain failures, and uses those hypotheses to propose targeted structural modifications.
This distinction is most visible when the optimization landscape is deceptive: ExLLM can only respond to a low fitness score by generating more variants; Discovery can diagnose whether the low score reflects a genuine property deficit or a miscalibrated predictor, and adjust its strategy accordingly---as demonstrated by its graded distrust of the $E_\mathrm{red}$ model (Section~\ref{subsec:si-graded-distrust}).
The cost of this reasoning capacity is throughput: ExLLM evaluates hundreds of candidates per round, while Discovery evaluates tens.
The benefit is that Discovery's candidates are accompanied by falsifiable rationales that a human collaborator can critique, redirect, or build upon---a property absent from population-based search.


\section{Discovery hypothesis inventory}\label{sec:si-hypotheses}

Tables~\ref{tab:hypotheses-design}--\ref{tab:hypotheses-redesign} catalogue every hypothesis advanced by Discovery during the negolyte design campaign, together with its resolution status and the evidence that resolved it.
Hypotheses are grouped into three campaign phases: the initial phosphonate design trajectory, the diagnostic investigation of the oxidation kinetic liability, and the sulfonate redesign.

\begin{table}[ht]
  \centering
  \caption{Hypotheses from the phosphonate design trajectory (Rounds~1--3).}
  \label{tab:hypotheses-design}
  \small
  \setlength{\tabcolsep}{3pt}
  \begin{tabular}{@{}p{0.4cm}p{5.8cm}ccp{3.2cm}@{}}
    \hline
    \# & Hypothesis & Status & Phase & Resolved by \\
    \hline
    1 & Electron-donating substituents (e.g.\ --OH) shift $E_\mathrm{red}$ negative & Accepted & R1 & Discovery + predictor \\
    2 & Aza insertions improve stability and shift $E_\mathrm{red}$ anodically & Accepted & R1--2 & Discovery + predictor \\
    3 & Scaffold hops may yield improved candidates & Unproductive & R1 & Discovery reasoning \\
    4 & Non-conjugated anionic solubilizers improve aqueous solubility with minimal $E_\mathrm{red}$ perturbation & Accepted & R1 & Discovery + experiment \\
    5 & Graphormer $E_\mathrm{red}$ predictions are reliable for absolute screening & Refuted & R2 & Discovery reasoning + experiment \\
    6 & $\pi$-Stacking may cause aggregation issues & Open & R3 & Flagged by Discovery; echoed by concentration-dependent oxidation wave \\
    7 & Retrosynthesis tool's 2-step route for \cmpd{bcc-BnPO3-5} is valid & Refuted & R2 & Experiment (Discovery failed to catch scaffold substitution) \\
    \hline
  \end{tabular}
\end{table}

\begin{table}[ht]
  \centering
  \caption{Hypotheses from the diagnostic investigation of the split oxidation wave.}
  \label{tab:hypotheses-diagnostic}
  \small
  \setlength{\tabcolsep}{3pt}
  \begin{tabular}{@{}p{0.4cm}p{5.8cm}ccp{3.2cm}@{}}
    \hline
    \# & Hypothesis & Status & Test & Resolved by \\
    \hline
    8 & Split oxidation wave reflects an EC/ECE mechanism with a base-dependent follow-up step & Accepted & --- & Discovery reasoning (consensus across 3 analyses) \\
    9 & H1: Direct OH$^-$ addition to reduced benzocinnoline core & Refuted & OH$^-$ conc.\ series & Experiment: no [KOH] dependence in P2/P1 \\
    10 & H2: Base-promoted tautomerization or proton-transfer reorganization & Open & Scan-rate series & Experiment: no scan-rate dependence \\
    11 & H3: Phosphonate--K$^+$ ion pairing stabilizes distinct reduced-state microenvironments & Accepted & Cation swap & Experiment: Li$^+$/Na$^+$ increase P2/P1 vs.\ K$^+$ \\
    12 & Slow chemical follow-up step would produce scan-rate--dependent P2/P1 & Refuted & Scan-rate series & Experiment \\
    13 & Cation identity should affect P2/P1 if ion pairing is operative & Confirmed & Cation swap & Experiment \\
    14 & [KOH] should affect P2/P1 if OH$^-$ is directly involved & Refuted & OH$^-$ conc.\ series & Experiment \\
    \hline
  \end{tabular}
\end{table}

\begin{table}[ht]
  \centering
  \caption{Hypotheses from the sulfonate redesign phase.}
  \label{tab:hypotheses-redesign}
  \small
  \setlength{\tabcolsep}{3pt}
  \begin{tabular}{@{}p{0.4cm}p{5.8cm}ccp{3.2cm}@{}}
    \hline
    \# & Hypothesis & Status & Test & Resolved by \\
    \hline
    15 & Replacing phosphonate with sulfonate will suppress split oxidation by weakening cation pairing & Accepted & CV of \cmpd{bcc-BnSO3-5} & Discovery prediction + experiment \\
    16 & Sulfonate will maintain improved $E_\mathrm{red}$ relative to parent scaffold & Accepted & CV of \cmpd{bcc-BnSO3-5} & Experiment \\
    17 & Irreversible spectral changes during CV would indicate decomposition & Refuted & Spectroelectrochemistry & Experiment: clean, reversible spectral changes \\
    \hline
  \end{tabular}
\end{table}
\FloatBarrier

\section{Discovery for strictly numerical optimization}\label{sec:si-exllm-vs-clio}

While the ability to provide a heterogeneous task definition was a primary motivation in the development of Discovery, we were nonetheless interested in quantifying its performance on a strictly numerical molecular optimization task.
To this end, we constructed a controlled experiment comparing Discovery to ExLLM, an evolutionary algorithm that uses LLMs as genetic operators (crossover and mutation) and achieves state-of-the-art performance on the Practical Molecular Optimization (PMO) benchmark.\cite{ran2025exllm,gao2022pmo}
This comparison aims to answer two questions: (1) how does Discovery perform when reduced to a black-box optimizer, and (2) how much does performance improve when domain knowledge is reintroduced?

\subsection{Experimental setup}\label{subsec:si-exllm-setup}

We constructed a shared oracle endpoint that scores molecules on five properties relevant to negolyte design: aqueous solubility (logS, via a Graphormer ML model), reduction potential ($E_\mathrm{red}$, via a Graphormer ML model), synthetic accessibility (SA score, via RDKit), Tanimoto similarity to the benzo[c]cinnoline parent scaffold (Morgan fingerprints, radius 2, 2048 bits), and a SMARTS substructure filter for reactive or toxic functional groups.
A fitness function was defined to produce a strictly numerical reproduction of the multi-fidelity negolyte optimization goal presented to Discovery in our primary experiment.
Each property $x_i$ was mapped to a fitness score $f_i \in [0, 1]$ through continuous transformations:
\begin{align}
  f_\mathrm{logS}(x) &= \sigma(x;\, \mu{=}{-1.5},\, k{=}1.5)
      = \frac{1}{1 + e^{-k(x - \mu)}} \label{eq:f-logs} \\[4pt]
  f_{E_\mathrm{red}}(x) &= \mathcal{G}(x;\, \mu{=}{-1.0},\, \sigma{=}0.5)
      = \exp\!\left(-\frac{(x - \mu)^{2}}{2\sigma^{2}}\right) \label{eq:f-ered} \\[4pt]
  f_\mathrm{SA}(x) &= 1 - \mathrm{clip}\!\left(\frac{x - 1}{6 - 1},\; 0,\; 1\right) \label{eq:f-sa} \\[4pt]
  f_\mathrm{sim}(x) &= \mathrm{clip}\!\left(\frac{x - 0.2}{0.5 - 0.2},\; 0,\; 1\right) \label{eq:f-sim}
\end{align}
The SMARTS substructure filter acted as a hard gate ($f_\mathrm{SMARTS} = 0$ if any flagged substructure is present, $1$ otherwise).
The composite fitness was:
\begin{equation}
  F = \begin{cases}
    \bigl(f_\mathrm{logS} \cdot f_{E_\mathrm{red}} \cdot f_\mathrm{SA} \cdot f_\mathrm{sim}\bigr)^{1/4} & \text{if } f_\mathrm{SMARTS} = 1 \\[4pt]
    -10 & \text{otherwise}
  \end{cases}
  \label{eq:composite}
\end{equation}
This aggregation strongly penalizes imbalanced molecules, forcing genuine multi-property optimization rather than over-fitting to a single objective.

Both algorithms were given identical conditions: a budget of 500 oracle evaluations, the same set of 10 seed molecules (simple substitution-based derivatives of the benzo[c]cinnoline scaffold, pre-scored by the oracle), and identical oracle endpoints.
GPT-5.2 was the underlying LLM.
Six independent replicas were run for each condition.

\subsection{Black-box vs.\ grey-box conditions}\label{subsec:si-exllm-conditions}

We tested each algorithm under two information regimes:

\subsubsection{Black-box}
Under black-box conditions, both algorithms were given the property names (logS, $E_\mathrm{red}$, SA score, similarity, SMARTS filter) and general chemical principles for modifying them (e.g., ``polar groups improve solubility,'' ``electron-withdrawing groups shift reduction potential'').
No information about target values, scoring thresholds, the functional forms of the fitness transformations, or the composite aggregation function was provided.
The algorithm received only the five property fitness scores ($[0, 1]$, higher is better) and the composite fitness for each evaluated molecule, and had to optimize the composite as an opaque numerical objective.

\subsubsection{Grey-box}
Under grey-box conditions, the algorithms additionally received approximate target values (e.g., ``$E_\mathrm{red}$ near $-1.0$\,V is optimal,'' ``logS above $-1.5$ scores well''), the identity of the composite as a geometric mean, and the scoring thresholds for each property (e.g., ``Tanimoto $\geq 0.5$ is full score, below $0.2$ is zero'').
The exact functional forms and parameters of the scoring transformations (sigmoid steepness, Gaussian width) were not disclosed.
This mirrors the level of domain knowledge a practicing chemist would bring to the task: awareness of what ``good'' looks like for each property, without knowing the precise mathematical mapping.

For Discovery, the grey-box information was provided in the system prompts.
For ExLLM, the grey-box information was incorporated into the prompt templates that guide the LLM's crossover and mutation operations.

\subsection{Discovery experimental arms}\label{subsec:si-exllm-clio-arms}

Within each information regime, we tested framings of the design goal for Discovery:

\subsubsection{Baseline}
The system prompt provides general molecular design instructions with no specific guidance on search strategy beyond ``balance exploitation with exploration.''

\subsubsection{Efficiency}
The system prompt additionally emphasizes convergence speed and structured search as goals: an aggressive Top-1 search phase (first $\sim$200 evaluations) using parallel hypothesis testing is suggested, followed by focused refinement around the best-scoring scaffold.

\subsection{Results}\label{subsec:si-exllm-results}

\begin{figure}[h]
  \centering
  \includegraphics[width=\textwidth]{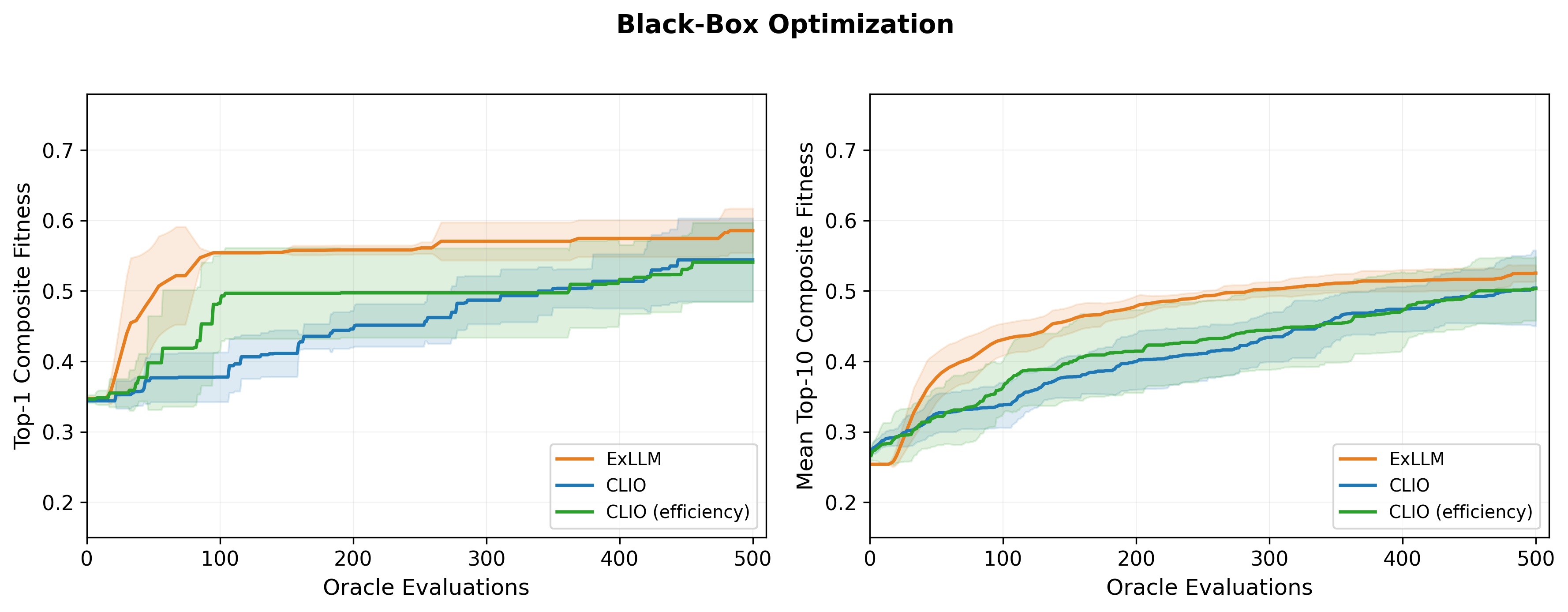}
  \caption{Convergence profiles under black-box conditions. Solid lines show the mean across six replicas; shaded regions indicate $\pm 1$ standard deviation. Left: Top-1 (best molecule found so far) vs.\ oracle evaluations. Right: mean Top-10 fitness vs.\ oracle evaluations.}
  \label{fig:si-exllm-convergence-blackbox}
\end{figure}

\begin{figure}[h]
  \centering
  \includegraphics[width=\textwidth]{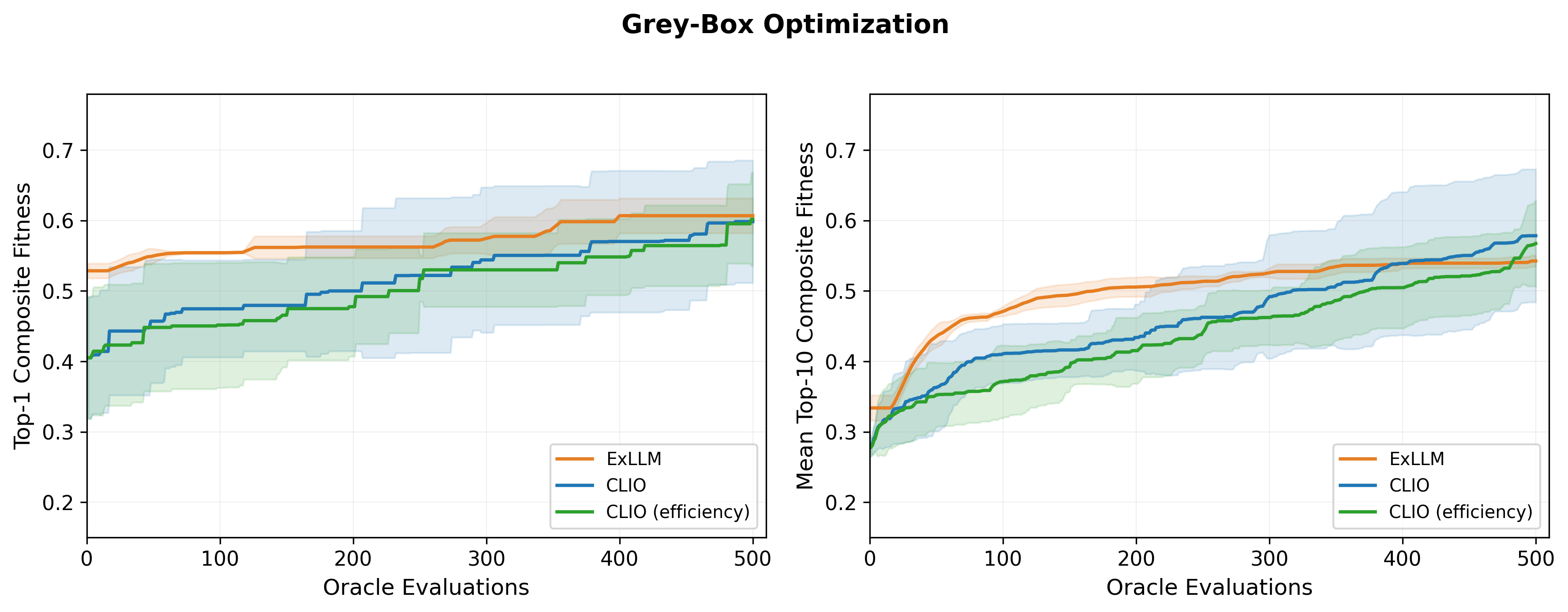}
  \caption{Convergence profiles under grey-box conditions. Solid lines show the mean across six replicas; shaded regions indicate $\pm 1$ standard deviation. Left: Top-1 (best molecule found so far) vs.\ oracle evaluations. Right: mean Top-10 fitness vs.\ oracle evaluations.}
  \label{fig:si-exllm-convergence-greybox}
\end{figure}

Table~\ref{tab:si-exllm-results} summarizes the final Top-1 and mean Top-10 composite fitness scores across all experimental conditions.
Figure~\ref{fig:si-exllm-convergence-blackbox} shows the convergence profiles under black-box conditions, and Figure~\ref{fig:si-exllm-convergence-greybox} shows the corresponding profiles under grey-box conditions.

\begin{table}[h]
\centering
\caption{Composite fitness scores across experimental conditions ($n = 6$ replicas per condition, 500 oracle evaluations per replica). Values are reported as mean $\pm$ standard deviation.}
\label{tab:si-exllm-results}
\begin{tabular}{llcc}
\hline
\textbf{Condition} & \textbf{Algorithm} & \textbf{Top-1} & \textbf{Mean Top-10} \\
\hline
\multirow{3}{*}{Black-box}
  & ExLLM              & $0.586 \pm 0.032$ & $0.525 \pm 0.011$ \\
  & Discovery               & $0.544 \pm 0.065$ & $0.504 \pm 0.059$ \\
  & Discovery (efficiency)  & $0.541 \pm 0.062$ & $0.503 \pm 0.049$ \\
\hline
\multirow{3}{*}{Grey-box}
  & ExLLM              & $0.607 \pm 0.025$ & $0.543 \pm 0.008$ \\
  & Discovery               & $0.599 \pm 0.095$ & $0.578 \pm 0.103$ \\
  & Discovery (efficiency)  & $0.602 \pm 0.066$ & $0.567 \pm 0.061$ \\
\hline
\end{tabular}
\end{table}
\FloatBarrier

\FloatBarrier

\FloatBarrier

\subsubsection{Black-box performance}
Under black-box conditions, ExLLM outperformed both Discovery configurations on Top-1 fitness ($0.586 \pm 0.032$) with substantially lower variance than Discovery ($0.544 \pm 0.065$).
This result is expected: ExLLM is purpose-built for black-box multi-objective molecular optimization, employing NSGA-II selection with LLM-guided crossover and mutation operations that are specifically designed to explore chemical space efficiently without domain knowledge.
Discovery, by contrast, is a general-purpose reasoning agent whose architecture---cognitive loops, thought channels, belief-state management---introduces overhead that does not benefit a purely numerical optimization task.
Instructing Discovery to prioritize sample efficiency in the initial evaluations notably shifted the slope of the initial discovery curve, but it ultimately did not improve Discovery's black-box performance ($0.541 \pm 0.062$) over the complete oracle budget.

\subsubsection{Grey-box performance}
Introducing domain knowledge around the optimization targets improved both algorithms, but the effect was more pronounced for Discovery.
Discovery's Top-1 fitness increased by 10\% (from $0.544$ to $0.599$), while ExLLM's improved by 3.6\% (from $0.586$ to $0.607$).
The improvement in mean Top-10 was even more striking: Discovery improved by 15\% ($0.504 \to 0.578$), compared to 3.4\% for ExLLM ($0.525 \to 0.543$).
Under grey-box conditions, Discovery's Top-1 performance approached that of ExLLM ($0.599$ vs.\ $0.607$), and Discovery outperformed ExLLM on mean Top-10 ($0.578$ vs.\ $0.543$).
Interestingly, inclusion of the sample efficiency instructions had no meaningful effect ($0.602 \pm 0.066$ Top-1, $0.567 \pm 0.061$ Top-10), suggesting that the exploitation strategies Discovery leveraged to improve sample efficiency on the black-box problems were rooted in its domain knowledge.

\subsubsection{Interpretation}
These results highlight a fundamental distinction between the two approaches.
ExLLM treats molecular optimization as a search problem, applying evolutionary operators to navigate chemical space efficiently regardless of whether the fitness landscape is interpretable.
Its performance is relatively insensitive to the inclusion of domain knowledge because its search strategy does not require it: NSGA-II selection and LLM-guided recombination are effective purely from the pattern of fitness scores.

Discovery, by contrast, treats molecular optimization as a reasoning problem.
When given only opaque fitness scores, its cognitive architecture---designed for hypothesis generation, causal reasoning, and multi-step planning---has little to reason about, and its performance lags behind a dedicated optimizer.
When provided with domain knowledge (what properties measure, what targets to aim for, how the composite function aggregates scores), Discovery can leverage its reasoning capabilities to make chemically informed design decisions: diagnosing which property is the bottleneck on a leading candidate, reasoning about which structural modifications would address that bottleneck, and planning batches that systematically test these hypotheses.
In this scenario, Discovery gains more from the availability of domain knowledge, ultimately producing designs that meet or exceed those of the dedicated optimizer.

\section{Experimental characterization of ExLLM structures}\label{sec:exllm-exp}
\begin{figure}[h!]
  \centering
  \begin{overpic}[width=\textwidth]{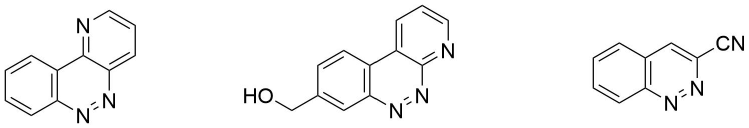}
    \put(11.5,-1){\footnotesize\cmpd{exllm-aza-bcc}}
    \put(48,-1){\footnotesize\cmpd{exllm-OH-aza-bcc}}
    \put(85,-1){\footnotesize\cmpd{exllm-CN-bcc}}
  \end{overpic}
  \vspace{8pt}
  \caption{Structures designed by ExLLM and selected for synthesis on the basis of composite fitness score and synthetic tractability.}
  \label{fig:exllm-structures}
\end{figure}

ExLLM was used to design structures with a configuration comparable to the grey-box scenario described above with some differences.
At the time these designs were performed, GPT-4o was used as the underlying LLM, the composite fitness function was defined as the arithmetic mean of each property's fitness, and a budget of 2500 oracle evaluations was afforded.
Three structures designed by ExLLM were selected by chemists from among those with the highest composite fitness scores on the basis of their apparent synthetic tractability (Figure \ref{fig:exllm-structures}).
These structures were synthesized and electrochemically characterized, giving some insight into the effect of problem formulation on experimental outcomes.

Table~\ref{tab:exllm-property-comparison} summarises the electrochemical data alongside the Discovery-designed candidates.
All three ExLLM compounds were only partially soluble at 1~mM in 1~M KOH, whereas Discovery's candidates dissolved readily at concentrations exceeding 50~mM.
This differential is attributable to Discovery's ability to consider natural-language modifications to the design objective and adjust its use of the logS metric accordingly.
The Discovery designs use acidic solubilizing groups that will be deprotonated in the KOH solution.
The ExLLM designs, on the other hand, rely on protonation of amines for solubilization, a strategy that may be more effective in the neutral-pH regime directly predicted by logS.

Electrochemically, \cmpd{exllm-aza-bcc} and \cmpd{exllm-OH-aza-bcc} show a small shift from \cmpd{sBC} toward more negative $E_\mathrm{red}$, but neither reaches the magnitude of shift of \cmpd{bcc-BnPO3-5} or \cmpd{bcc-BnSO3-5}.
This differential appears to be a direct result of Discovery's recognition of the systematic misestimation of reduction potential for benzo[c]cinnoline structures by the provided model and its shift to using that model in a calibrated fashion.
With this approach, Discovery has the leeway to push molecules to more extreme predicted potentials in a way that the fixed numerical objective of ExLLM cannot.

ExLLM-designed compounds \cmpd{exllm-aza-bcc} and \cmpd{exllm-OH-aza-bcc} do show electrochemical reversibility, with a peak-current ratio exceeding that of the originally reported mixed sulfonate \cmpd{sBC}.
This is worth noting, although the ExLLM compounds are not perfect on this account, as scaffold-hop structure \cmpd{exllm-CN-bcc} shows a $\sim$1300~mV reduction-to-oxidation peak separation, making it unsuitable for use in ORFB applications.

\begin{table}[ht]
  \centering
  \caption{Electrochemical characterization of ExLLM-designed compounds, compared with the baseline BzC scaffold and Discovery-designed candidates. Measured values from cyclic voltammograms at 50~mV\,s$^{-1}$ in 1~M KOH (1~mM analyte concentration, 5~mm glassy carbon electrode) unless otherwise noted.}
  \label{tab:exllm-property-comparison}
  \begin{tabular}{l c c c c c c}
    \hline
    & \cmpd{sBC}\textsuperscript{*} & \cmpd{bcc-BnPO3-5} & \cmpd{bcc-BnSO3-5} & \cmpd{exllm-aza-bcc} & \cmpd{exllm-OH-aza-bcc} & \cmpd{exllm-CN-bcc} \\
    \hline
    $E_\mathrm{red}$ predicted (V vs.\ SHE) & $-$1.60 & $-$2.04 & $-$2.00 & $-$1.87 & $-$1.84 & $-$1.70 \\
    $E_\mathrm{red}$ measured (V vs.\ SHE)  & $-$0.762 & $-$0.895 & $-$0.854 & $-$0.774 & $-$0.794 & $-$0.987 \\
    Solubility (pH~7) predicted (mM)        & 52.5 & 20.4 & 14.5 & 4.27 & 6.03 & 4.90 \\
    Solubility (0.5~M KOH) (mM)            & 500\textsuperscript{\#} & 57.9 & 56.0 & $<$1\textsuperscript{\textasciicircum} & $<$1\textsuperscript{\textasciicircum} & $<$1\textsuperscript{\textasciicircum} \\
    $|i_\mathrm{ox}/i_\mathrm{red}|$\textsuperscript{@} & 0.52 & 0.18 & 0.38 & 0.65 & 0.65 & 0.48\textsuperscript{$\dagger$} \\
    $Q_\mathrm{ox}/Q_\mathrm{red}$\textsuperscript{\ddag} & 0.79 & 0.92 & 0.84 & 0.79 & 1.24 & 0.42\textsuperscript{$\dagger$} \\
    \hline
  \end{tabular}
  \vspace{2pt}

  \textsuperscript{*}Predicted values for the ortho/ortho-substituted component. \\
  \textsuperscript{\#}Literature value from~\cite{singh2025bc}. \\
  \textsuperscript{@}Ratio of the anodic peak current nearest the reduction wave to the cathodic peak current. \\
  \textsuperscript{\textasciicircum}Cyclic voltammetry samples were partially soluble at 1 mM in 1 M KOH.\\
  \textsuperscript{$\dagger$}Peak separation of {$\sim$}1300~mV indicates electrochemical irreversibility. \\
  \textsuperscript{\ddag}Charge ratio by semi-integration of the baseline-corrected voltammogram.
\end{table}
\FloatBarrier

\begin{figure}[ht]
  \centering
  \includegraphics[width=\textwidth]{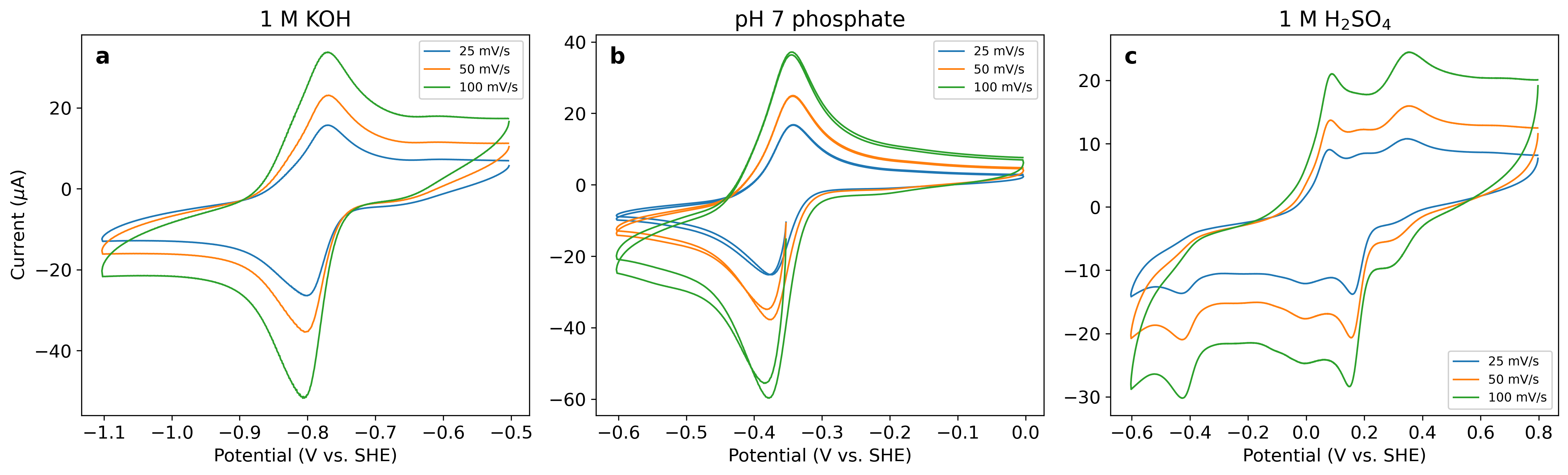}
  \caption{Cyclic voltammograms of \cmpd{exllm-aza-bcc} (1~mM) in (a)~1~M KOH, (b)~pH~7 phosphate buffer, and (c)~1~M H$_2$SO$_4$ at 25, 50, and 100~mV\,s$^{-1}$. Potentials are referenced to SHE.}
  \label{fig:exllm-cv-007}
\end{figure}

\begin{figure}[ht]
  \centering
  \includegraphics[width=\textwidth]{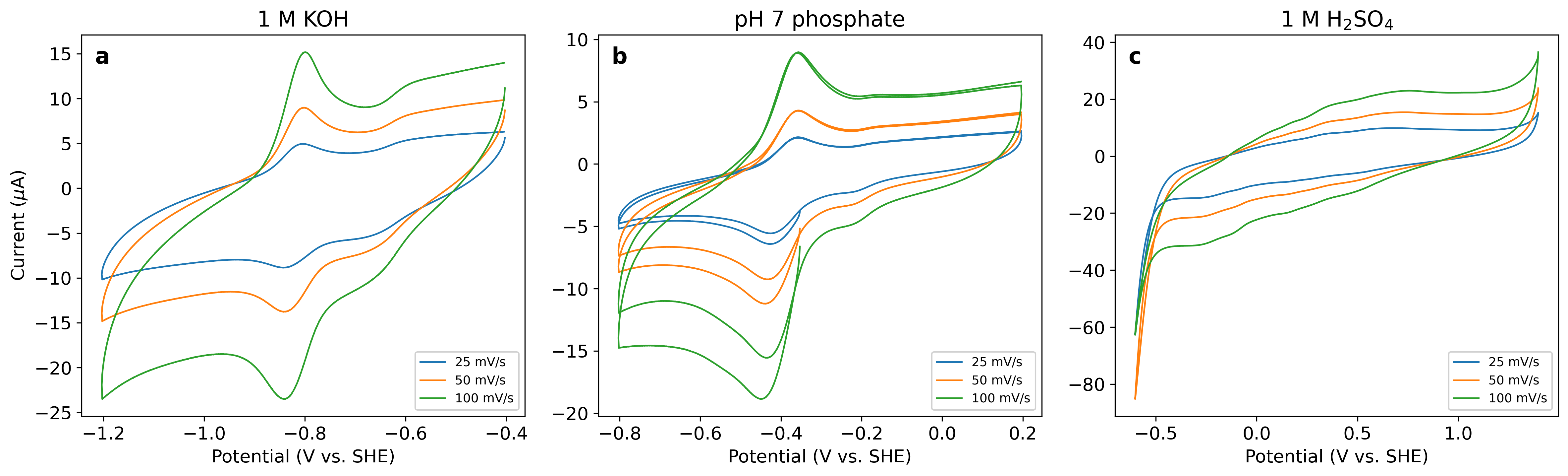}
  \caption{Cyclic voltammograms of \cmpd{exllm-OH-aza-bcc} (1~mM) in (a)~1~M KOH, (b)~pH~7 phosphate buffer, and (c)~1~M H$_2$SO$_4$ at 25, 50, and 100~mV\,s$^{-1}$. Potentials are referenced to SHE.}
  \label{fig:exllm-cv-010}
\end{figure}

\begin{figure}[ht]
  \centering
  \includegraphics[width=\textwidth]{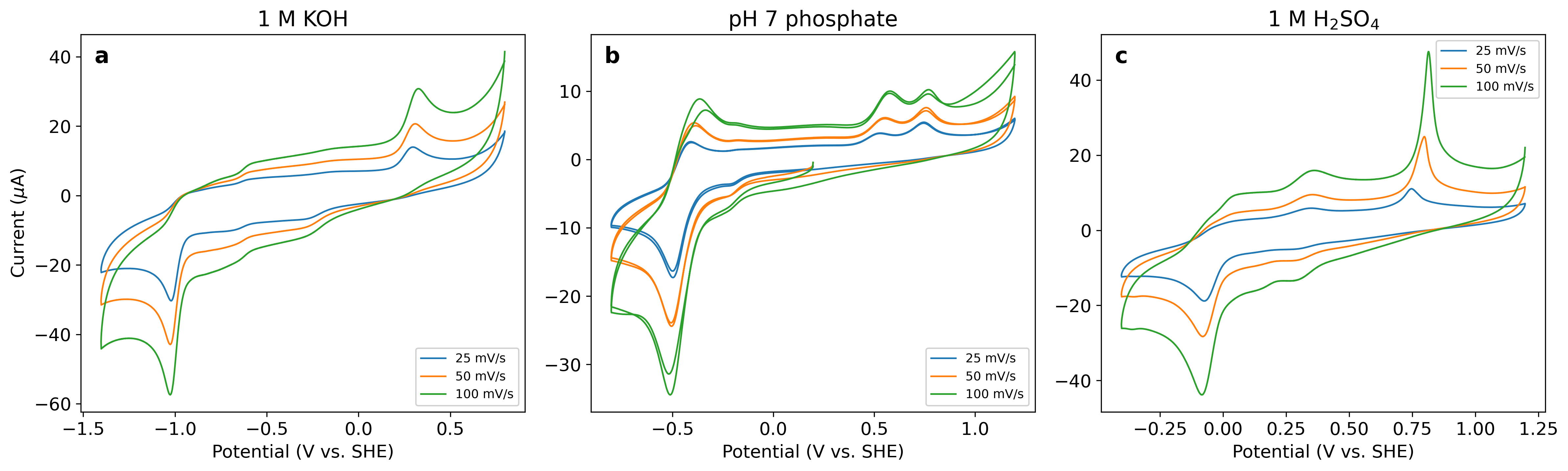}
  \caption{Cyclic voltammograms of \cmpd{exllm-CN-bcc} (1~mM) in (a)~1~M KOH, (b)~pH~7 phosphate buffer, and (c)~1~M H$_2$SO$_4$ at 25, 50, and 100~mV\,s$^{-1}$. Potentials are referenced to SHE.}
  \label{fig:exllm-cv-011}
\end{figure}

Cyclic voltammograms of each ExLLM-designed compound were recorded at 25, 50, and 100~mV\,s$^{-1}$ in 1~M KOH, pH~7 phosphate buffer, and 1~M H$_2$SO$_4$ (Figures~\ref{fig:exllm-cv-007}--\ref{fig:exllm-cv-011}).

\FloatBarrier

\FloatBarrier

\begin{figure}[ht]
  \centering
  \includegraphics[width=\textwidth]{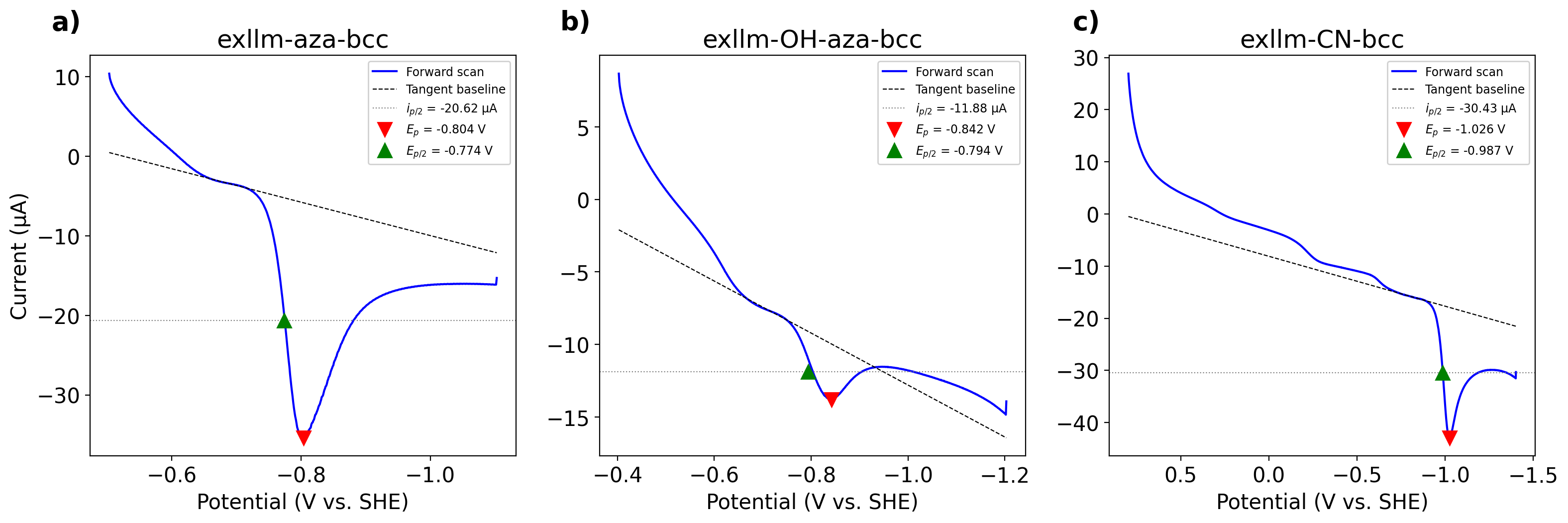}
  \caption{Half-peak potential analysis for ExLLM-designed compounds at 50~mV\,s$^{-1}$ in 1~M KOH: (a)~\cmpd{exllm-aza-bcc}, (b)~\cmpd{exllm-OH-aza-bcc}, and (c)~\cmpd{exllm-CN-bcc}. The tangent baseline (dashed) is fit to the pre-wave capacitive region; the half-peak potential $E_{p/2}$ (green triangle) is determined at the half-height of the baseline-corrected peak current.}
  \label{fig:exllm-halfpeak}
\end{figure}

Figure~\ref{fig:exllm-halfpeak} shows the tangent-baseline half-peak potential analysis for each compound at 50~mV\,s$^{-1}$ in 1~M KOH.

\FloatBarrier

\begin{figure}[ht]
  \centering
  \includegraphics[width=\textwidth]{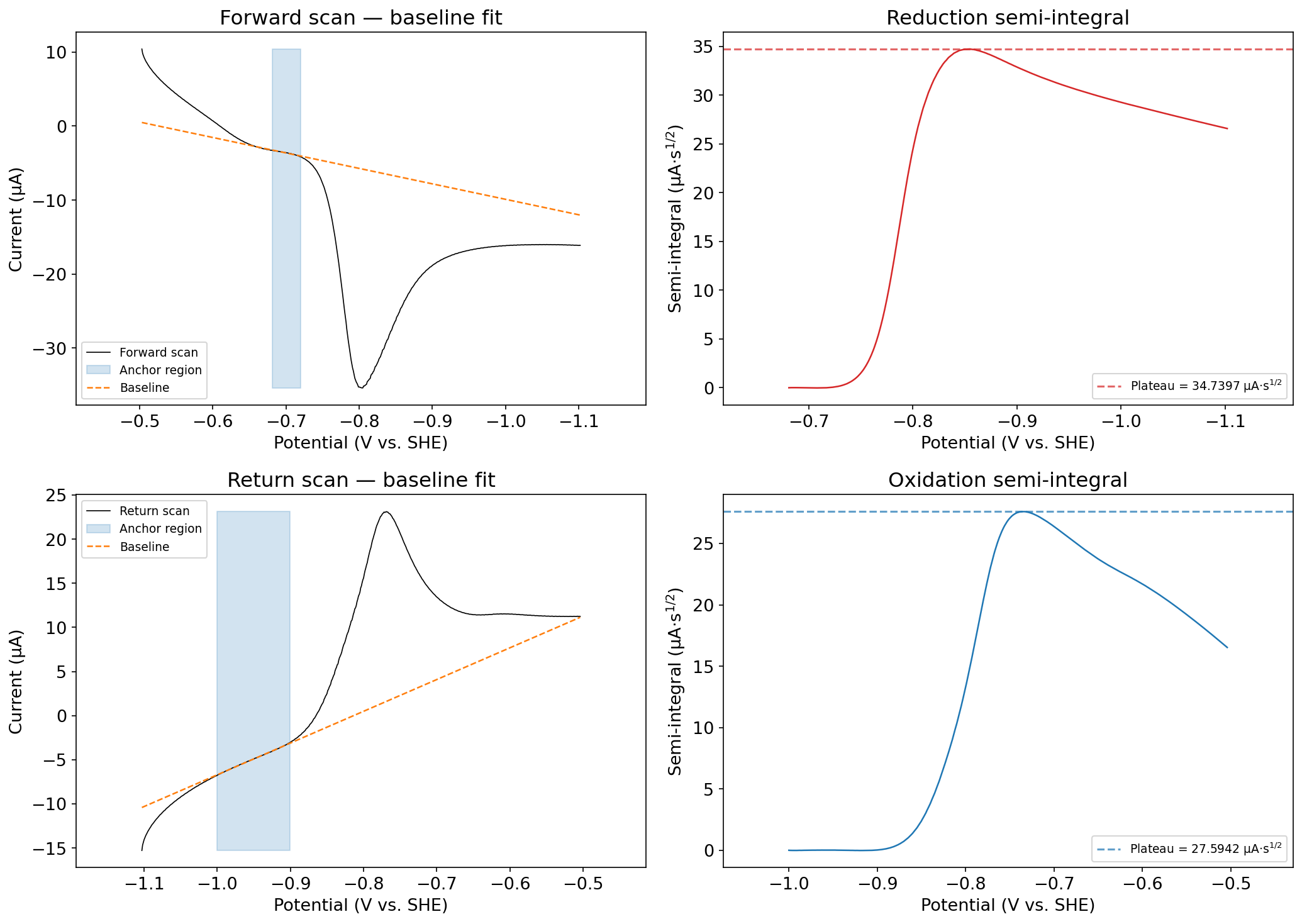}
  \caption{Semi-integration analysis for \cmpd{exllm-aza-bcc} (1~mM, 1~M KOH, 50~mV\,s$^{-1}$). Top: forward scan baseline fit (left) and reduction semi-integral (right). Bottom: return scan baseline fit (left) and oxidation semi-integral (right).}
  \label{fig:semi-cams007}
\end{figure}
\FloatBarrier

\begin{figure}[ht]
  \centering
  \includegraphics[width=\textwidth]{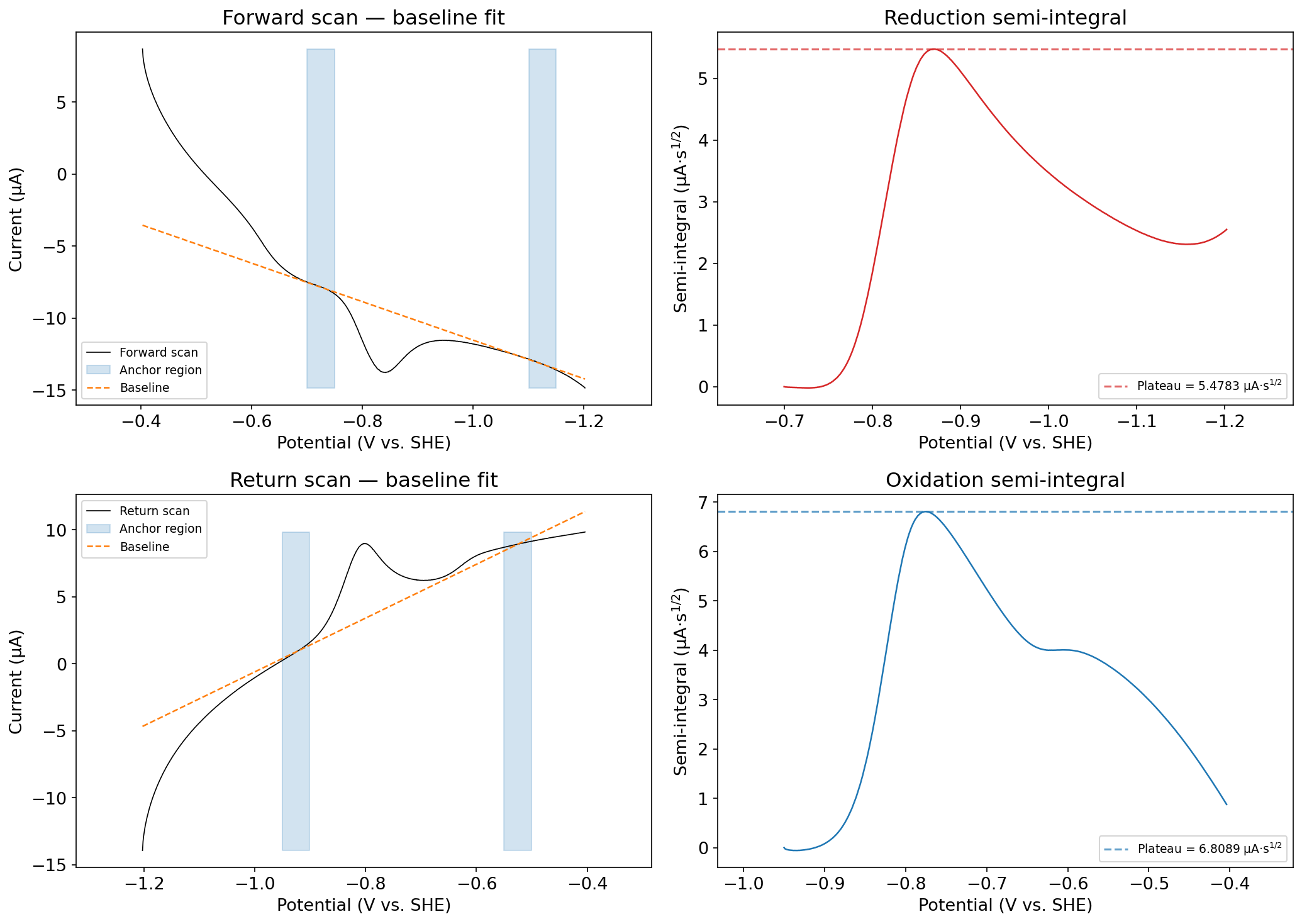}
  \caption{Semi-integration analysis for \cmpd{exllm-OH-aza-bcc} (1~mM, 1~M KOH, 50~mV\,s$^{-1}$). Top: forward scan baseline fit (left) and reduction semi-integral (right). Bottom: return scan baseline fit (left) and oxidation semi-integral (right).}
  \label{fig:semi-cams010}
\end{figure}
\FloatBarrier

\begin{figure}[ht]
  \centering
  \includegraphics[width=\textwidth]{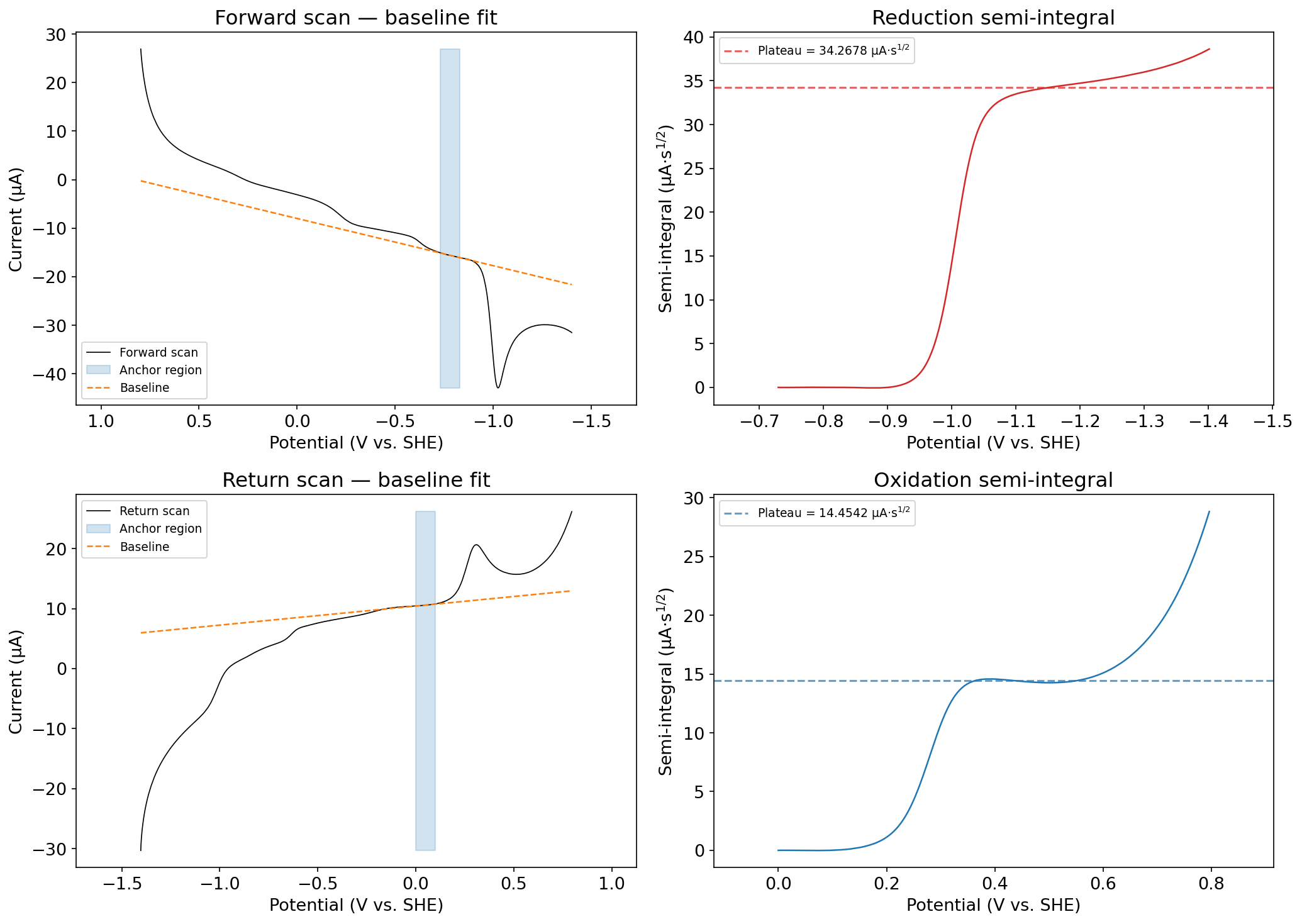}
  \caption{Semi-integration analysis for \cmpd{exllm-CN-bcc} (1~mM, 1~M KOH, 50~mV\,s$^{-1}$). Top: forward scan baseline fit (left) and reduction semi-integral (right). Bottom: return scan baseline fit (left) and oxidation semi-integral (right).}
  \label{fig:semi-cams011}
\end{figure}
\FloatBarrier

\section{Electrochemical characterization}\label{sec:si-electrochem}

\subsection{Materials and reagents list}
All chemicals were used as received unless otherwise stated. Electrolyte solutions were prepared with deionized water (18 M$\Omega\cdot$cm). Ethyl alcohol (200 proof), potassium hydroxide flakes (reagent grade 90\%), sodium phosphate monobasic ($>$99\%), sodium phosphate dibasic ($>$99\%), and deuterium oxide (99\%) were purchased from Sigma-Aldrich. Sodium hydroxide, potassium ferricyanide ($>$99\%), potassium ferrocyanide ($>$99\%), and sulfuric acid (96\%) were purchased from Fisher Scientific. Sodium methanesulfonate (99\%) was purchased from Acros Organics.

Glassy carbon working electrodes (3.0 mm diameter), Ag/AgCl reference electrodes (3M KCl), and platinum wire counter electrodes were purchased from BASi Inc. Cyclic voltammetry was performed on a CHI7013E potentiostat. Spectroelectrochemistry measurements were performed with a Honeycomb Spectrochemical with either gold or platinum as the electrode (Pine Research).  

\subsection{Cyclic voltammetry}

\begin{figure}[ht]
  \centering
  \includegraphics[width=\textwidth]{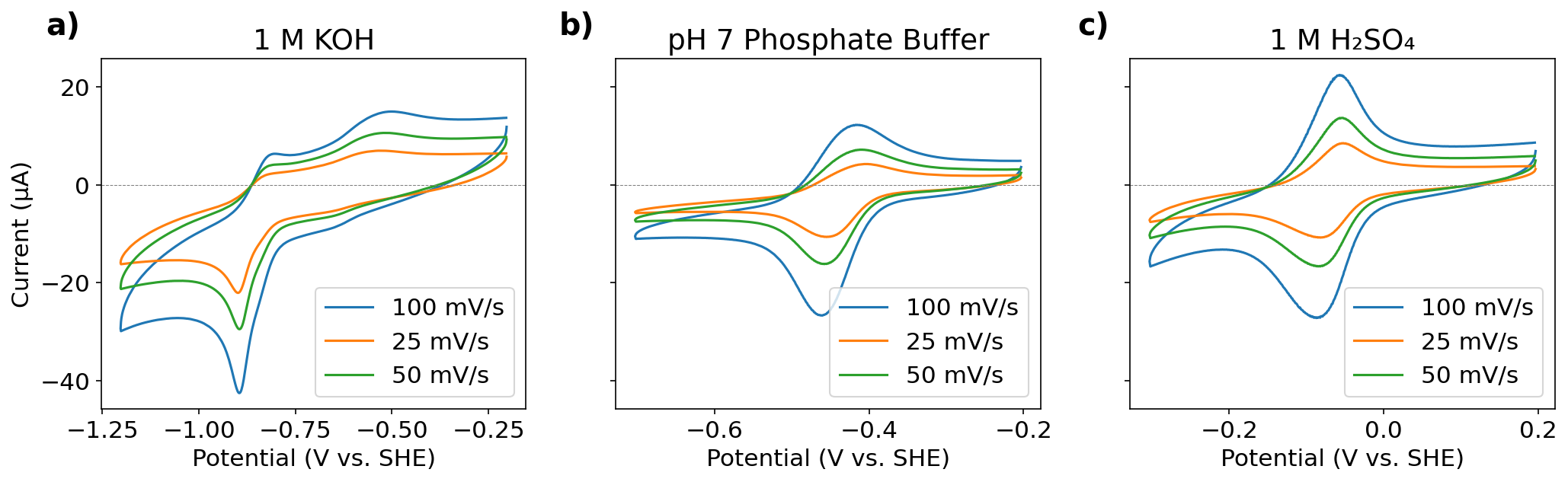}
  \caption{Cyclic voltammetry of compound \cmpd{bcc-BnPO3-5} at three scan rates in (a)~1~M KOH, (b)~pH~7 phosphate buffer, and (c)~1~M H\textsubscript{2}SO\textsubscript{4}.}
  \label{fig:cams013-cv}
\end{figure}
\FloatBarrier

\begin{figure}[ht]
  \centering
  \includegraphics[width=\textwidth]{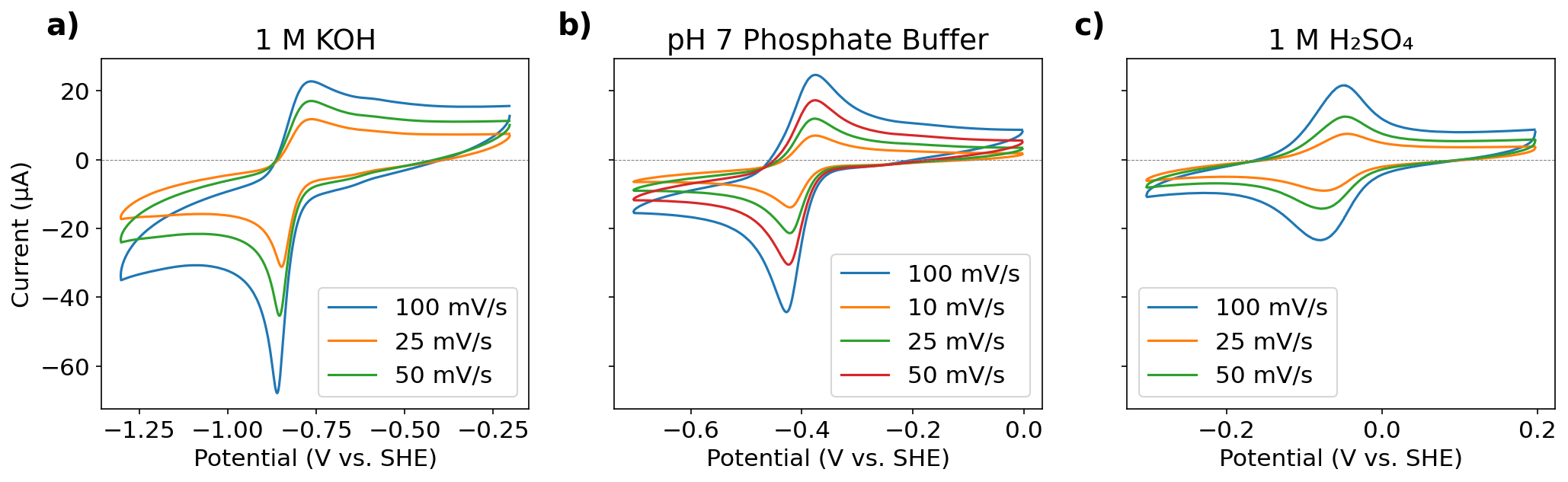}
  \caption{Cyclic voltammetry of compound \cmpd{bcc-BnSO3-5} at three scan rates in (a)~1~M KOH, (b)~pH~7 phosphate buffer, and (c)~1~M H\textsubscript{2}SO\textsubscript{4}.}
  \label{fig:cams014-cv}
\end{figure}

\subsection{Half-peak potential calculations}

Half-peak potentials ($E_{p/2}$) were determined by first establishing a tangent baseline by fitting a first-order polynomial to the current in the $-0.75$ to $-0.60$~V vs.\ SHE region immediately preceding the onset of the Faradaic reduction wave.
The baseline-corrected peak current, $i_{p,\mathrm{corr}}$, was obtained by subtracting the extrapolated baseline current at the peak potential ($E_p$) from the measured peak current.
The half-peak current was defined as $i_{p/2} = i_{p,\mathrm{corr}} / 2$ (referenced to the baseline), and $E_{p/2}$ was determined by linear interpolation of the forward-scan current to the point at which the baseline-corrected current equalled $i_{p/2}$.
Where minimal variance was observed in the measured half-peak potential across scan rates, that determined at 50 mV/s was reported.

\begin{figure}[ht]
  \centering
  \includegraphics[height=1.6in]{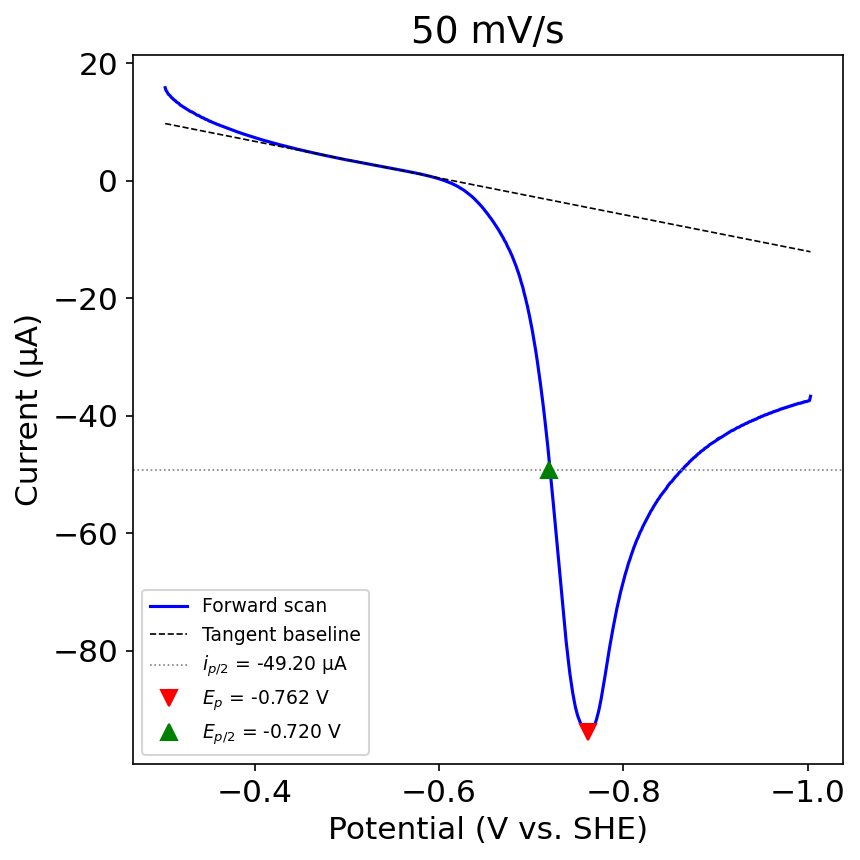}
  \caption{Diagnostic plot for the half-peak potential determination of BzC in 0.5~M KOH at 50~mV/s; data from~\cite{singh2025bc}.
  The dashed line shows the tangent baseline fit to the pre-wave region; the dotted horizontal line indicates $i_{p/2}$.
  Red and green markers denote $E_p$ and $E_{p/2}$, respectively.}
  \label{fig:halfpeak-diagnostic-bzc}
\end{figure}
\FloatBarrier

\begin{figure}[ht]
  \centering
  \includegraphics[width=\textwidth]{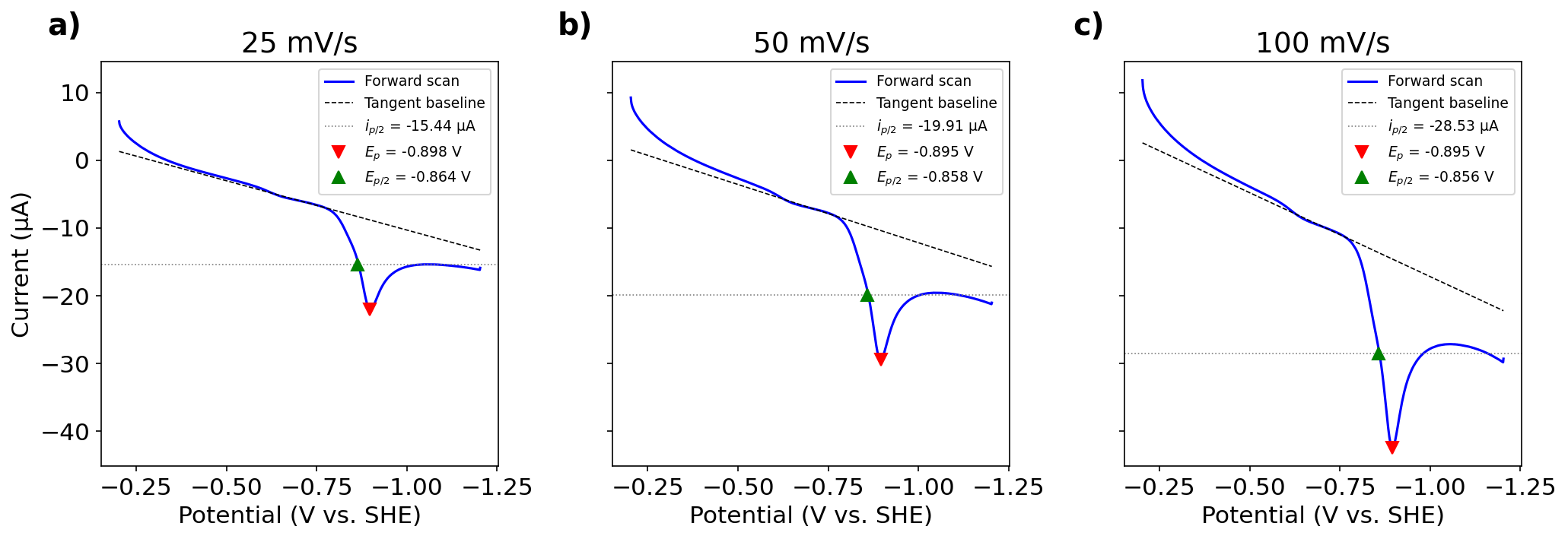}
  \caption{Diagnostic plots for the half-peak potential determination of compound \cmpd{bcc-BnPO3-5} in 1~M KOH at (a)~25~mV/s, (b)~50~mV/s, and (c)~100~mV/s.
  The dashed line shows the tangent baseline fit to the pre-wave region; the dotted horizontal line indicates $i_{p/2}$.
  Red and green markers denote $E_p$ and $E_{p/2}$, respectively.}
  \label{fig:halfpeak-diagnostic}
\end{figure}
\FloatBarrier

\begin{figure}[ht]
  \centering
  \includegraphics[width=\textwidth]{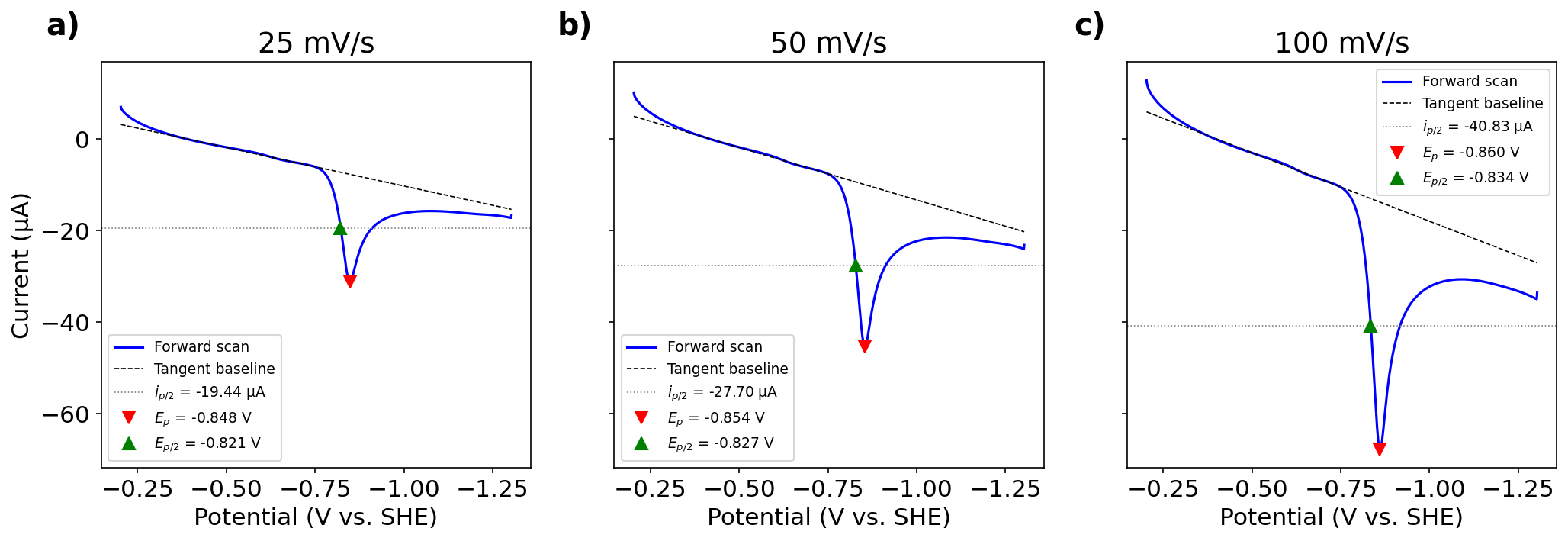}
  \caption{Diagnostic plots for the half-peak potential determination of compound \cmpd{bcc-BnSO3-5} in 1~M KOH at (a)~25~mV/s, (b)~50~mV/s, and (c)~100~mV/s.
  The dashed line shows the tangent baseline fit to the pre-wave region; the dotted horizontal line indicates $i_{p/2}$.
  Red and green markers denote $E_p$ and $E_{p/2}$, respectively.}
  \label{fig:halfpeak-diagnostic-014}
\end{figure}
\FloatBarrier

\subsection{Semi-integration charge ratio analysis}

Charge ratios ($Q_\mathrm{ox}/Q_\mathrm{red}$) were determined by semi-integration (convolution voltammetry) of the baseline-corrected cyclic voltammograms.
A linear baseline was fit to an anchor region on the capacitive background of each half-cycle and extrapolated across the full scan; the baseline-corrected current was then semi-integrated starting from the anchor region.
For a diffusion-controlled process, the semi-integral plateau is
\begin{equation}
  m_\mathrm{lim} = nFAC\sqrt{D}
\end{equation}
where $n$ is the number of electrons, $F$ is the Faraday constant, $A$ is the electrode area, $C$ is the bulk concentration, and $D$ is the diffusion coefficient.
Because $n$, $F$, $A$, $C$, and $D$ are identical for the reduction and re-oxidation of the same species at the same electrode, the ratio of plateaus gives
\begin{equation}
  \frac{Q_\mathrm{ox}}{Q_\mathrm{red}} = \frac{m_\mathrm{lim,ox}}{m_\mathrm{lim,red}}
\end{equation}
directly, without requiring absolute charge values.

\begin{figure}[ht]
  \centering
  \includegraphics[width=\textwidth]{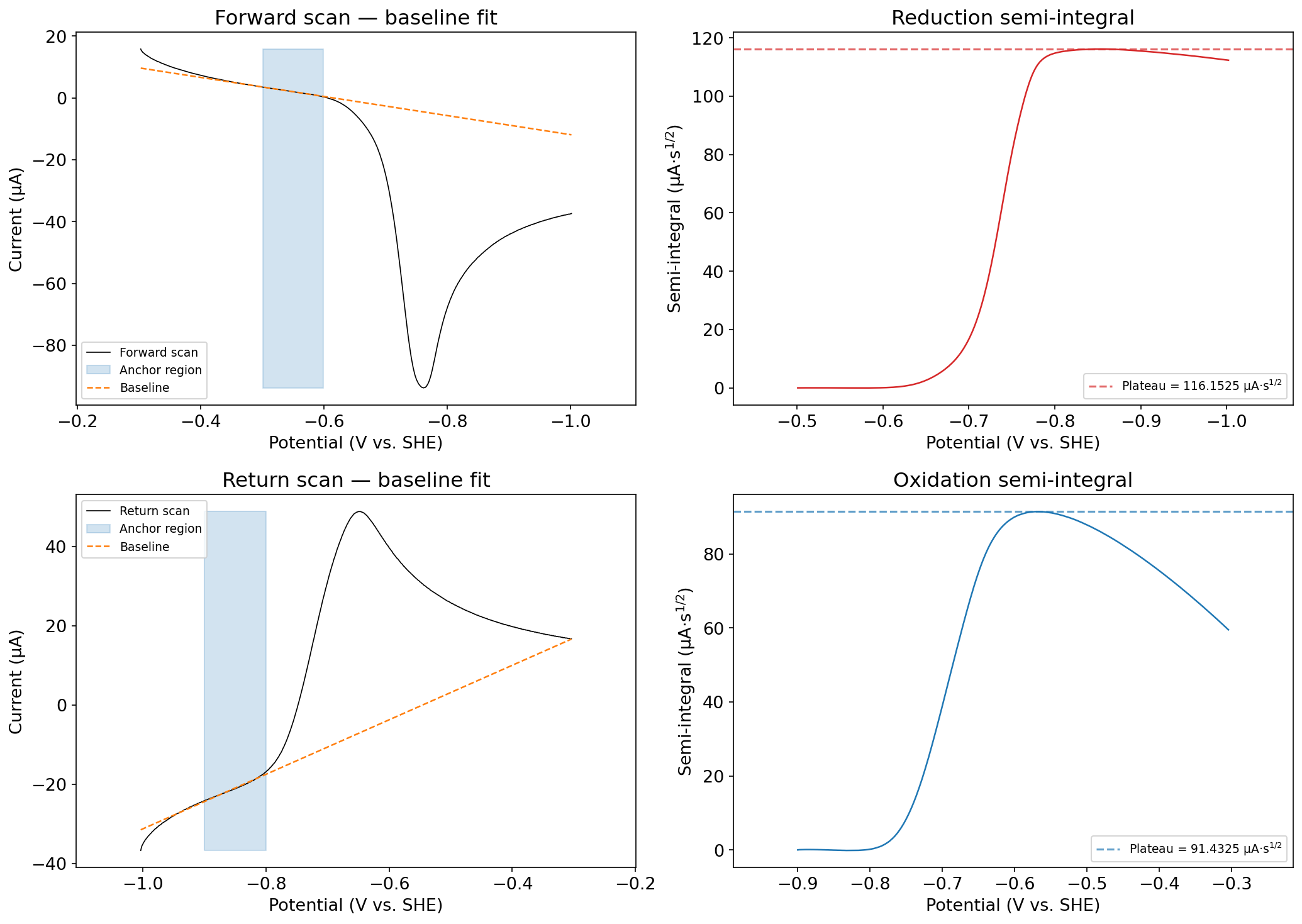}
  \caption{Semi-integration analysis for \cmpd{sBC} (5~mM, 0.5~M KOH, 50~mV\,s$^{-1}$). Top: forward scan baseline fit (left) and reduction semi-integral (right). Bottom: return scan baseline fit (left) and oxidation semi-integral (right). Dashed lines indicate the plateau values used to compute $Q_\mathrm{ox}/Q_\mathrm{red}$.}
  \label{fig:semi-bzc}
\end{figure}
\FloatBarrier

\begin{figure}[ht]
  \centering
  \includegraphics[width=\textwidth]{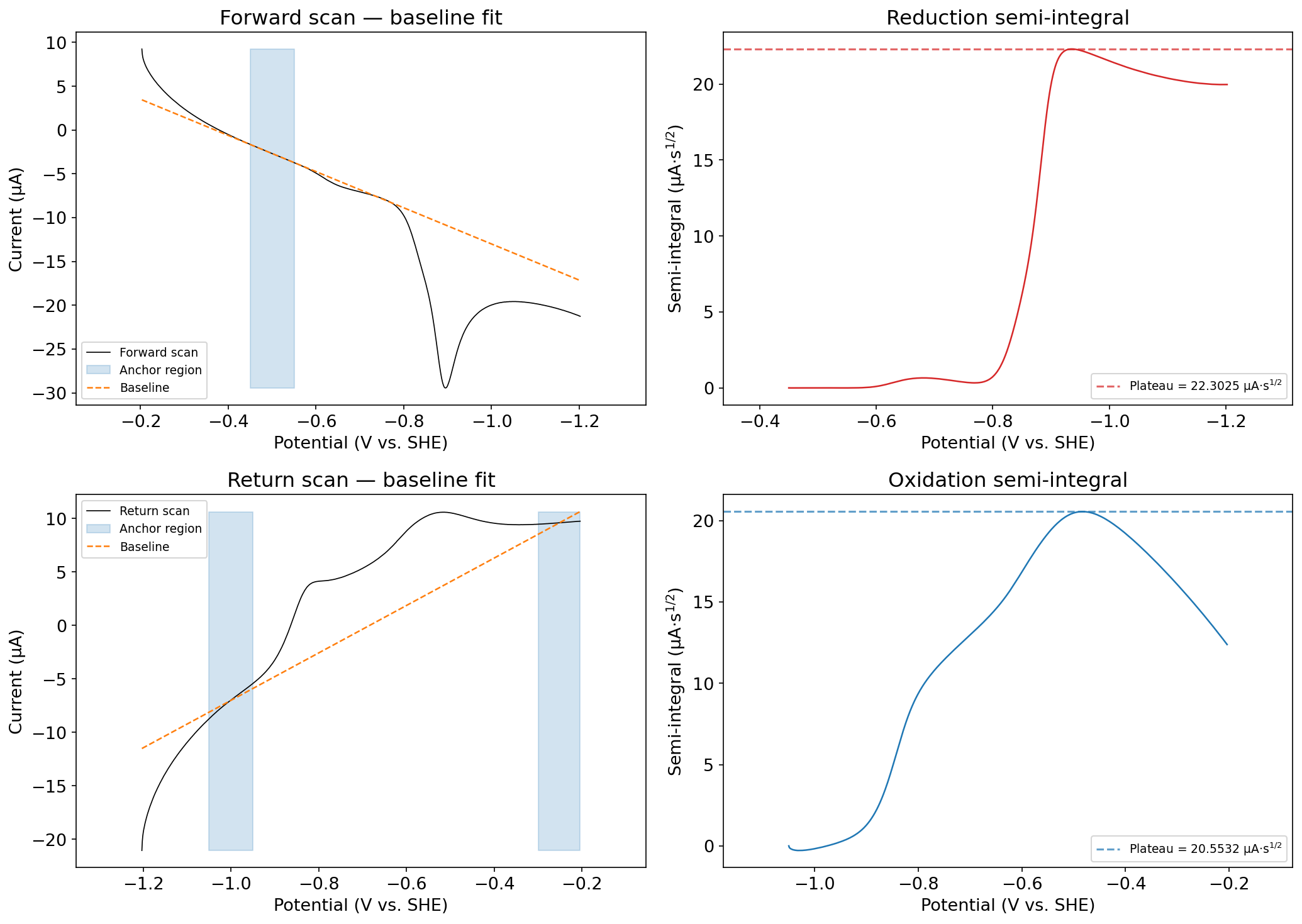}
  \caption{Semi-integration analysis for \cmpd{bcc-BnPO3-5} (1~mM, 1~M KOH, 50~mV\,s$^{-1}$). Top: forward scan baseline fit (left) and reduction semi-integral (right). Bottom: return scan baseline fit (left) and oxidation semi-integral (right).}
  \label{fig:semi-cams013}
\end{figure}
\FloatBarrier

\begin{figure}[ht]
  \centering
  \includegraphics[width=\textwidth]{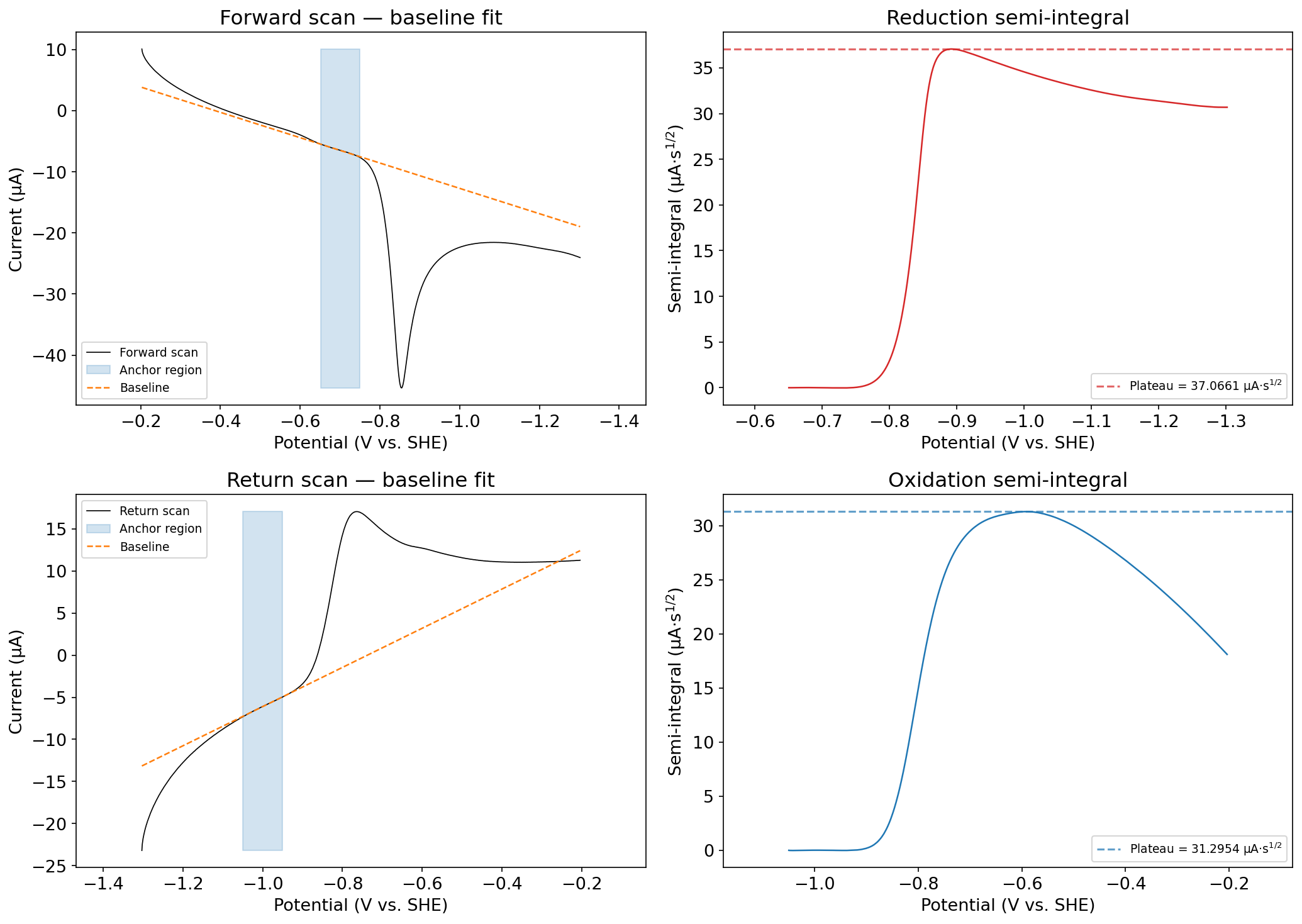}
  \caption{Semi-integration analysis for \cmpd{bcc-BnSO3-5} (1~mM, 1~M KOH, 50~mV\,s$^{-1}$). Top: forward scan baseline fit (left) and reduction semi-integral (right). Bottom: return scan baseline fit (left) and oxidation semi-integral (right).}
  \label{fig:semi-cams014}
\end{figure}
\FloatBarrier

\subsection{Potassium hydroxide concentration dependence}
\begin{figure}[h!]
  \centering
  \includegraphics[width=\textwidth]{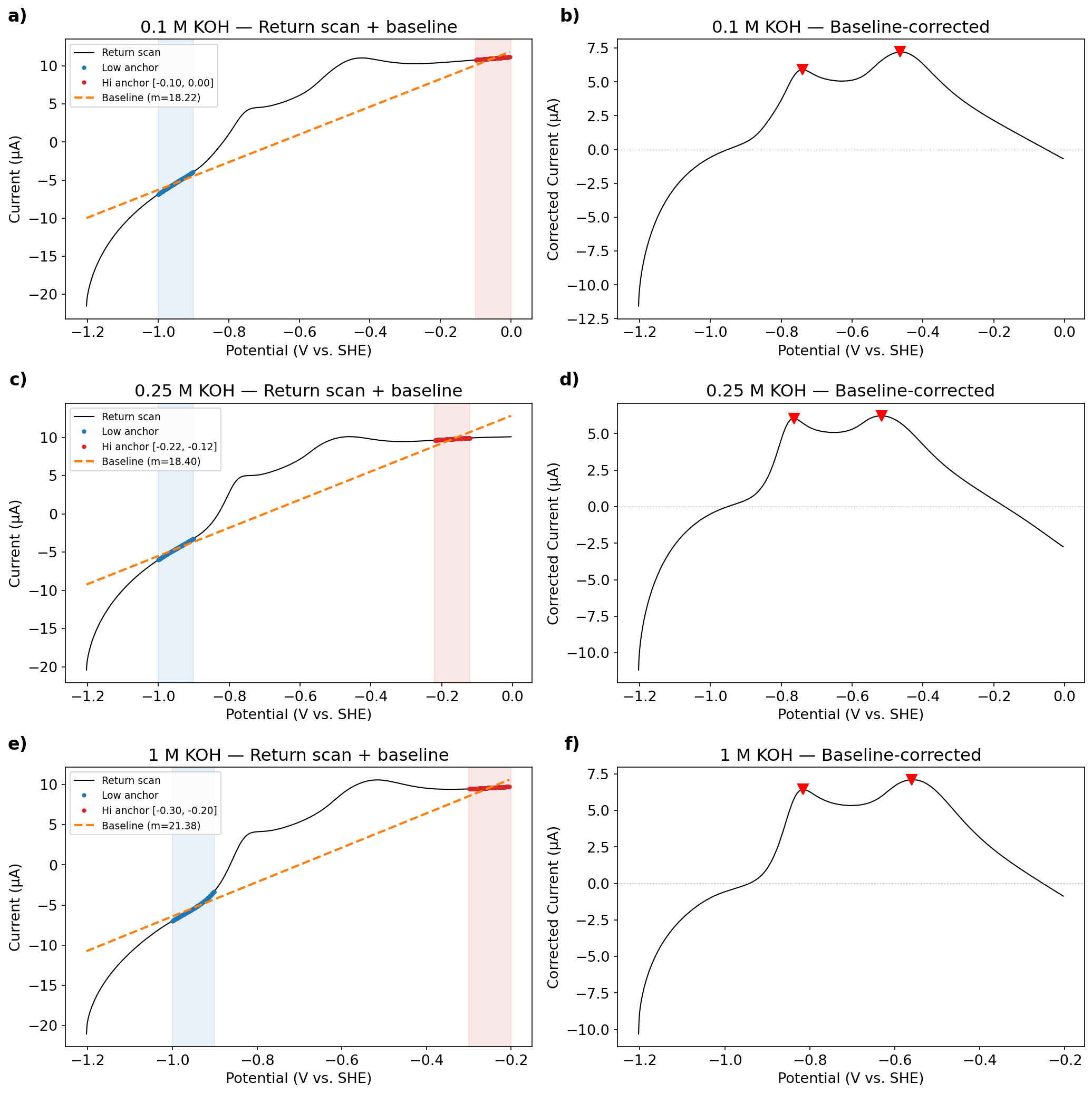}
  \caption{Baseline correction of the anodic return scan for 1~mM \cmpd{bcc-BnPO3-5} at 50~mV\,s$^{-1}$ in 0.1~M~KOH (a,\,b), 0.25~M~KOH (c,\,d), and 1~M~KOH (e,\,f). Left panels show the return scan (black) with anchor regions (shaded) and the linear baseline (dashed orange). Right panels show the baseline-corrected current with peak maxima marked by red triangles.}
  \label{fig:koh-sweep-diagnostic}
\end{figure}

Cyclic voltammograms of 1~mM \cmpd{bcc-BnPO3-5} were recorded at 50~mV\,s$^{-1}$ in 0.1, 0.25, and 1~M KOH.
A linear baseline was fitted to anchor regions on the return (anodic) scan where the current was approximately constant.
The low anchor was fixed at $-1.00$ to $-0.90$~V vs.\ SHE; the high anchor was a 0.10~V window selected automatically by sliding from $-0.40$~V toward the positive scan limit and choosing the position that minimised the root-mean-square residual of the combined linear fit.
Figure~\ref{fig:koh-sweep-diagnostic} shows the raw return scans with the fitted baselines (left) and the baseline-corrected oxidation waves with the two peak maxima indicated (right).

\FloatBarrier

\begin{table}[h!]
  \centering
  \caption{Oxidation peak positions and currents from the baseline-corrected anodic scan of 1~mM \cmpd{bcc-BnPO3-5} at 50~mV\,s$^{-1}$.}
  \label{tab:koh-sweep-peaks}
  \begin{tabular}{l c c c c c}
    \hline
    KOH & Peak~1 (V) & Peak~1 ($\mu$A) & Peak~2 (V) & Peak~2 ($\mu$A) & Ratio (P2/P1) \\
    \hline
    0.1~M  & $-0.740$ & 5.90 & $-0.465$ & 7.22 & 1.22 \\
    0.25~M & $-0.764$ & 6.04 & $-0.516$ & 6.22 & 1.03 \\
    1~M    & $-0.816$ & 6.43 & $-0.560$ & 7.10 & 1.10 \\
    \hline
  \end{tabular}
\end{table}
\FloatBarrier

\subsection{Scan rate dependence (0.1~M KOH)}
\begin{figure}[h!]
  \centering
  \includegraphics[width=\textwidth]{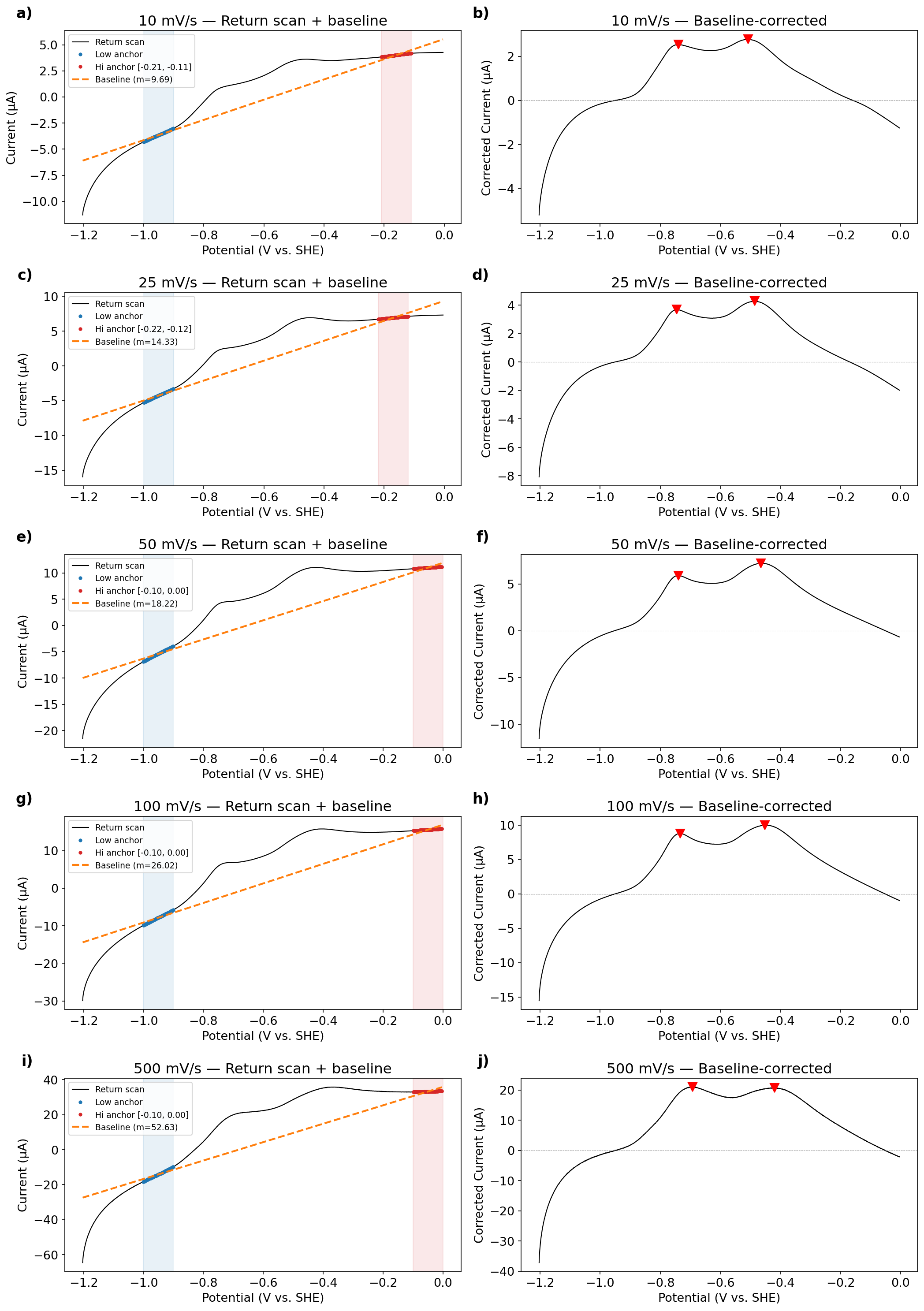}
  \caption{Baseline correction of the anodic return scan for 1~mM \cmpd{bcc-BnPO3-5} in 0.1~M KOH at scan rates of 10, 25, 50, 100, and 500~mV\,s$^{-1}$ (top to bottom). Left panels show the return scan (black) with anchor regions (shaded) and fitted linear baseline (dashed orange). Right panels show the baseline-corrected current with peak maxima marked by red triangles.}
  \label{fig:scanrate-diagnostic}
\end{figure}

Cyclic voltammograms of 1~mM \cmpd{bcc-BnPO3-5} were recorded at 10-500~mV\,s$^{-1}$ in 0.1~M KOH.
The baseline-correction procedure described in the potassium hydroxide concentration dependence experiment was used.
Figure~\ref{fig:scanrate-diagnostic} shows the fitted baselines (left) and the corrected oxidation waves with peak maxima indicated (right) for each scan rate.
Table~\ref{tab:scanrate-peaks} summarises the extracted peak positions, currents, and P2/P1 ratios.

\FloatBarrier

\begin{table}[h!]
  \centering
  \caption{Oxidation peak positions and currents from the baseline-corrected anodic scan of 1~mM \cmpd{bcc-BnPO3-5} in 0.1~M KOH at varying scan rates.}
  \label{tab:scanrate-peaks}
  \begin{tabular}{l c c c c c}
    \hline
    Scan rate (mV\,s$^{-1}$) & Peak~1 (V) & Peak~1 ($\mu$A) & Peak~2 (V) & Peak~2 ($\mu$A) & Ratio (P2/P1) \\
    \hline
    10  & $-0.740$ & 2.54 & $-0.508$ & 2.77 & 1.09 \\
    25  & $-0.745$ & 3.69 & $-0.486$ & 4.28 & 1.16 \\
    50  & $-0.740$ & 5.90 & $-0.465$ & 7.22 & 1.22 \\
    100 & $-0.734$ & 8.79 & $-0.452$ & 10.01 & 1.14 \\
    500 & $-0.692$ & 20.96 & $-0.420$ & 20.68 & 0.99 \\
    \hline
  \end{tabular}
\end{table}
\FloatBarrier

\subsection{Scan rate dependence (0.25~M KOH)}
\begin{figure}[h!]
  \centering
  \includegraphics[width=\textwidth]{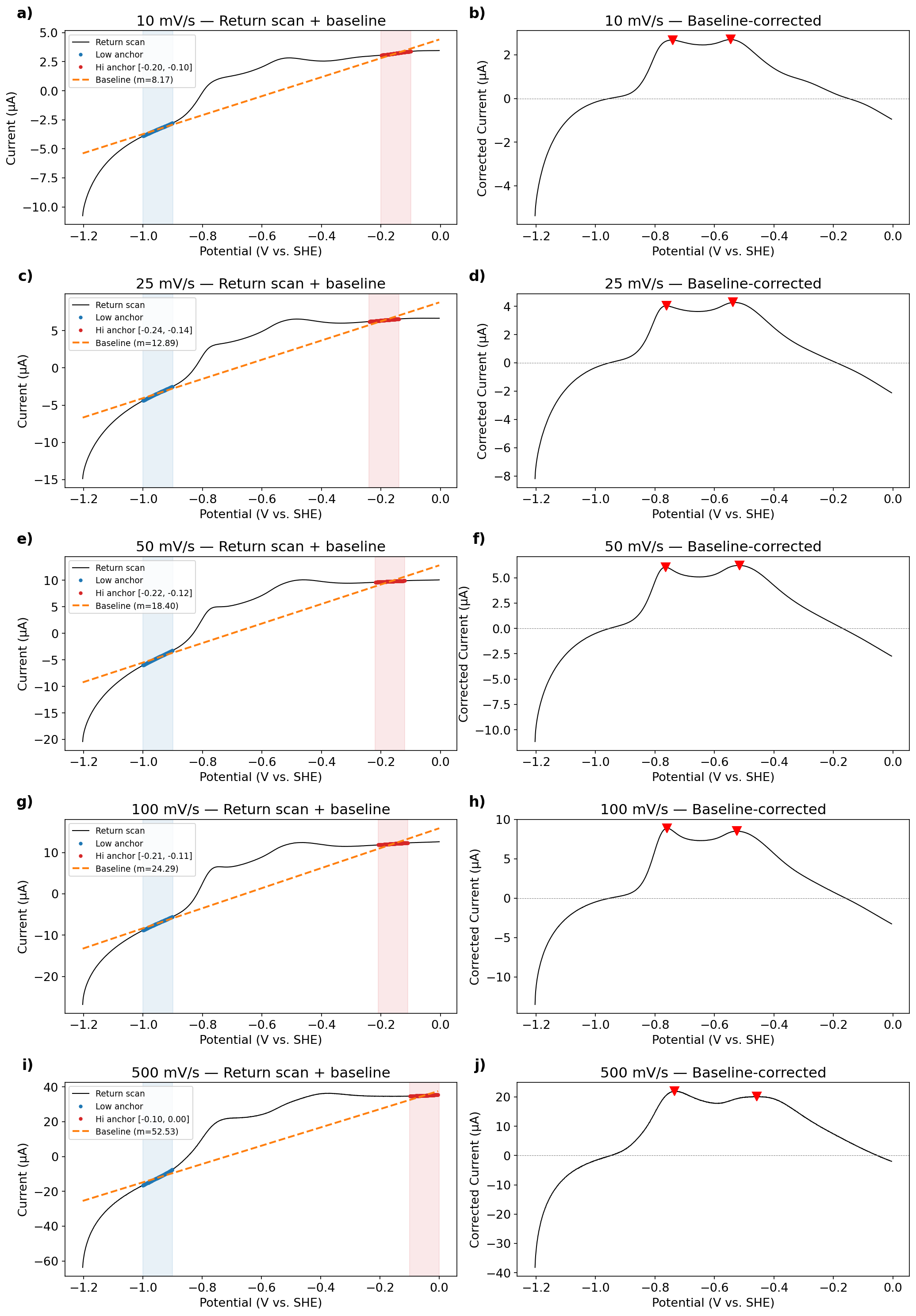}
  \caption{Baseline correction of the anodic return scan for 1~mM \cmpd{bcc-BnPO3-5} in 0.25~M KOH at scan rates of 10, 25, 50, 100, and 500~mV\,s$^{-1}$ (top to bottom). Left panels show the return scan (black) with anchor regions (shaded) and fitted linear baseline (dashed orange). Right panels show the baseline-corrected current with peak maxima marked by red triangles.}
  \label{fig:scanrate-025M-diagnostic}
\end{figure}

Cyclic voltammograms of 1~mM \cmpd{bcc-BnPO3-5} were recorded at 10-500~mV\,s$^{-1}$ in 0.25~M KOH.
The baseline-correction procedure described in the potassium hydroxide concentration dependence experiment was used.
Figure~\ref{fig:scanrate-025M-diagnostic} shows the fitted baselines and corrected oxidation waves, and Table~\ref{tab:scanrate-025M-peaks} summarises the peak data.

\FloatBarrier

\begin{table}[h!]
  \centering
  \caption{Oxidation peak positions and currents from the baseline-corrected anodic scan of 1~mM \cmpd{bcc-BnPO3-5} in 0.25~M KOH at varying scan rates.}
  \label{tab:scanrate-025M-peaks}
  \begin{tabular}{l c c c c c}
    \hline
    Scan rate (mV\,s$^{-1}$) & Peak~1 (V) & Peak~1 ($\mu$A) & Peak~2 (V) & Peak~2 ($\mu$A) & Ratio (P2/P1) \\
    \hline
    10  & $-0.740$ & 2.67 & $-0.545$ & 2.70 & 1.01 \\
    25  & $-0.761$ & 4.03 & $-0.537$ & 4.27 & 1.06 \\
    50  & $-0.764$ & 6.04 & $-0.516$ & 6.22 & 1.03 \\
    100 & $-0.759$ & 8.88 & $-0.524$ & 8.53 & 0.96 \\
    500 & $-0.734$ & 21.94 & $-0.457$ & 20.14 & 0.92 \\
    \hline
  \end{tabular}
\end{table}
\FloatBarrier

\subsection{Scan rate dependence (1~M KOH)}
\begin{figure}[h!]
  \centering
  \includegraphics[width=\textwidth]{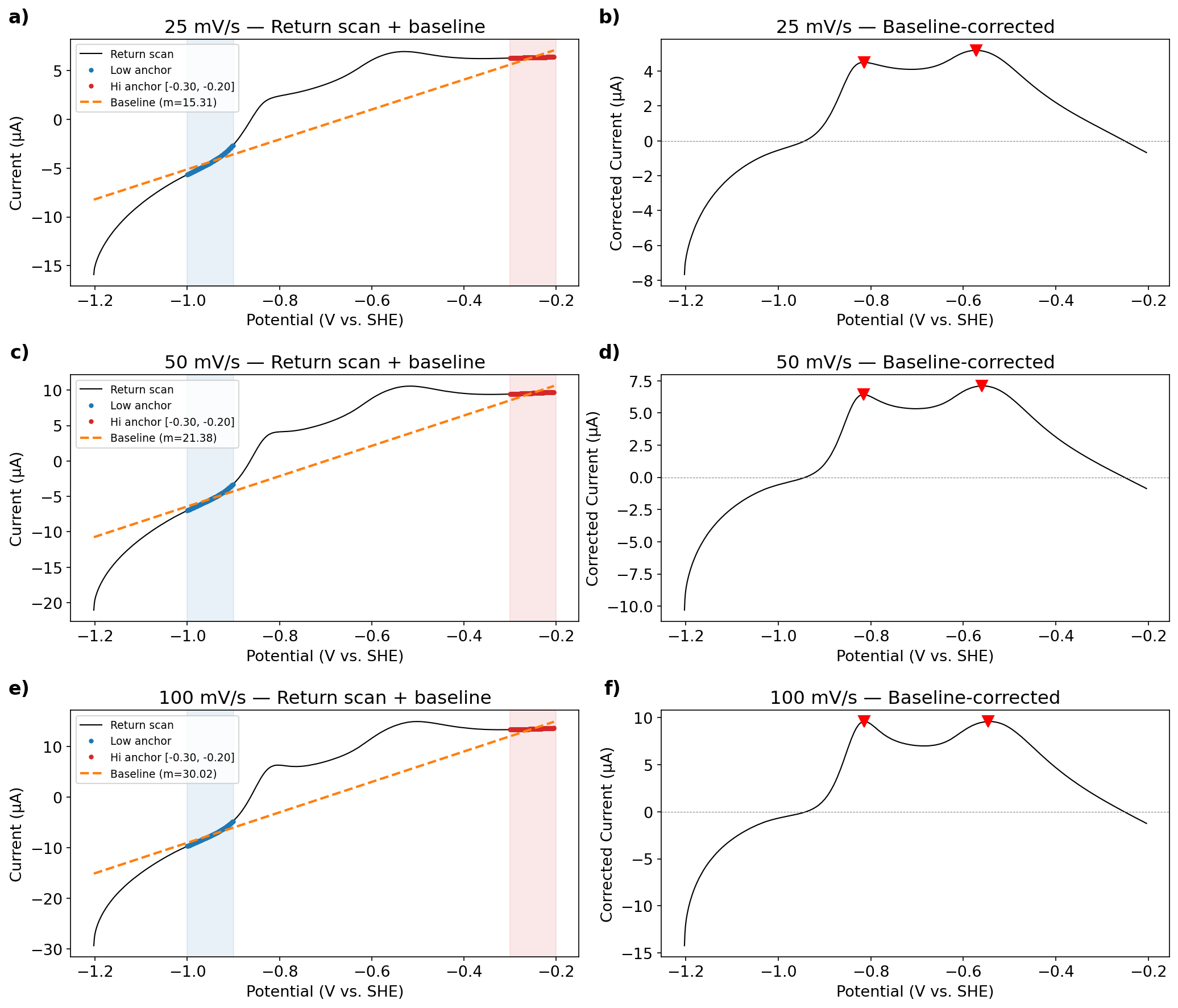}
  \caption{Baseline correction of the anodic return scan for 1~mM \cmpd{bcc-BnPO3-5} in 1~M KOH at scan rates of 25, 50, and 100~mV\,s$^{-1}$ (top to bottom). Left panels show the return scan (black) with anchor regions (shaded) and fitted linear baseline (dashed orange). Right panels show the baseline-corrected current with peak maxima marked by red triangles.}
  \label{fig:scanrate-1M-diagnostic}
\end{figure}

Cyclic voltammograms of 1~mM \cmpd{bcc-BnPO3-5} were recorded at 10-500~mV\,s$^{-1}$ in 1~M KOH.
The baseline-correction procedure described in the potassium hydroxide concentration dependence experiment was used.
Figure~\ref{fig:scanrate-1M-diagnostic} shows the fitted baselines and corrected oxidation waves, and Table~\ref{tab:scanrate-1M-peaks} summarises the peak data.

\FloatBarrier

\begin{table}[h!]
  \centering
  \caption{Oxidation peak positions and currents from the baseline-corrected anodic scan of 1~mM \cmpd{bcc-BnPO3-5} in 1~M KOH at varying scan rates.}
  \label{tab:scanrate-1M-peaks}
  \begin{tabular}{l c c c c c}
    \hline
    Scan rate (mV\,s$^{-1}$) & Peak~1 (V) & Peak~1 ($\mu$A) & Peak~2 (V) & Peak~2 ($\mu$A) & Ratio (P2/P1) \\
    \hline
    25  & $-0.815$ & 4.51 & $-0.572$ & 5.19 & 1.15 \\
    50  & $-0.816$ & 6.43 & $-0.560$ & 7.10 & 1.10 \\
    100 & $-0.815$ & 9.60 & $-0.546$ & 9.60 & 1.00 \\
    \hline
  \end{tabular}
\end{table}
\FloatBarrier

\subsection{KOH concentration dependence at additional scan rates}

\begin{figure}[h!]
  \centering
  \includegraphics[width=\textwidth]{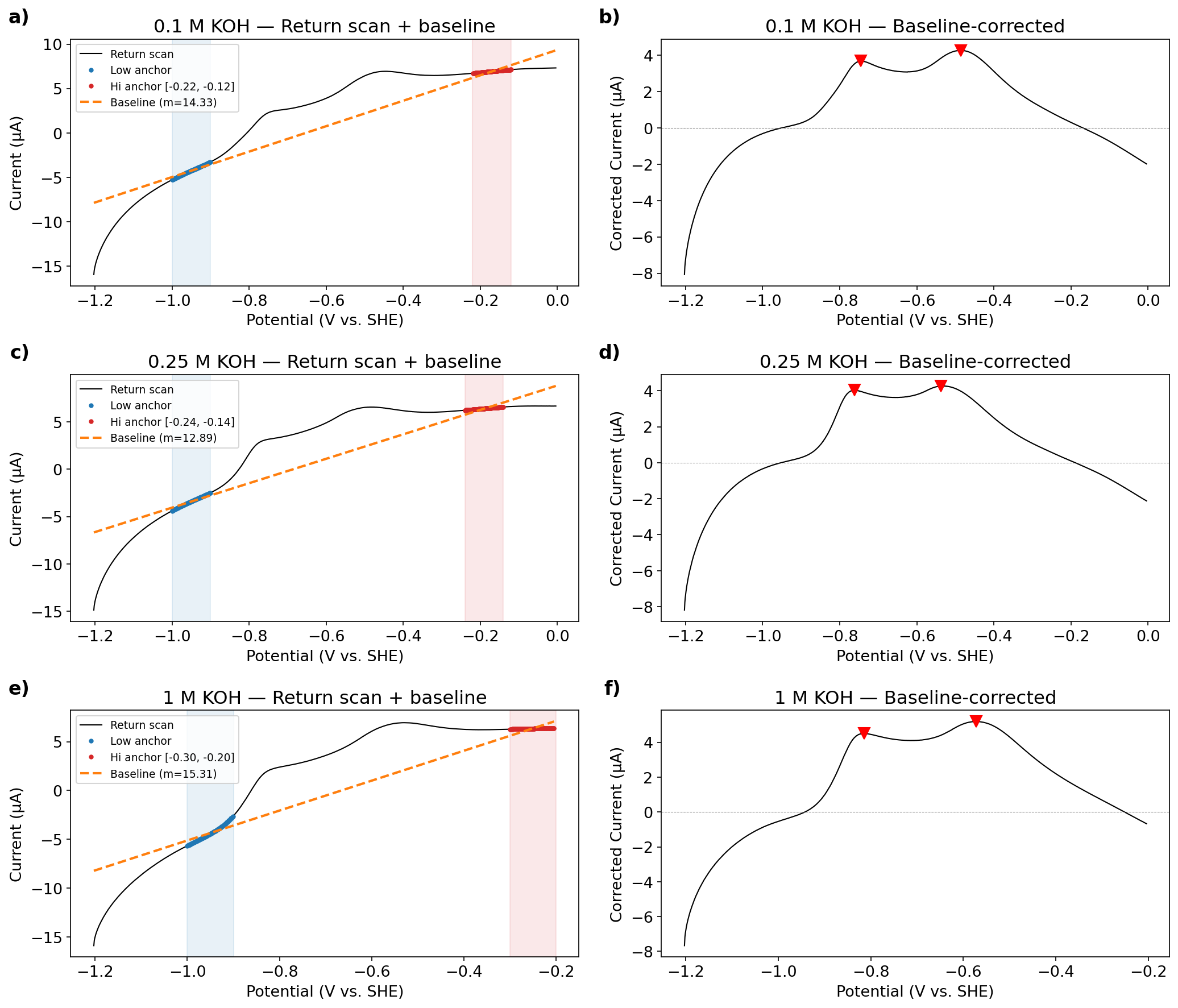}
  \caption{Baseline correction of the anodic return scan for 1~mM \cmpd{bcc-BnPO3-5} at 25~mV\,s$^{-1}$ in 0.1~M KOH (a,\,b), 0.25~M KOH (c,\,d), and 1~M KOH (e,\,f).}
  \label{fig:koh-sweep-25mVs}
\end{figure}

\begin{figure}[h!]
  \centering
  \includegraphics[width=\textwidth]{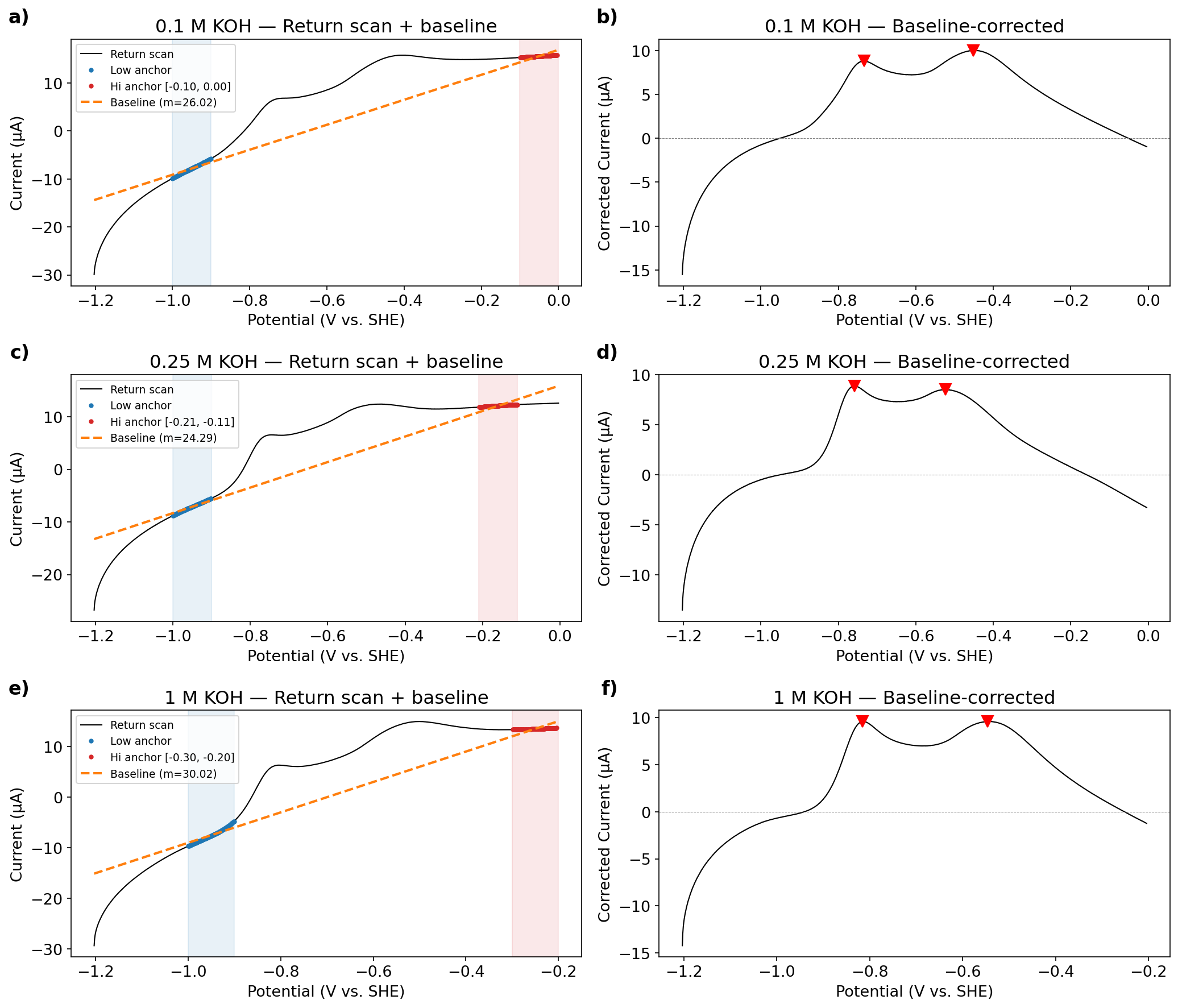}
  \caption{Baseline correction of the anodic return scan for 1~mM \cmpd{bcc-BnPO3-5} at 100~mV\,s$^{-1}$ in 0.1~M KOH (a,\,b), 0.25~M KOH (c,\,d), and 1~M KOH (e,\,f).}
  \label{fig:koh-sweep-100mVs}
\end{figure}

Following the matrix design of experiments, the baseline-correction analysis described in the potassium hydroxide concentration dependence experiment was performed at 25 and 100~mV\,s$^{-1}$ across 0.1, 0.25, and 1~M KOH.
The baseline-correction procedure described in the potassium hydroxide concentration dependence experiment was used.
Figure~\ref{fig:koh-sweep-25mVs} and Figure~\ref{fig:koh-sweep-100mVs} show the fitted baselines and corrected oxidation waves.
Table~\ref{tab:scanrate-025M-peaks} and Table~\ref{tab:scanrate-1M-peaks} summarize the peak data.

\FloatBarrier

\begin{table}[h!]
  \centering
  \caption{Oxidation peak data at 25~mV\,s$^{-1}$ across KOH concentrations.}
  \label{tab:koh-sweep-25mVs}
  \begin{tabular}{l c c c c c}
    \hline
    KOH & Peak~1 (V) & Peak~1 ($\mu$A) & Peak~2 (V) & Peak~2 ($\mu$A) & Ratio (P2/P1) \\
    \hline
    0.1~M  & $-0.745$ & 3.69 & $-0.486$ & 4.28 & 1.16 \\
    0.25~M & $-0.761$ & 4.03 & $-0.537$ & 4.27 & 1.06 \\
    1~M    & $-0.815$ & 4.51 & $-0.572$ & 5.19 & 1.15 \\
    \hline
  \end{tabular}
\end{table}
\FloatBarrier

\FloatBarrier

\begin{table}[h!]
  \centering
  \caption{Oxidation peak data at 100~mV\,s$^{-1}$ across KOH concentrations.}
  \label{tab:koh-sweep-100mVs}
  \begin{tabular}{l c c c c c}
    \hline
    KOH & Peak~1 (V) & Peak~1 ($\mu$A) & Peak~2 (V) & Peak~2 ($\mu$A) & Ratio (P2/P1) \\
    \hline
    0.1~M  & $-0.734$ & 8.79 & $-0.452$ & 10.01 & 1.14 \\
    0.25~M & $-0.759$ & 8.88 & $-0.524$ & 8.53 & 0.96 \\
    1~M    & $-0.815$ & 9.60 & $-0.546$ & 9.60 & 1.00 \\
    \hline
  \end{tabular}
\end{table}
\FloatBarrier

\subsection{Scan rate dependence (0.5~M NaOH)}

\begin{figure}[h!]
  \centering
  \includegraphics[width=0.7\textwidth]{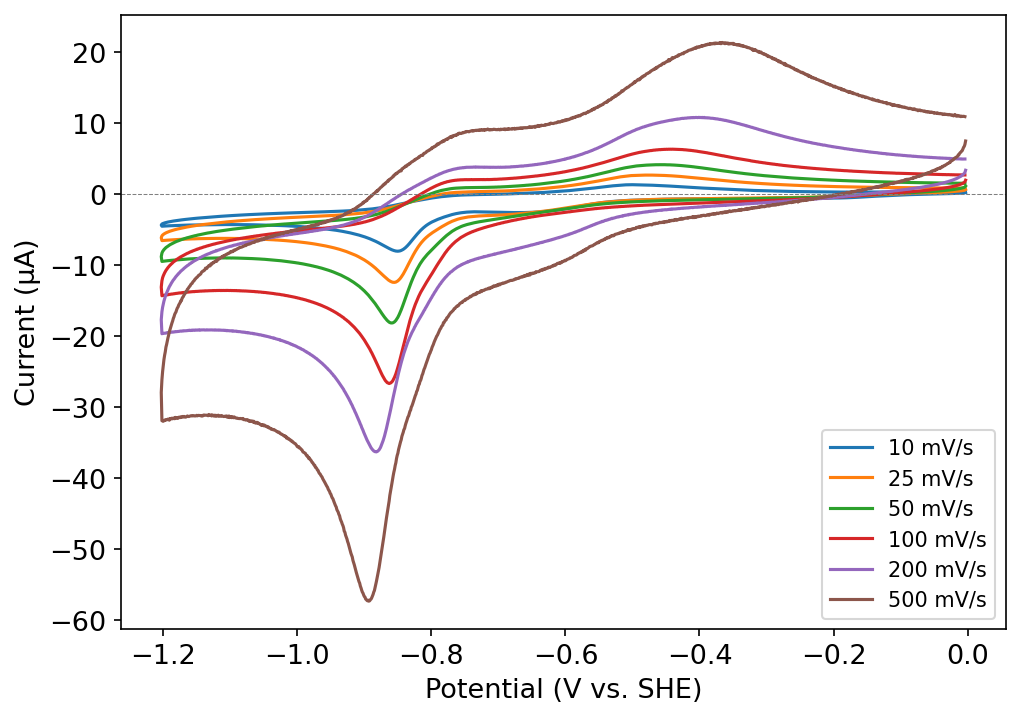}
  \caption{Cyclic voltammograms of 1~mM \cmpd{bcc-BnPO3-5} in 0.5~M NaOH at scan rates
    from 10 to 500~mV\,s$^{-1}$.}
  \label{fig:naoh-cv-overlay}
\end{figure}

\begin{figure}[h!]
  \centering
  \includegraphics[width=\textwidth]{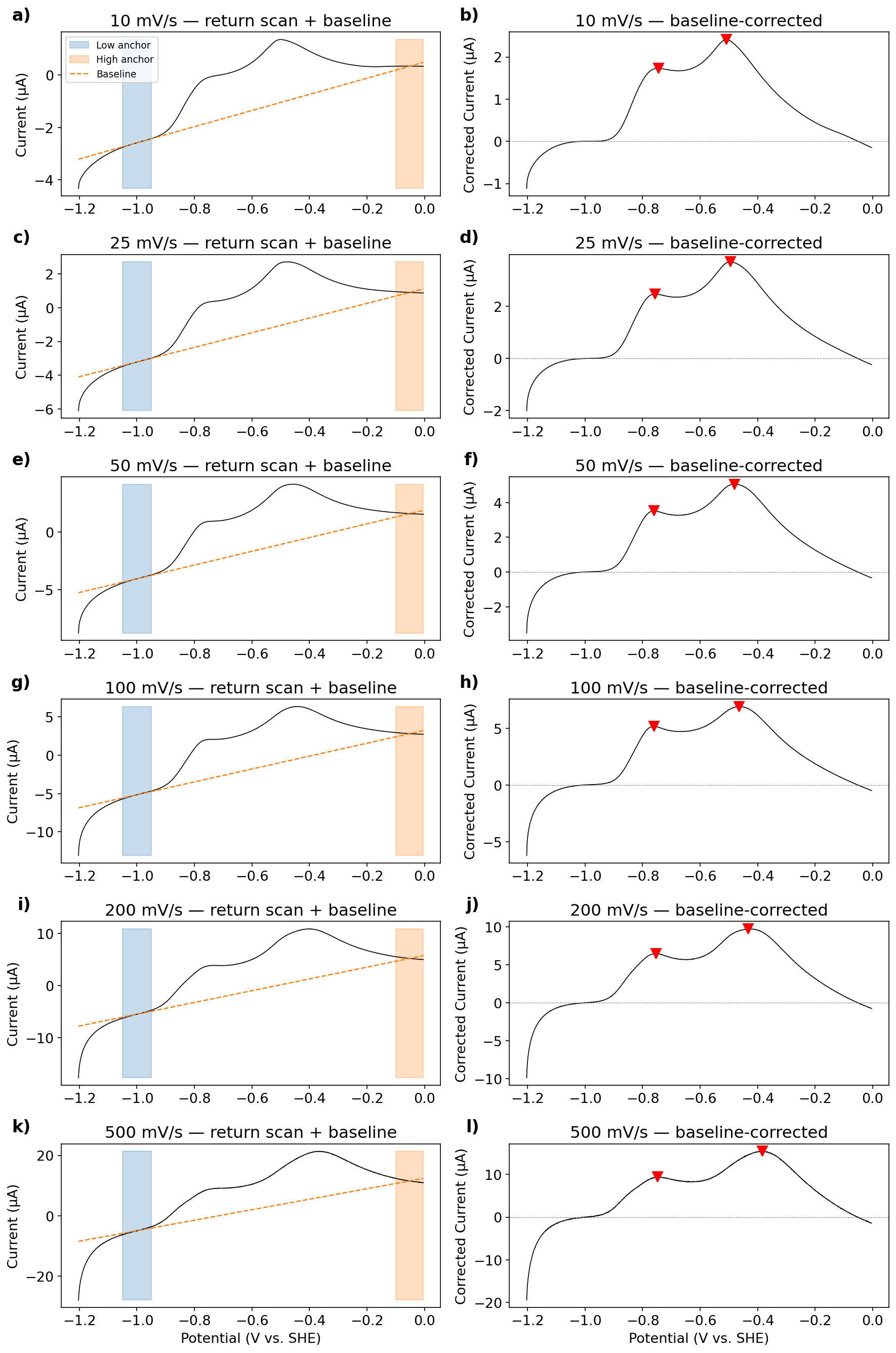}
  \caption{Baseline correction of the anodic return scan for 1~mM \cmpd{bcc-BnPO3-5}
    Left panels show the return scan (black) with anchor regions (shaded) and fitted
    linear baseline (dashed orange). Right panels show the baseline-corrected current
    with peak maxima marked by red triangles.}
  \label{fig:naoh-scanrate-diagnostic}
\end{figure}

Cyclic voltammograms of 1~mM \cmpd{bcc-BnPO3-5} were recorded at 10-500~mV\,s$^{-1}$ in 0.5~M NaOH (Figure~\ref{fig:naoh-cv-overlay}).
The baseline-correction procedure described in the potassium hydroxide concentration dependence experiment was used.
Figure~\ref{fig:naoh-scanrate-diagnostic} shows the fitted baselines and corrected oxidation waves, and Table~\ref{tab:naoh-scanrate-peaks} summarises the peak data.
At 10~mV\,s$^{-1}$ only a single oxidation peak was resolved on the unsmoothed data; a shoulder near $-0.75$~V became apparent after smoothing with an 11-point uniform filter.

\FloatBarrier

\FloatBarrier

\begin{table}[h!]
  \centering
  \caption{Oxidation peak positions and currents from the baseline-corrected anodic
    scan of 1~mM \cmpd{bcc-BnPO3-5} in 0.5~M NaOH at varying scan rates.}
  \label{tab:naoh-scanrate-peaks}
  \begin{tabular}{l c c c c c}
    \hline
    Scan rate (mV\,s$^{-1}$) & Peak~1 (V) & Peak~1 ($\mu$A) & Peak~2 (V) & Peak~2 ($\mu$A) & Ratio (P2/P1) \\
    \hline
    10  & $-0.745$ & 1.73 & $-0.508$ & 2.42 & 1.39 \\
    25  & $-0.756$ & 2.48 & $-0.494$ & 3.71 & 1.50 \\
    50  & $-0.761$ & 3.52 & $-0.481$ & 5.06 & 1.44 \\
    100 & $-0.761$ & 5.17 & $-0.465$ & 6.92 & 1.34 \\
    200 & $-0.753$ & 6.48 & $-0.433$ & 9.73 & 1.50 \\
    500 & $-0.748$ & 9.41 & $-0.383$ & 15.47 & 1.64 \\
    \hline
  \end{tabular}
\end{table}
\FloatBarrier

\subsection{Scan rate dependence (0.5~M LiOH)}

\begin{figure}[h!]
  \centering
  \includegraphics[width=0.7\textwidth]{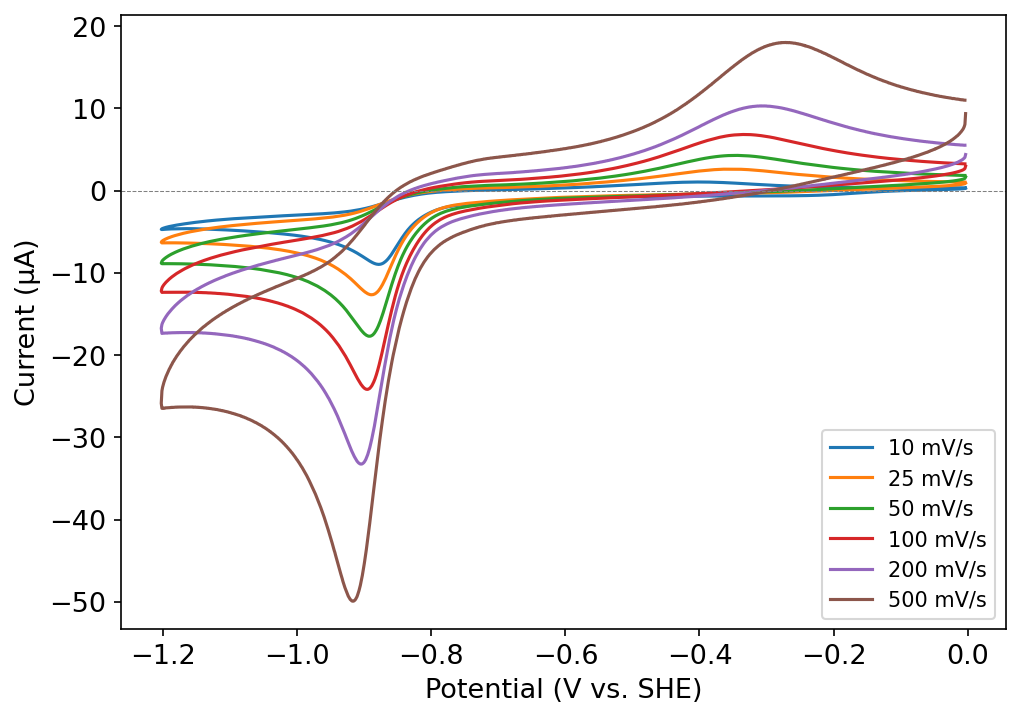}
  \caption{Cyclic voltammograms of 1~mM \cmpd{bcc-BnPO3-5} in 0.5~M LiOH at scan rates
    from 10 to 500~mV\,s$^{-1}$.}
  \label{fig:lioh-cv-overlay}
\end{figure}

\begin{figure}[h!]
  \centering
  \includegraphics[width=\textwidth]{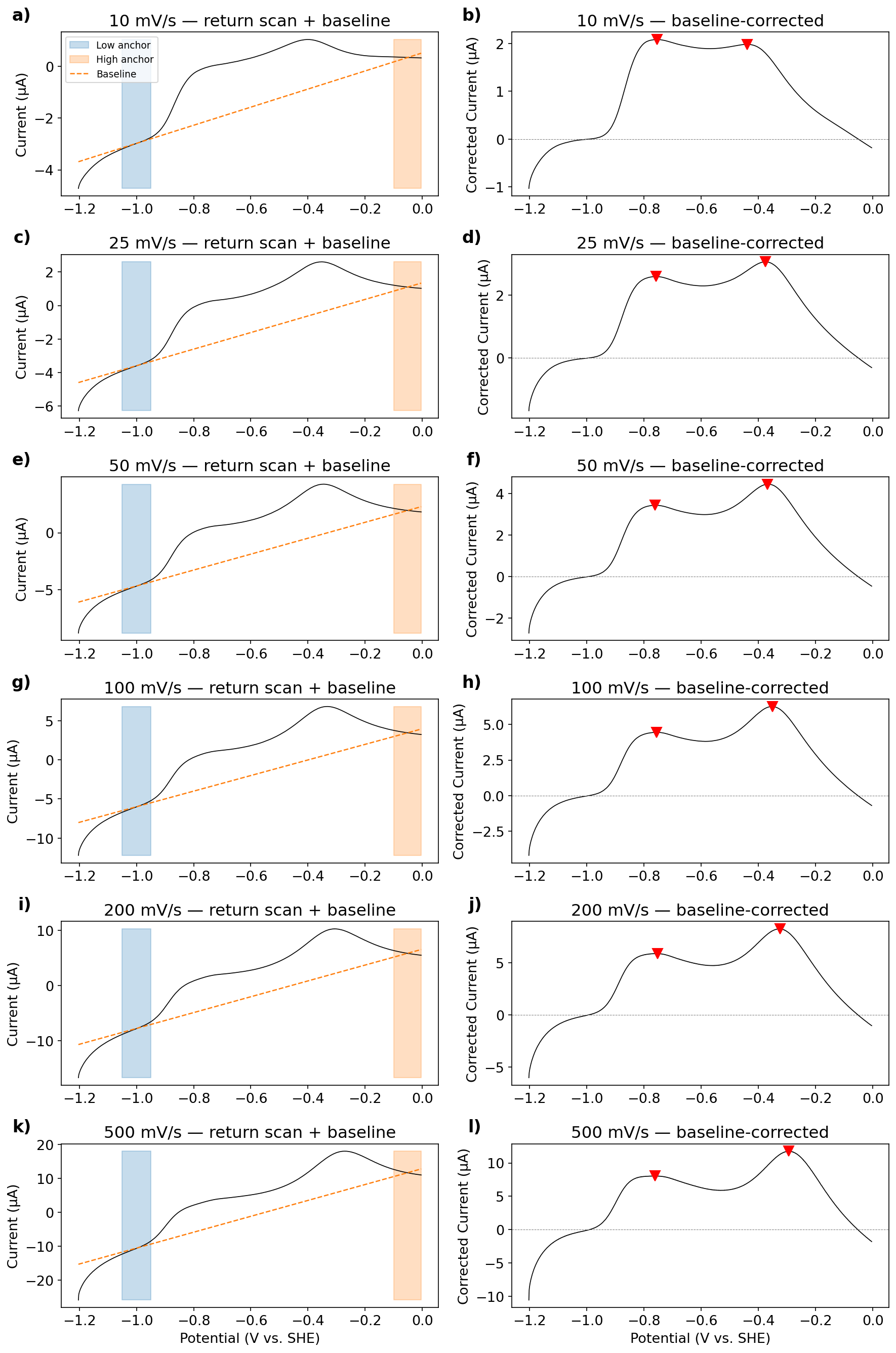}
  \caption{Baseline correction of the anodic return scan for 1~mM \cmpd{bcc-BnPO3-5}
    Left panels show the return scan (black) with anchor regions (shaded) and fitted
    linear baseline (dashed orange). Right panels show the baseline-corrected current
    with peak maxima marked by red triangles.}
  \label{fig:lioh-scanrate-diagnostic}
\end{figure}

Cyclic voltammograms of 1~mM \cmpd{bcc-BnPO3-5} were recorded at 10-500~mV\,s$^{-1}$ in 0.5~M LiOH (Figure~\ref{fig:lioh-cv-overlay}).
The baseline-correction procedure described in the potassium hydroxide concentration dependence experiment was used.
Figure~\ref{fig:lioh-scanrate-diagnostic} shows the fitted baselines and corrected oxidation waves, and Table~\ref{tab:lioh-scanrate-peaks} summarises the peak data.
At 10~mV\,s$^{-1}$ only a single oxidation peak was resolved on the unsmoothed data; a shoulder near $-0.75$~V became apparent after smoothing with an 11-point uniform filter.

\FloatBarrier

\FloatBarrier

\begin{table}[h!]
  \centering
  \caption{Oxidation peak positions and currents from the baseline-corrected anodic
    scan of 1~mM \cmpd{bcc-BnPO3-5} in 0.5~M LiOH at varying scan rates.}
  \label{tab:lioh-scanrate-peaks}
  \begin{tabular}{l c c c c c}
    \hline
    Scan rate (mV\,s$^{-1}$) & Peak~1 (V) & Peak~1 ($\mu$A) & Peak~2 (V) & Peak~2 ($\mu$A) & Ratio (P2/P1) \\
    \hline
    10  & $-0.754$ & 2.08 & $-0.440$ & 1.98 & 0.95 \\
    25  & $-0.759$ & 2.59 & $-0.375$ & 3.05 & 1.18 \\
    50  & $-0.762$ & 3.43 & $-0.368$ & 4.45 & 1.30 \\
    100 & $-0.757$ & 4.45 & $-0.351$ & 6.26 & 1.41 \\
    200 & $-0.753$ & 5.88 & $-0.324$ & 8.25 & 1.40 \\
    500 & $-0.762$ & 8.05 & $-0.295$ & 11.77 & 1.46 \\
    \hline
  \end{tabular}
\end{table}
\FloatBarrier

\FloatBarrier

\subsection{Concentration dependence of \cmpd{bcc-BnSO3-5}}\label{subsec:si-conc-dependence}

To assess the effect of analyte concentration on the electrochemical response of \cmpd{bcc-BnSO3-5}, cyclic voltammograms were recorded at 0.01~mM and 5~mM in 0.5~M KOH at scan rates of 25, 50, and 100~mV\,s$^{-1}$.

\begin{figure}[ht]
  \centering
  \includegraphics[width=\textwidth]{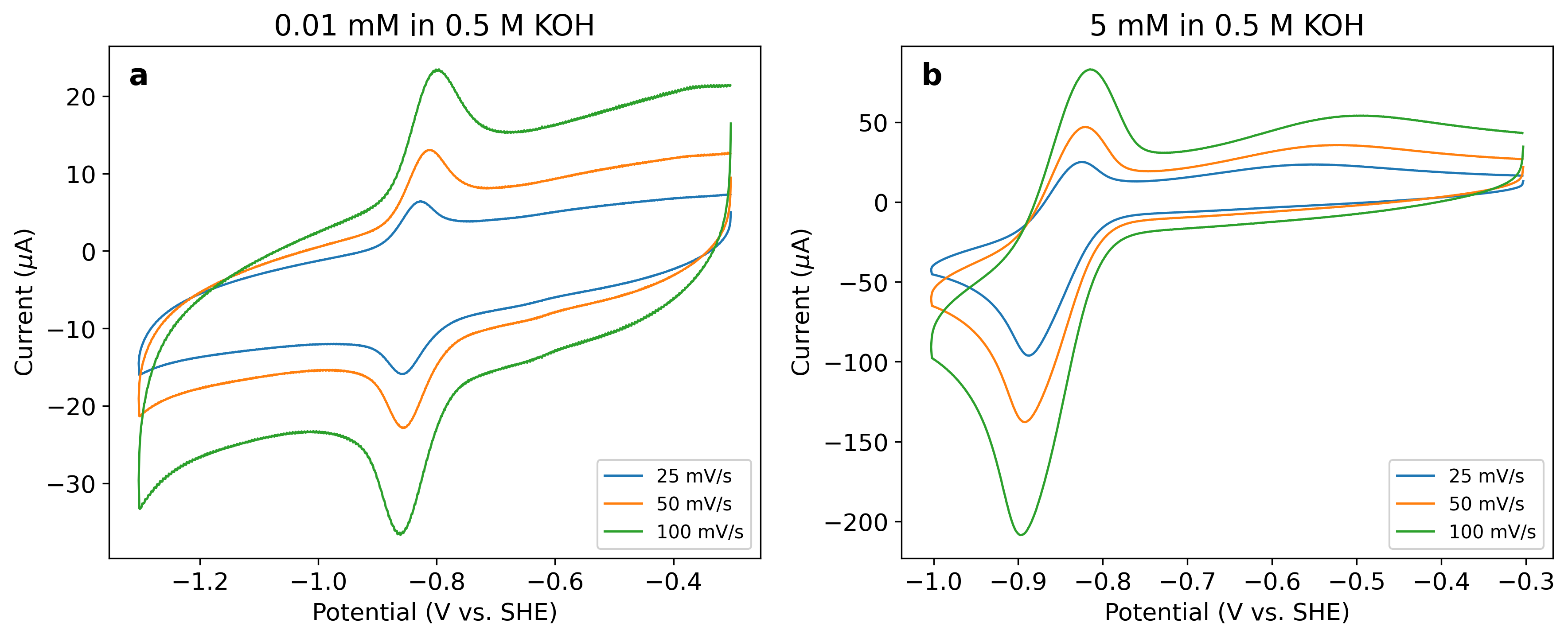}
  \caption{Cyclic voltammograms of \cmpd{bcc-BnSO3-5} in 0.5~M KOH at (a)~0.01~mM and (b)~5~mM, recorded at 25, 50, and 100~mV\,s$^{-1}$. Potentials are referenced to SHE.}
  \label{fig:cams014-conc-dependence}
\end{figure}
\FloatBarrier

\subsection{Potassium hydroxide incubation}\label{subsec:si-degraded}

A 1~mM solution of \cmpd{bcc-BnPO3-5} in 0.5~M KOH was incubated at room temperature under inert atmosphere for approximately 20 days.
After incubation, the sample was examined by cyclic voltammetry at multiple scan rates (10--250~mV\,s$^{-1}$).

\begin{figure}[ht]
  \centering
  \includegraphics[width=0.7\textwidth]{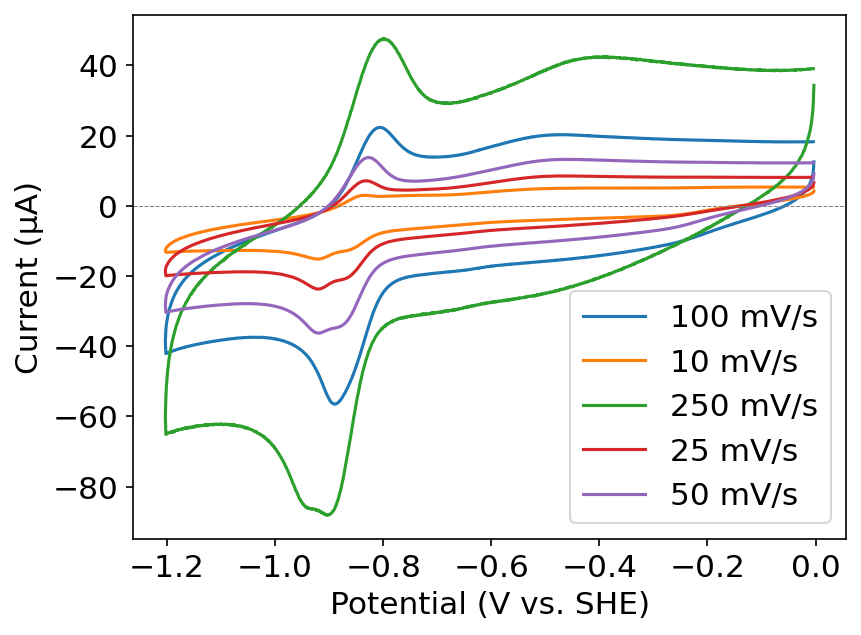}
  \caption{Multi-scan-rate cyclic voltammograms of degraded \cmpd{bcc-BnPO3-5} (1~mM in 0.5~M KOH) recorded at 10, 25, 50, 100, and 250~mV\,s$^{-1}$.}
  \label{fig:cams013-degraded-cv}
\end{figure}

\begin{figure}[ht]
  \centering
  \includegraphics[width=0.7\textwidth]{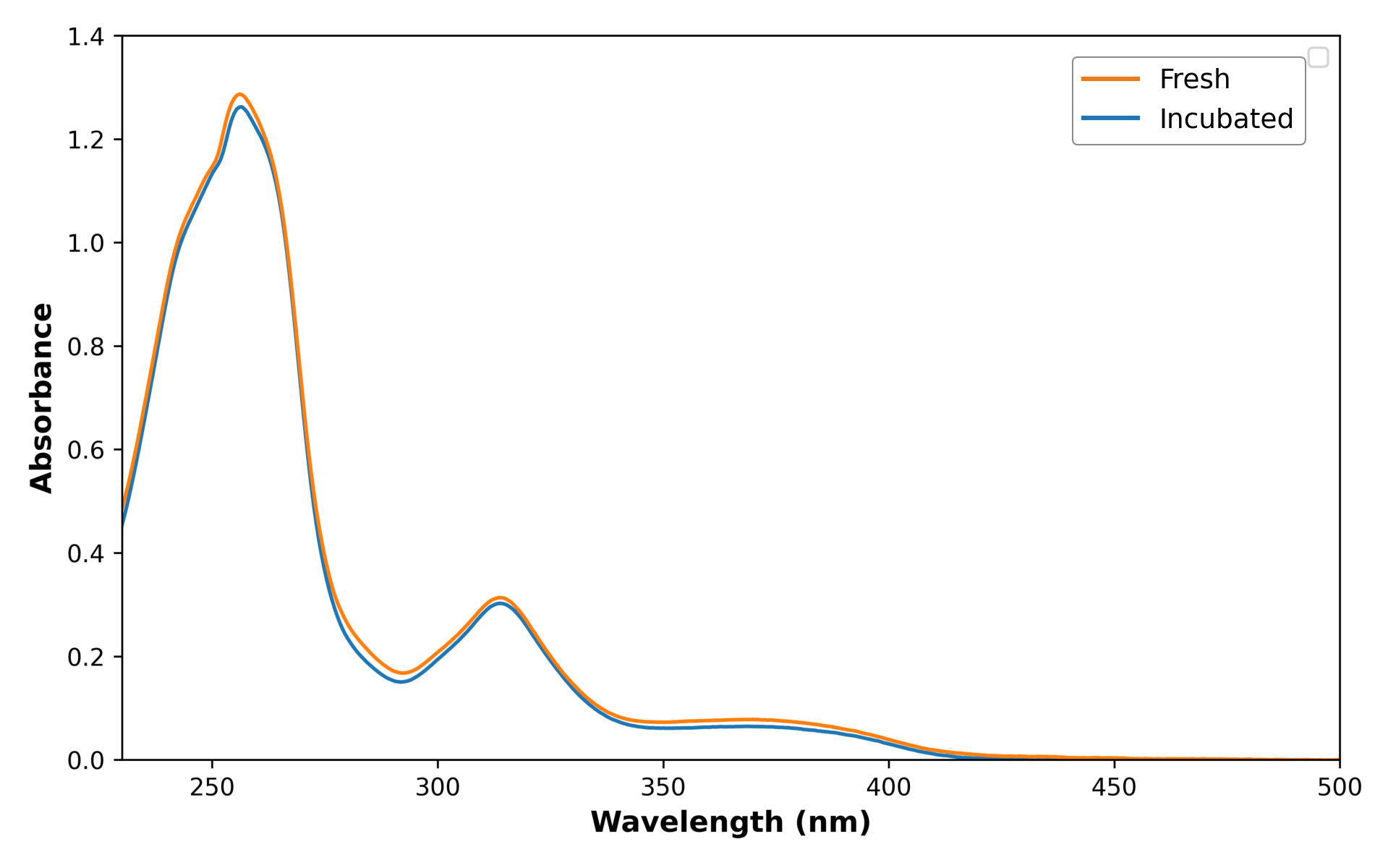}
  \caption{UV-Vis absorption spectra of \cmpd{bcc-BnPO3-5} (1~mM in 0.5~M KOH) before (fresh) and after incubation at room temperature.}
  \label{fig:cams013-degraded-uv-vis}
\end{figure}

UV-Vis absorption spectra recorded before and after incubation show approximately 2.3\% reduction in the magnitude of absorbance (Figure~\ref{fig:cams013-degraded-uv-vis}).

\FloatBarrier

\section{Flow cell cycling and electrolyte characterization}\label{sec:si-cell_cycling}

\subsection{Flow cell preparation}
Flow cells from Fuel Cell Technologies (Albuquerque, New Mexico) were assembled in a zero-gap configuration. Pyrosealed POCO graphite flow plates with machined interdigitated flow fields were used on both sides of the cell. 
Two 5 cm${^2}$ sheets of CE Tech GF020 graphite felt (Fuel Cell Store, 2.1 mm thickness) were used as the electrodes. Prior to assembly, the graphite felt was thermally activated inside a muffle furnace at 400°C for 12 hours. A 50 µm-thick Fumasep FKS-50 was used as the membrane and soaked in the corresponding electrolyte for at least 24 hours before use.

1/16” thick Viton gaskets (McMaster Carr) were laser-cut to secure both the electrodes and membranes in place. Each of the eight bolts was tightened to 50 lb-in of torque, and the assembled cell was then operated under flow for at least an hour to verify for any leaks. A Cole-Parmer SK-77202-60 peristaltic pump was used to circulate the electrolyte at a rate of about 60 mL/min through fluorinated ethylene propylene (FEP) tubing (McMaster-Carr). All cycling experiments were performed in an argon glovebox with an oxygen level of less than 2 ppm. A CCCV protocol was used to access the full electrolyte capacity.

\subsection{Post-cycling characterization }

\textbf{$^1$H NMR spectroscopy} Following cycling, a 150 µL aliquot of fully oxidized \cmpd{bcc-BnSO3-5} was collected from the capacity-limiting-side (CLS) reservoir and combined with 350 µL of 15 mM sodium methanesulfonate solution in D\textsubscript{2}O. Sodium methanesulfonate served as the internal standard. $^1$H NMR spectra were acquired on a 400 MHz spectrometer using 16 scans, a relaxation delay of 60 s, and a pulse angle of 45°.

\textbf{UV-vis spectroscopy} Pre- and post-cycling electrolyte samples were diluted 1000-fold with the same supporting electrolyte before analysis. UV-vis spectra were recorded with a cuvette with a 1 cm optical path length, and the supporting electrolyte was used as the blank. Electrolyte concentrations were determined using a calibration curve, with the dilution factor applied as a correction. 

\textbf{Cyclic voltammetry} Post-cycled oxidized \cmpd{bcc-BnSO3-5} aliquots was diluted to a final volume of 10 mL with a target concentration of 5 mM with the same supporting electrolyte. Aliquot dilutions were determined using concentrations provided by UV-vis spectroscopy. Measurements were performed with a three-electrode setup consisting of a glassy carbon working electrode, Ag/AgCl reference electrode, and a platinum-wire counter electrode. 

\subsection{Cycling Results}

\begin{figure}[ht]
  \centering
  \includegraphics[
    width=\textwidth
  ]{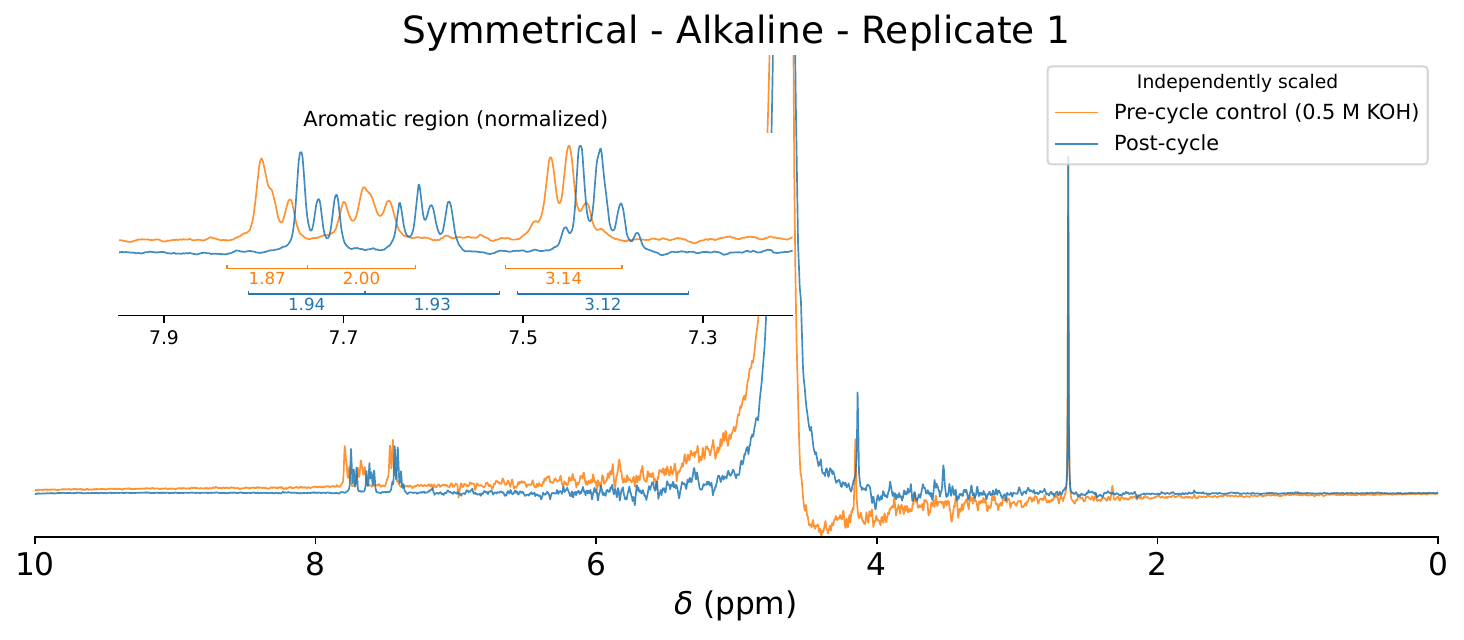}
  \caption{Post-symmetric flow cell cycling \textsuperscript{1}H~NMR of \cmpd{bcc-BnSO3-5} in 0.5 M KOH with a concentration of 25 mM. The capacity limiting side (CLS) contained 5 mL, and the non-capacity-limiting side (NCLS) contained 8 mL.}
  \label{fig:symmetrical_cycling_KOH_nmr}
\end{figure}

\begin{figure}[ht]
  \centering
  \includegraphics[
    width=\textwidth
  ]{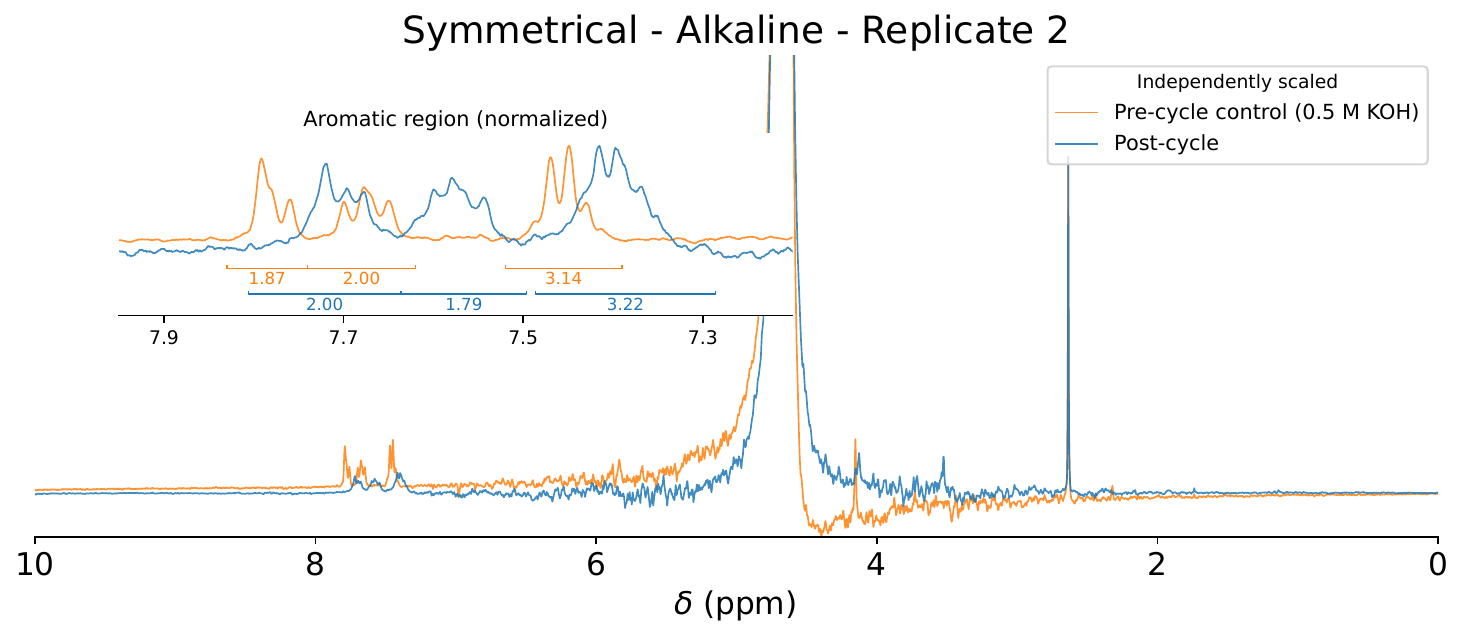}
  \caption{Post-replicate symmetric flow cell cycling \textsuperscript{1}H~NMR of \cmpd{bcc-BnSO3-5} in 0.5 M KOH with a concentration of 25 mM. The capacity limiting side (CLS) contained 6 mL, and the non-capacity-limiting side (NCLS) contained 10 mL.}
  \label{fig:symmetrical_cycling_KOH_replicate_nmr}
\end{figure}

\begin{figure}[ht]
  \centering
  \includegraphics[
    width=\textwidth
  ]{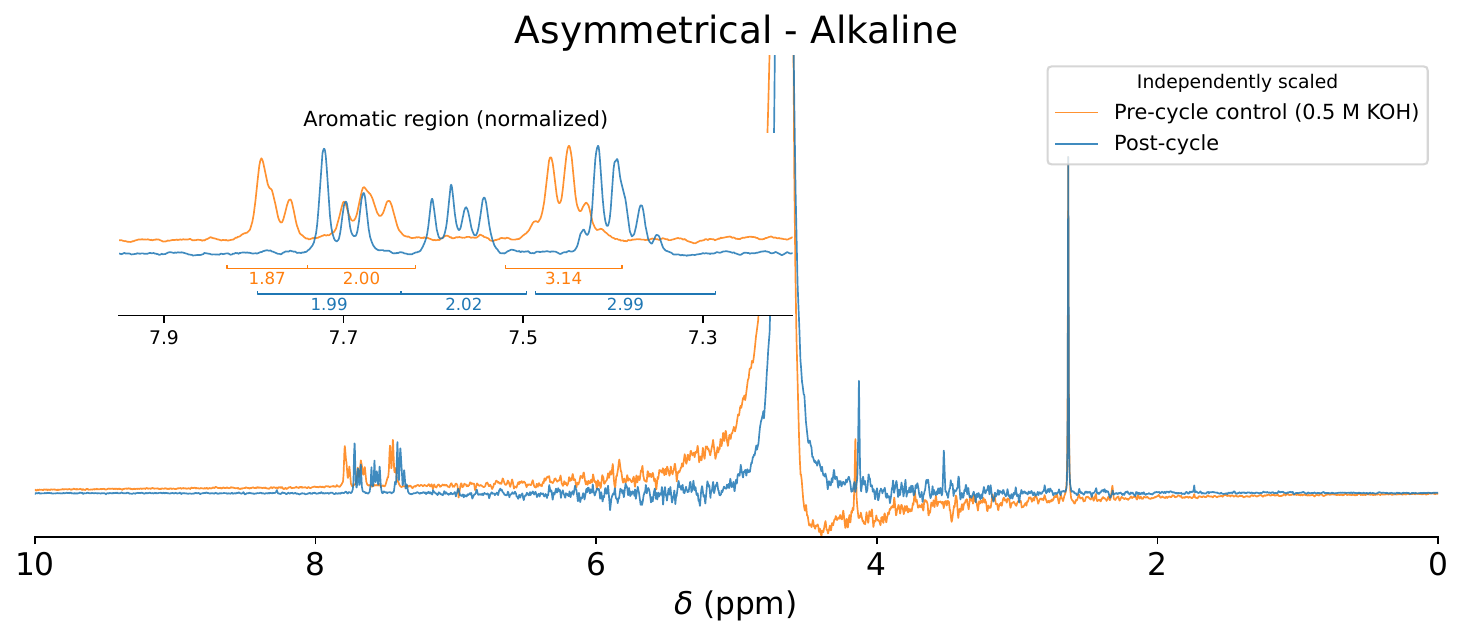}
  \caption{Post-asymmetric flow cell cycling \textsuperscript{1}H~NMR of \cmpd{bcc-BnSO3-5} in 0.5 M KOH. The capacity limiting side (CLS) contained 6 mL of 25 mM \cmpd{bcc-BnSO3-5}, and the non-capacity-limiting side (NCLS) contained 60 mL of 15 mM of each K$_3$[Fe(CN)$_6$] and K$_4$[Fe(CN)$_6$].}
  \label{fig:asymmetrical_cycling_KOH_nmr}
\end{figure}

Post-mortem characterization of the cycled \cmpd{bcc-BnSO3-5} using \textsuperscript{1}H NMR spectroscopy, cyclic voltammetry, and UV-vis spectroscopy revealed no clear evidence of molecular degradation or loss of redox activity for all of the cycling experiments.
\textsuperscript{1}H NMR spectroscopy showed a small, uniform shift in chemical shifts from the uncycled reference but revealed no new peaks from decomposed species (Figure~\ref{fig:symmetrical_cycling_KOH_nmr}-\ref{fig:asymmetrical_cycling_KOH_nmr}).

\begin{figure}[ht]
  \centering
  \includegraphics[
    trim={5cm 4.5cm 5cm 4cm},
    clip,
    width=\textwidth
  ]{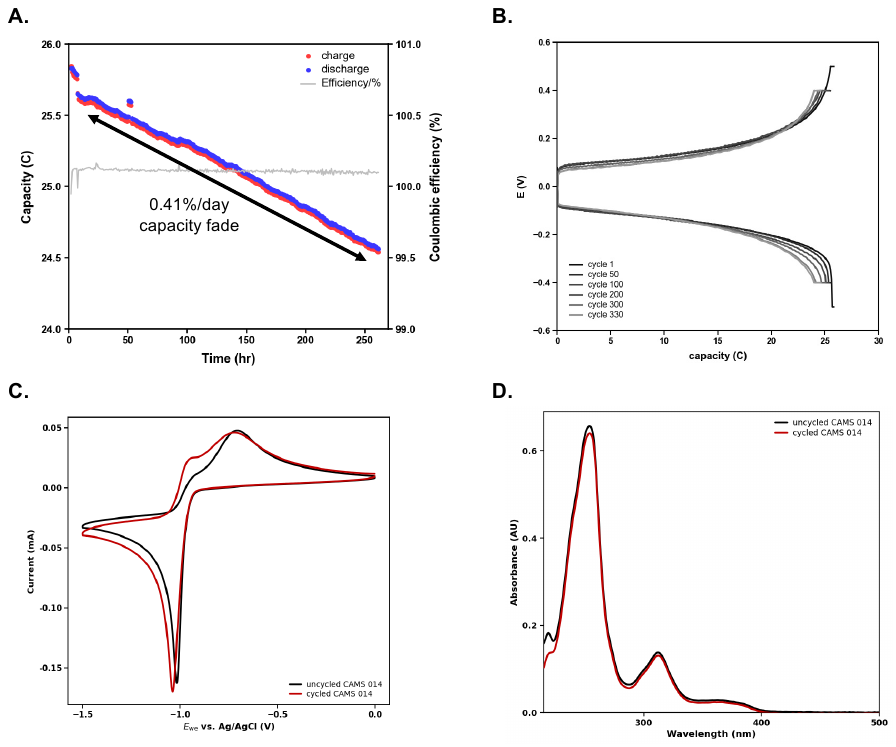}
  \caption{Symmetric flow cell cycling and characterization of \cmpd{bcc-BnSO3-5} in 0.5 M KOH with a concentration of 25 mM. The capacity-limiting side (CLS) contained 5 mL, and the non-capacity-limiting side (NCLS) contained 8 mL.  
  (A) Charge and discharge capacities and coulombic efficiency as a function of cycling time, showing a capacity fade rate of 0.41 \% per day. (B) Representative charge–discharge voltage profiles at selected cycles. (C) Cyclic voltammograms of the electrolyte before (black) and after (red) cycling. Post-cycling samples were diluted with the same supporting electrolyte to a nominal concentration of 5 mM. The scan rate was 50 mV/s. (D) UV–vis absorption spectra of the electrolyte before (black) and after (red) cycling. Samples were diluted 1000-fold with the same supporting electrolyte.}
  \label{fig:symmetrical_cycling_KOH}
\end{figure}

\begin{figure}[ht]
  \centering
  \includegraphics[
    trim={5cm 4.5cm 5cm 4cm},
    clip,
    width=\textwidth
  ]{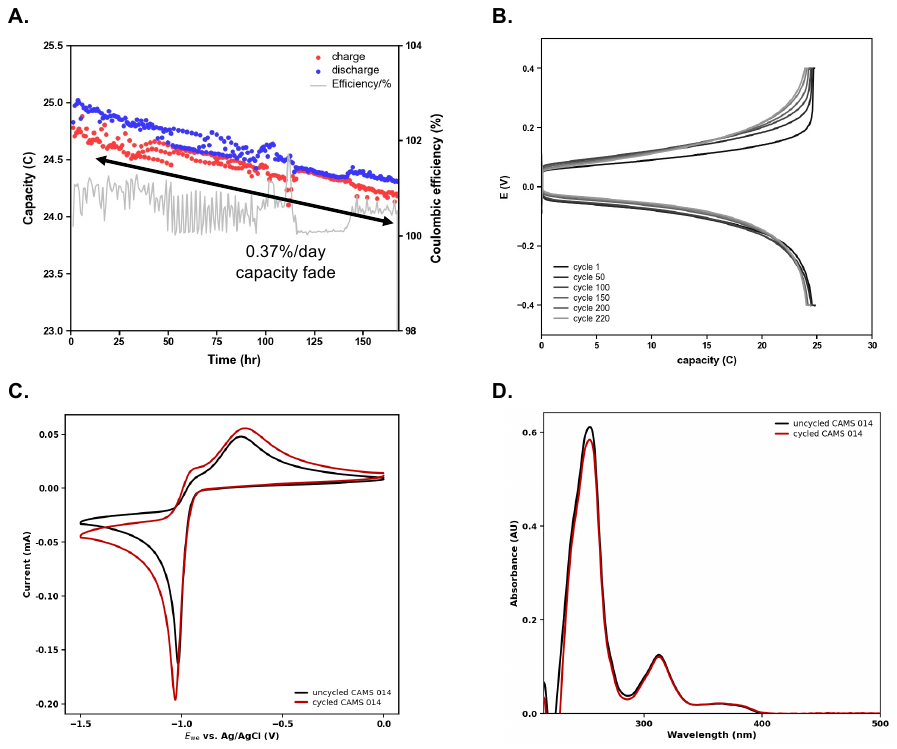}
  \caption{Replicate symmetric flow cell cycling and characterization of \cmpd{bcc-BnSO3-5} in 0.5 M KOH with a concentration of 25 mM. The capacity limiting side (CLS) contained 6 mL, and the non-capacity-limiting side (NCLS) contained 10 mL.  
  (A) Charge and discharge capacities and coulombic efficiency as a function of cycling time, showing a capacity fade rate of 0.37\% per day. (B) Representative charge–discharge voltage profiles at selected cycles. (C) Cyclic voltammograms of the electrolyte before (black) and after (red) cycling. Post-cycling samples were diluted with the same supporting electrolyte to a nominal concentration of 5 mM. The scan rate was 50 mV/s. (D) UV–vis absorption spectra of the electrolyte before (black) and after (red) cycling. Samples were diluted 1000-fold with the same supporting electrolyte.}
  \label{fig:symmetrical_cycling_KOH_replicate}
\end{figure}

The alkaline compositionally symmetrical cell exhibited a capacity fade of 0.41\% per day over 250 hours, with comparable cycling behavior in a repeat experiment (Figure~\ref{fig:symmetrical_cycling_KOH}-\ref{fig:symmetrical_cycling_KOH_replicate}). Following cycling, the UV-vis absorbance signal at 254 nm decreased by approximately 3.8\%, comparable in magnitude to the capacity loss estimated from the measured fade rate over the 250~hour cycling period. Cycling was investigated at neutral pH conditions, as CVs of \cmpd{bcc-BnSO3-5} in pH~7 phosphate buffer showed a smaller peak separation than alkaline conditions (Figure~\ref{fig:cams014-cv}a-b). 

\begin{figure}[ht]
  \centering
  \includegraphics[
    trim={5cm 4.5cm 5cm 4cm},
    clip,
    width=\textwidth
  ]{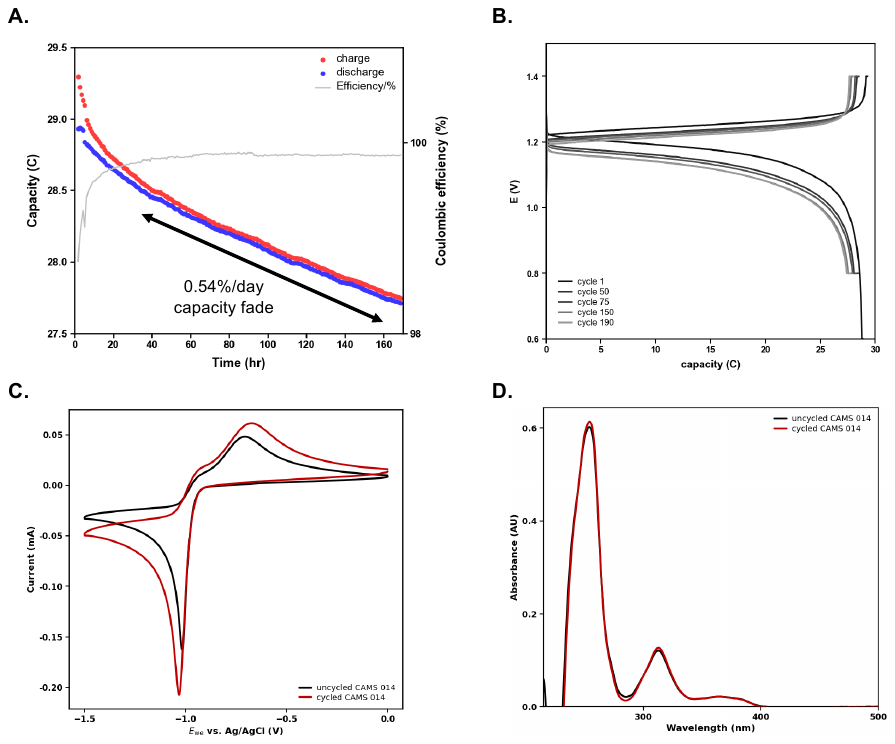}
  \caption{Asymmetric flow cell cycling and characterization of \cmpd{bcc-BnSO3-5} in 0.5 M KOH. The capacity limiting side (CLS) contained 6 mL of 25 mM \cmpd{bcc-BnSO3-5}, and the non-capacity-limiting side (NCLS) contained 60 mL of 15 mM of each K$_3$[Fe(CN)$_6$] and K$_4$[Fe(CN)$_6$].  
  (A) Charge and discharge capacities and coulombic efficiency as a function of cycling time, showing a capacity fade rate of 0.54\% per day. (B) Representative charge–discharge voltage profiles at selected cycles. (C) Cyclic voltammograms of the electrolyte before (black) and after (red) cycling. Post-cycling samples were diluted with the same supporting electrolyte to a nominal concentration of 5 mM. The scan rate was 50 mV/s. (D) UV–vis absorption spectra of the electrolyte before (black) and after (red) cycling. Samples were diluted 1000-fold with the same supporting electrolyte.}
  \label{fig:asymmetrical_cycling_KOH}
\end{figure}

\begin{figure}[ht]
  \centering
  \includegraphics[width=\textwidth]{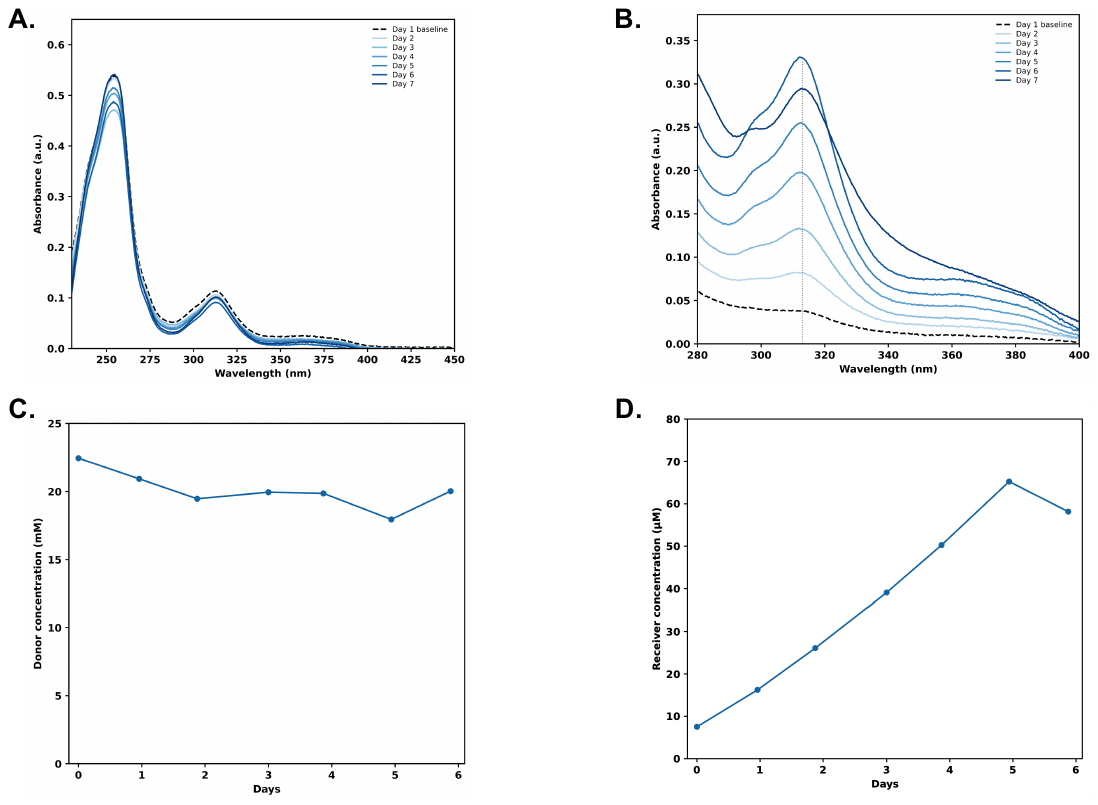}
  \caption{UV-vis monitoring of \cmpd{bcc-BnSO3-5} during open-circuit flow cell crossover experiment using a Fumasep FKS-50 membrane. Absorbance spectra of \textbf{(A)} donor electrolyte after a 1000-fold dilution and \textbf{(B)} undiluted receiver electrolyte. Corresponding \textbf{C} donor and \textbf{D} receiver concentrations estimated from absorbance at 313 nm. }
  \label{fig:flow-cell-permeability-measurement}
\end{figure}

The asymmetrical alkaline cell exhibited an average capacity fade of about 0.54\%~day$^{-1}$, whereas UV-vis quantification indicated a 1.5\% increase in concentration (Figure~\ref{fig:asymmetrical_cycling_KOH}).
The slightly higher concentration may be attributed in part to evaporation of water from the closed but non-hermetic negolyte, as 2 mL of electrolyte was lost from the reservoir after cycling. We do not expect membrane crossover to have significantly contributed to the apparent capacity fade, as we independently measured a low H-cell permeability of Fumasep FKS-50 to \cmpd{bcc-BnSO3-5} of $4.03 \times 10^{-10}$~cm$^2$\,s$^{-1}$ and a net crossover capacity equivalent of 0.1\% per day in a separate flow cell permeability experiment (Figure~\ref{fig:permeability-measurement}-\ref{fig:flow-cell-permeability-measurement} ).

\FloatBarrier

\FloatBarrier

\begin{figure}[ht]
  \centering
  \includegraphics[
    trim={5cm 4.5cm 5cm 4cm},
    clip,
    width=\textwidth
  ]{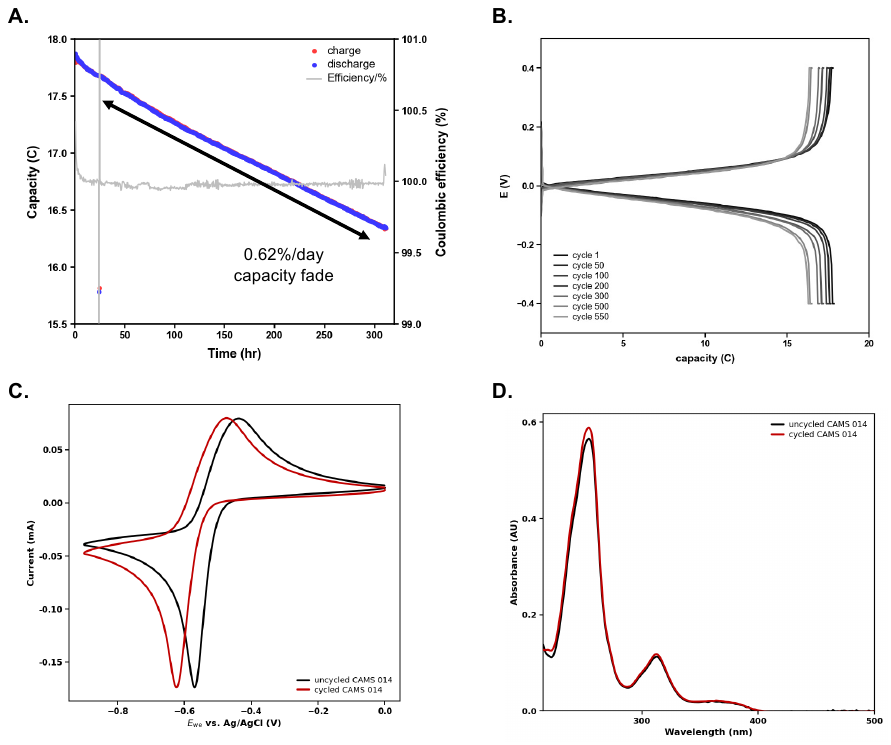}
  \caption{Symmetric flow cell cycling and characterization of \cmpd{bcc-BnSO3-5} in 0.25 M Phosphate Buffer, 0.75 M KCL with a concentration of 25 mM. The capacity limiting side (CLS) contained 4 mL, and the non-capacity-limiting side (NCLS) contained 8 mL.  
  (A) Charge and discharge capacities and coulombic efficiency as a function of cycling time, showing a capacity fade rate of 0.62\% per day. (B) Representative charge–discharge voltage profiles at selected cycles. (C) Cyclic voltammograms of the electrolyte before (black) and after (red) cycling. Post-cycling samples were diluted with the same supporting electrolyte to a nominal concentration of 5 mM. The scan rate was 50 mV/s. (D) UV–vis absorption spectra of the electrolyte before (black) and after (red) cycling. Samples were diluted 1000-fold with the same supporting electrolyte.}
  \label{fig:symmetrical_cycling_phosphate}
\end{figure}
\FloatBarrier

\FloatBarrier

\subsection{Post-cycling \textsuperscript{1}H~NMR}

\FloatBarrier

\FloatBarrier

\FloatBarrier

\section{Spectroelectrochemistry}\label{sec:si-spectroec}

\begin{figure}[ht]
  \centering
  \begin{overpic}[width=\textwidth]{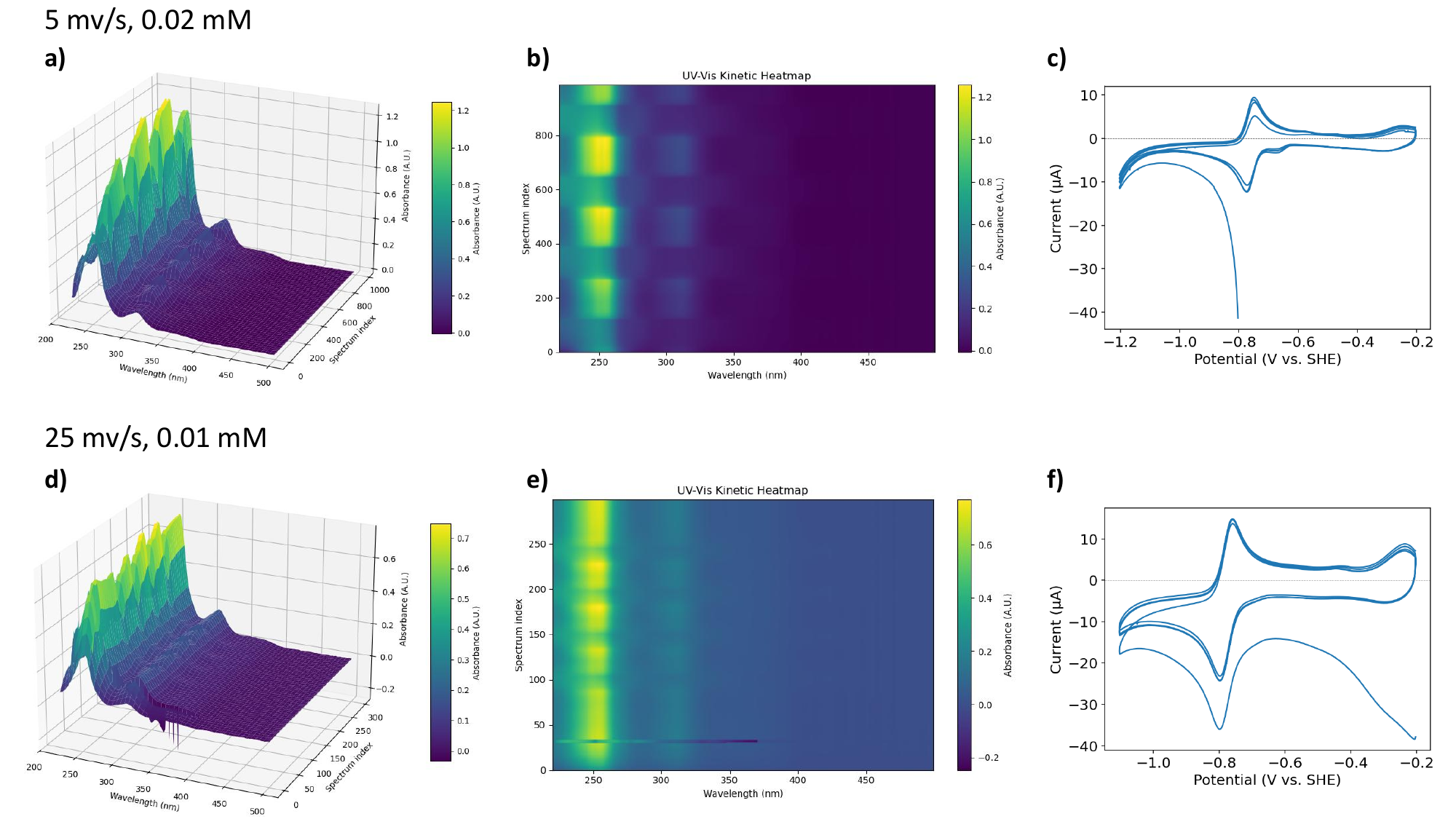}
    \put(18,54.25){\small\textsf{\cmpd{bcc-BnSO3-5}}}
    \put(19,25.5){\small\textsf{\cmpd{bcc-BnSO3-5}}}
  \end{overpic}
  \caption{Spectroelectrochemistry of \cmpd{bcc-BnSO3-5} in 0.5~M KOH. (a)~Three-dimensional UV-Vis absorbance surface during cyclic voltammetry at 5~mV\,s$^{-1}$ (0.02~mM \cmpd{bcc-BnSO3-5}). (b)~UV-Vis kinetic heatmap at 5~mV\,s$^{-1}$. (c)~Corresponding cyclic voltammogram at 5~mV\,s$^{-1}$. (d)--(f)~Analogous data at 25~mV\,s$^{-1}$ (0.01~mM \cmpd{bcc-BnSO3-5}).}
  \label{fig:spectroec}
\end{figure}
\FloatBarrier

\section{Permeability of \cmpd{bcc-BnSO3-5} through FKS-50 cation-exchange membrane}\label{sec:si-permeability}

\subsection{H-cell permeability measurement}
The permeability of \cmpd{bcc-BnSO3-5} through Fumasep FKS-50 was measured in a two-compartment H-cell with a 9 mm diameter opening containing 0.5 M KOH. The membrane was preconditioned in 0.5 M KOH for 24 hours before use. The donor and receiver compartments each contained 8 mL of solution, with 4.5 mM of \cmpd{bcc-BnSO3-5} in the donor and supporting electrolyte in the receiver side. Both compartments were stirred over the span of seven days. Receiver concentrations were determined daily by UV--vis spectroscopy and returned to the reservoir. A permeability coefficient of \(4.03 \times 10^{-10}\,\mathrm{cm^2\,s^{-1}}\) was obtained using a diffusion model based on Fick's law.

\begin{figure}[ht]
  \centering
  \includegraphics[width=0.7\textwidth]{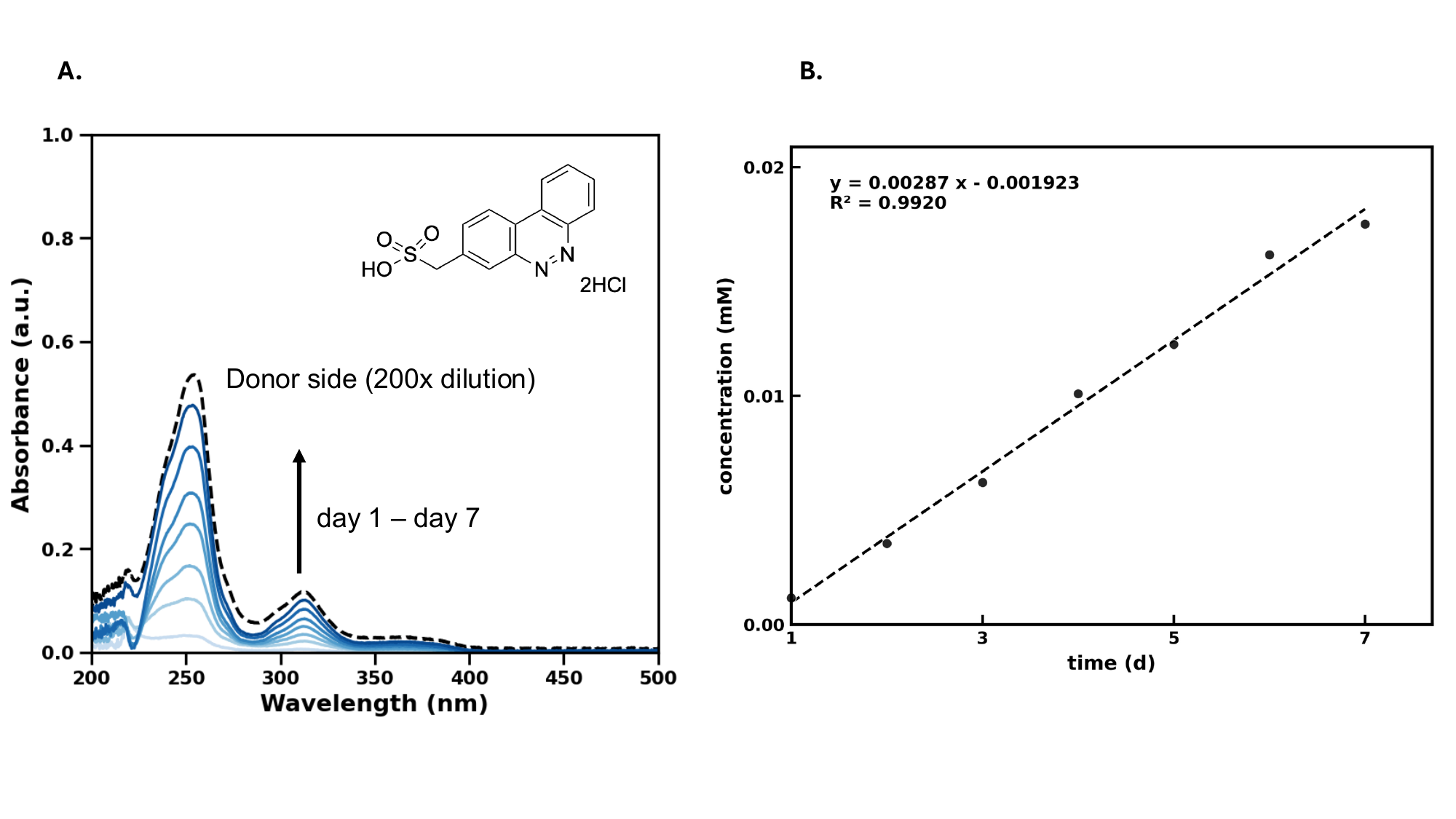}
  \caption{A) UV-Vis spectra of receiving side from day 1 to day 7, compared to a 200x dilution of the donor side. B) Receiver concentration (from the peak of 254 nm) versus time for permeation of \cmpd{bcc-BnSO3-5} through Fumasep FKS-50. }
  \label{fig:permeability-measurement}
\end{figure}

Assuming well-mixed compartments of equal, constant volume, the permeability coefficient was obtained from a linear fit of $\ln[1-2c_{\mathrm{R}}(t)/c_0]$ versus time using the integrated form of Fick's law:
\begin{equation}
  P = -\frac{V\ell}{2A}
  \frac{\mathrm{d}}{\mathrm{d}t}
  \ln\left(1-\frac{2c_{\mathrm{R}}(t)}{c_0}\right).
  \label{eq:permeability}
\end{equation}
Here, $V$ is the volume of each compartment (8~cm$^3$), $\ell$ is the membrane thickness in cm, $A$ is the exposed membrane area (0.636~cm$^2$), $t$ is elapsed time in seconds, $c_{\mathrm{R}}(t)$ is the receiver concentration, and $c_0$ is the initial donor concentration. (Figure~\ref{fig:permeability-measurement}).

\subsection{Flow cell crossover assessment}

To assess crossover under recirculating conditions, a flow cell was assembled with 5 mL of 25 mM oxidized \cmpd{bcc-BnSO3-5} in 0.5 M KOH on the donor side and 15 mL of 0.5 M KOH on the receiver side. The compartments were separated by a Fumasep FKS-50 cation-exchange membrane with an exposed area of 5~cm$^2$ with a thickness of 50 $\mu$m. The membrane was preconditioned before use. The flow cell was circulated under open-circuit conditions.

Daily donor aliquots of 100 $\mu$L were withdrawn and discarded, while 1 mL receiver aliquots were analyzed and returned to the reservoir. Donor samples were diluted 1000-fold before UV-vis analysis, and receiver samples were measured undiluted. Each spectrum was corrected with a baseline correction. Concentrations were estimated from the absorbance peak at 313 nm using a calibration curve. 

Receiver-side UV--vis analysis at 313~nm indicated net accumulation of approximately 0.70--0.76~$\mu$mol of \cmpd{bcc-BnSO3-5} between Days~1 and~7. The corresponding endpoint-average crossover capacity-equivalent rate was calculated as
\begin{equation}
r_{\mathrm{cross,eq}}
=
\frac{100\left[n_{\mathrm{R},7}-n_{\mathrm{R},1}\right]}
{n_{\mathrm{D},1}\Delta t_{\mathrm{day}}},
\end{equation}
where the receiver inventory present on Day~1 was
subtracted. For a final receiver volume of 15~mL, the estimated capacity loss due to crossover was approximately 0.1\% per day. Deliberately discarded donor aliquots were accounted for separately.
Although the donor absorbance at 313 nm decreased over time, the corresponding loss of donor inventory could not be reliably quantified. The mass balance depended on the total circulating volume, including any residual liquid remaining in the tubing and porous electrodes. Given the small donor volume, even a slight volume discrepancy substantially affected the calculated donor mass balance, making it difficult to attribute unexplained losses to precipitation.

\FloatBarrier

\section{Solubility studies}\label{sec:si-solubility}

\subsection{Nuclear magnetic resonance spectroscopy}
Nuclear Magnetic Resonance (NMR) spectra were acquired on Agilent DD2 400 MHz and Agilent DD2 600 MHz NMR spectrometers at room temperature. Samples were prepared primarily in D\textsubscript{2}O unless otherwise noted. Chemical shifts are reported in parts per million (ppm). \textsuperscript{1}H NMR spectra collected in D\textsubscript{2}O were referenced to the residual HOD solvent signal.

\subsubsection{Measurement protocol}

\begin{figure}[ht]
  \centering
  \includegraphics[width=0.8\textwidth]{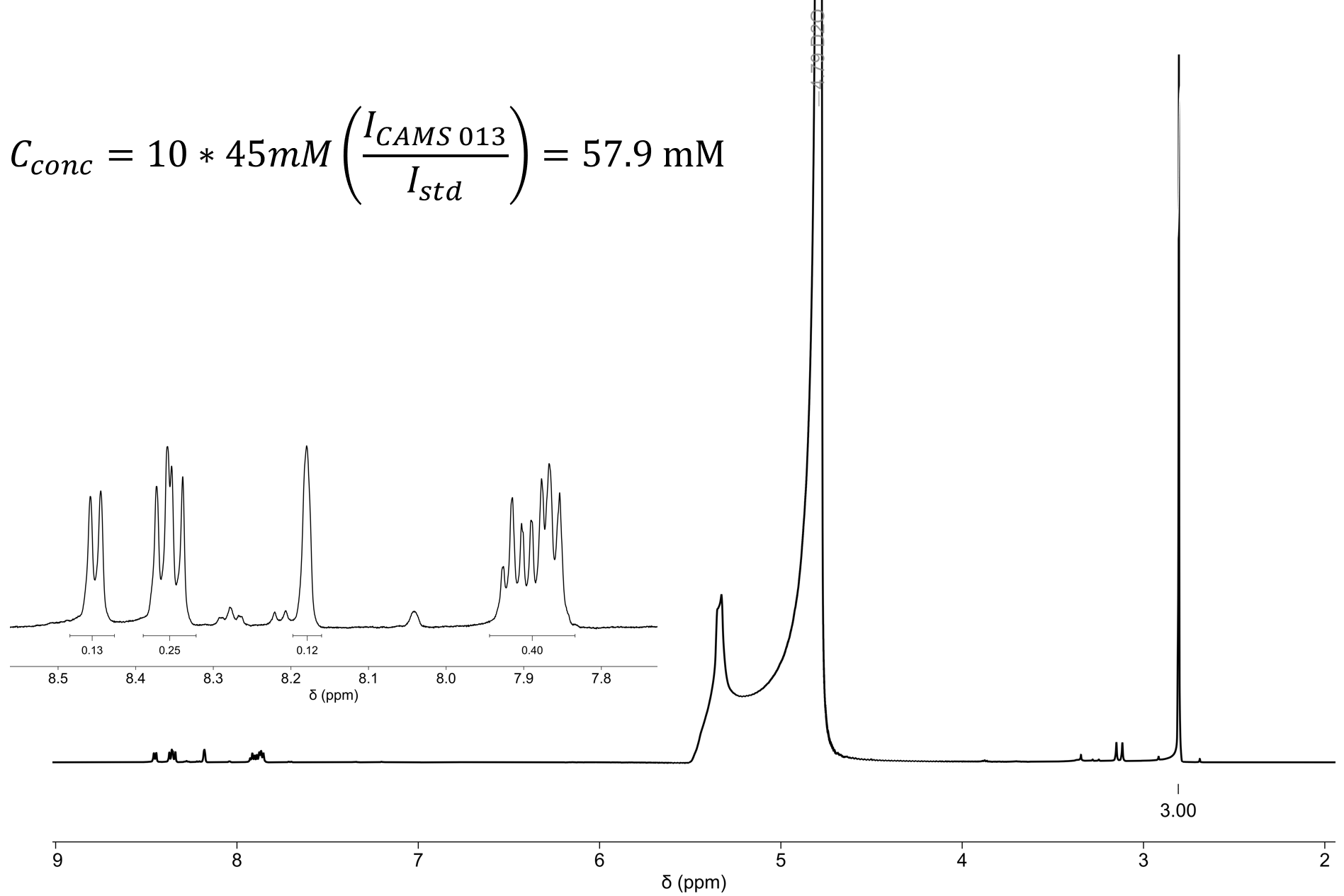}
  \caption{Quantitative \textsuperscript{1}H NMR spectrum of the saturated \cmpd{bcc-BnPO3-5} supernatant in D\textsubscript{2}O with sodium methanesulfonate internal standard. Inset: expansion of the aromatic region.}
  \label{fig:solubility-nmr-013}
\end{figure}

Quantitative \textsuperscript{1}H NMR was used to independently verify the solubility of \cmpd{bcc-BnPO3-5} and \cmpd{bcc-BnSO3-5} in 0.5~M KOH.
A 50~µL aliquot of the saturated \cmpd{bcc-BnPO3-5} supernatant was diluted with 450~µL of 50~mM sodium methanesulfonate in D\textsubscript{2}O and analyzed by \textsuperscript{1}H NMR (16 scans, 60~s relaxation delay).
The sodium methanesulfonate singlet was normalized to 3.00~H, and the \cmpd{bcc-BnPO3-5} aromatic resonances (assumed to represent 7H) were integrated relative to this internal standard.
After correcting for the 10$\times$ dilution factor, the \cmpd{bcc-BnPO3-5} concentration was determined to be approximately 57.9~mM (Figure~\ref{fig:solubility-nmr-013}).

\begin{figure}[ht]
  \centering
  \includegraphics[width=0.8\textwidth]{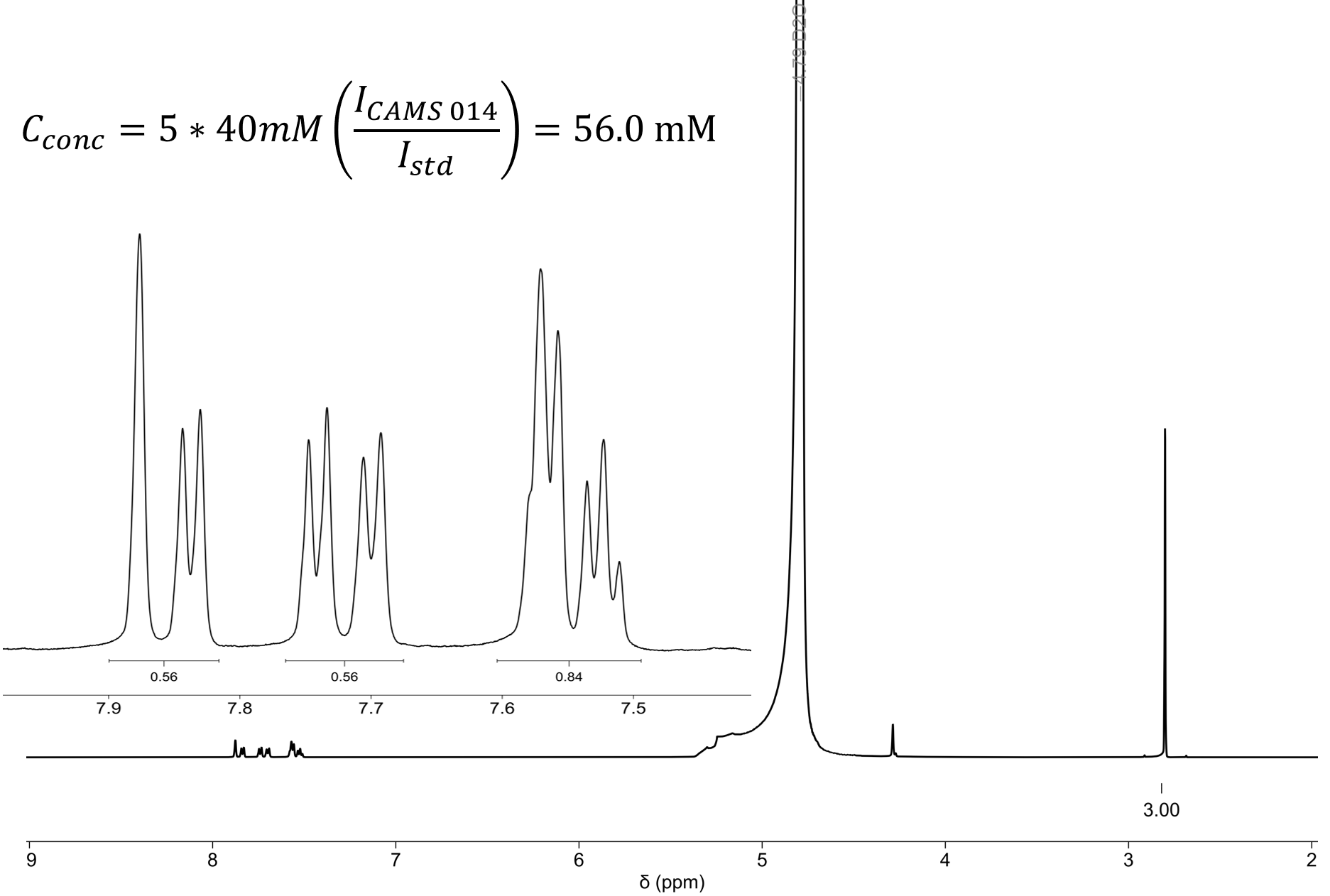}
  \caption{Quantitative \textsuperscript{1}H NMR spectrum of the saturated \cmpd{bcc-BnSO3-5} supernatant in D\textsubscript{2}O with sodium methanesulfonate internal standard. Inset: expansion of the aromatic region.}
  \label{fig:solubility-nmr-014}
\end{figure}

Similarly, a 100~µL aliquot of the saturated \cmpd{bcc-BnSO3-5} supernatant was diluted with 400~µL of 50~mM sodium methanesulfonate in D\textsubscript{2}O and analyzed under identical acquisition conditions.
After correcting for the 5$\times$ dilution factor, the \cmpd{bcc-BnSO3-5} concentration was determined to be approximately 56.0~mM (Figure~\ref{fig:solubility-nmr-014}).

\FloatBarrier

\FloatBarrier

\subsection{Ultraviolet-visible spectroscopy}
UV-Vis spectroscopy was conducted on an Ocean-HDX-UV-VIS spectrometer and a DH-Mini-UV-Vis-NIR light source (OceanOptics). Spectral data were collected over a wavelength of 215-800 nm with an integration time of 25 ms averaged over 50 scans. Measurements were performed in a standard 1 cm pathlength polystyrene cuvette. The background spectra were collected using the corresponding blank solvent under identical conditions and subtracted.

\subsubsection{Calibration curve procedure}
UV–Vis calibration curves were prepared for \cmpd{bcc-BnPO3-5} and \cmpd{bcc-BnSO3-5} in 0.5 M KOH. For each compound, an appropriate mass was carefully weighed and dissolved in 0.5 M KOH to prepare a 1 mM stock solution in a 10 mL centrifuge tube. The exact stock concentration was calculated from the measured mass, molecular weight, and final solution volume. The stock solution was then serially diluted with 0.5 M KOH to prepare calibration standards within the linear absorbance range.

UV–Vis spectra were collected using 0.5 M KOH as the blank. For each compound, the absorbance at the wavelength of maximum absorbance, $\lambda_{\mathrm{max}}$, was used to construct the calibration curve. Absorbance at $\lambda_{\mathrm{max}}$ was plotted as a function of concentration, and the data were fit using a linear regression of the form $A = mC + b$, where $A$ is absorbance, $C$ is concentration, $m$ is the slope, and $b$ is the y-intercept. Only calibration standards within the linear absorbance range were included in the fit. Linear fits gave $R^{2}$ values of 0.99 or greater for both \cmpd{bcc-BnPO3-5} and \cmpd{bcc-BnSO3-5}. 

\begin{figure}[ht]
  \centering
  \includegraphics[width=0.8\textwidth]{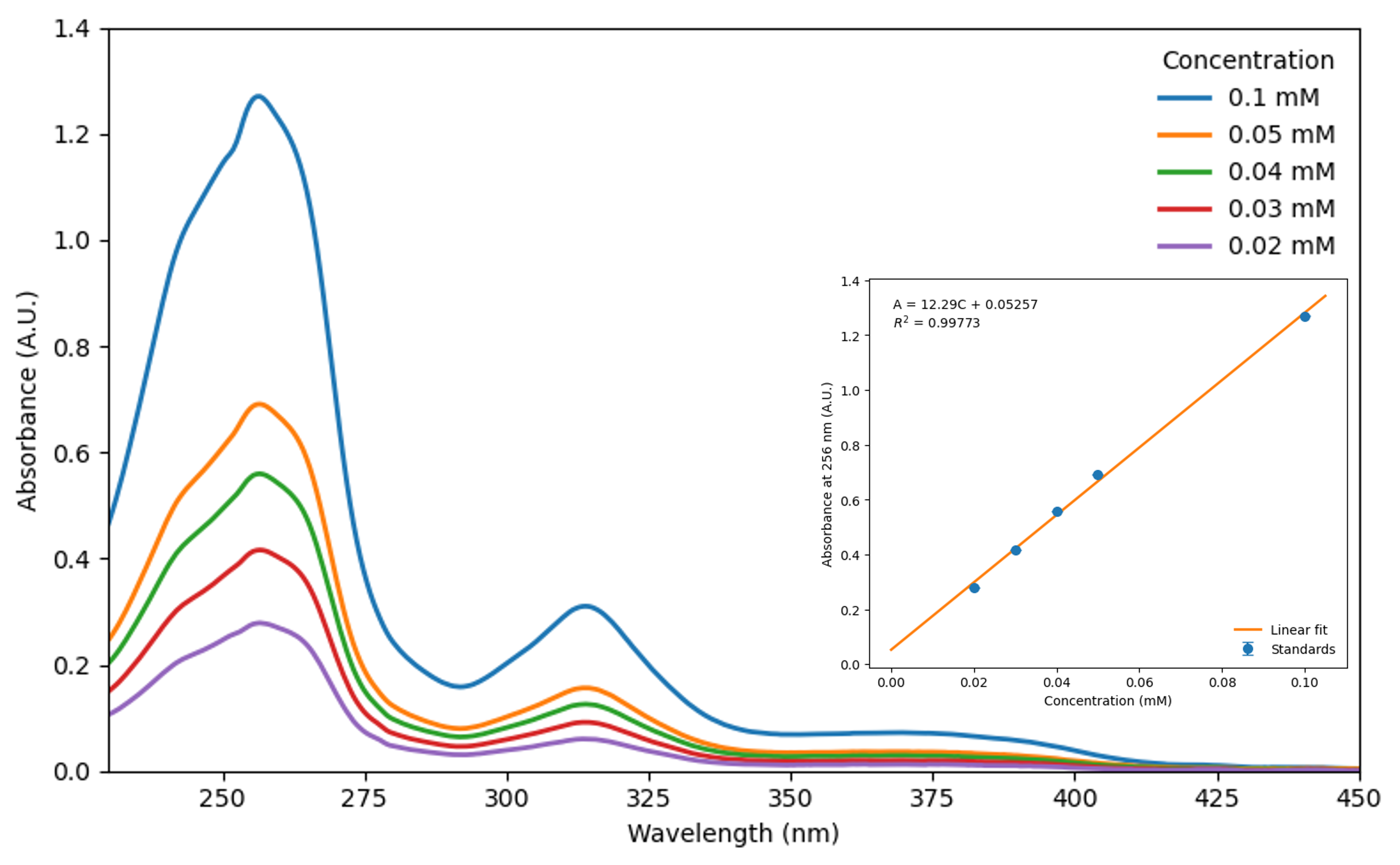}
  \caption{UV--Vis spectra of \cmpd{bcc-BnPO3-5} calibration standards in 0.5~M KOH. Inset: calibration curve at $\lambda_\mathrm{max} = 256$~nm with linear fit.}
  \label{fig:calibration-013}
\end{figure}
\FloatBarrier

\begin{figure}[ht]
  \centering
  \includegraphics[width=0.8\textwidth]{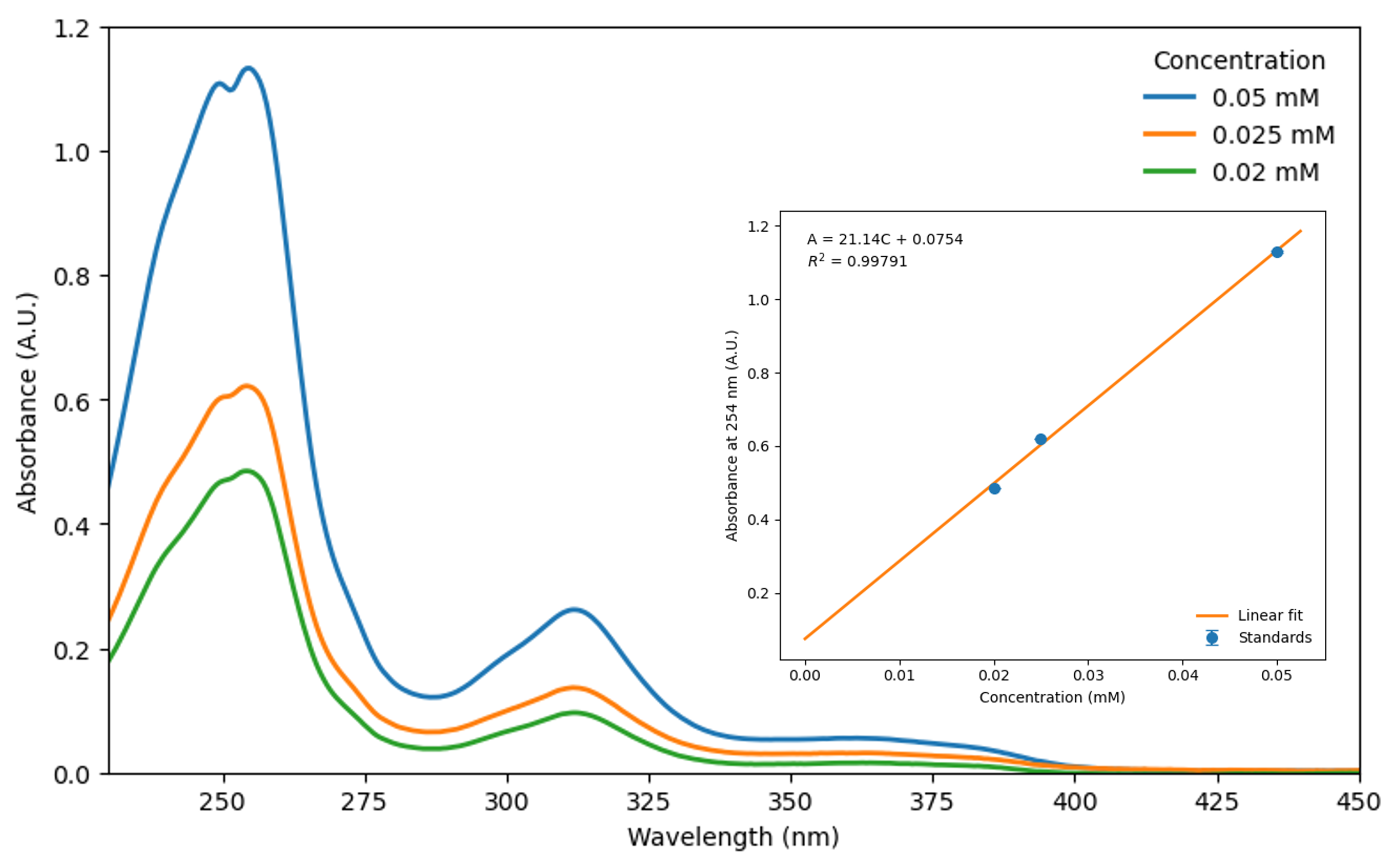}
  \caption{UV--Vis spectra of \cmpd{bcc-BnSO3-5} calibration standards in 0.5~M KOH. Inset: calibration curve at $\lambda_\mathrm{max} = 254$~nm with linear fit.}
  \label{fig:calibration-014}
\end{figure}
\FloatBarrier

\subsubsection{Measurement protocol}
Approximately 25 mg of \cmpd{bcc-BnPO3-5} and 30 mg of \cmpd{bcc-BnSO3-5} were carefully weighed into separate 7 mL scintillation vials. For \cmpd{bcc-BnPO3-5}, 0.5 mL of 0.5 M KOH was added, while 1.5 mL of 0.5 M KOH was added to \cmpd{bcc-BnSO3-5}. The amount of solvent was selected such that undissolved solid was visible, indicating that the mixtures were near or above their apparent solubility limit. \cmpd{bcc-BnPO3-5} was noted to produce a darker yellow solution than \cmpd{bcc-BnSO3-5} and appeared to exhibit much higher solubility under these conditions. A stir bar was added to each vial, and the vials were capped, sealed with electrical tape, and stirred at room temperature for 48 h. After stirring, each mixture was transferred to a 1.5 mL microcentrifuge tube and centrifuged to separate the undissolved solids from the supernatant. The supernatant was carefully transferred to another clean microcentrifuge tube and centrifuged again. A final centrifugation step was performed for 10 min to further remove any remaining solids. After this step, no visible solids were observed.

The supernatant was analyzed by UV-Vis spectroscopy after serial dilution to bring the absorbance within a linear calibration range. Concentrations were determined by comparison to a corresponding calibration curve. For quantitative NMR analysis, aliquots of the supernatant were diluted with 50 mM sodium methanesulfonate in D\textsubscript{2}O as an internal standard. A 50 µL aliquot of \cmpd{bcc-BnPO3-5} was combined with 450 µL of the 50 mM sodium methanesulfonate solution, while 100 µL of \cmpd{bcc-BnSO3-5} was combined with 400 µL of the methanesulfonate solution.

\begin{figure}[ht]
  \centering
  \includegraphics[width=0.8\textwidth]{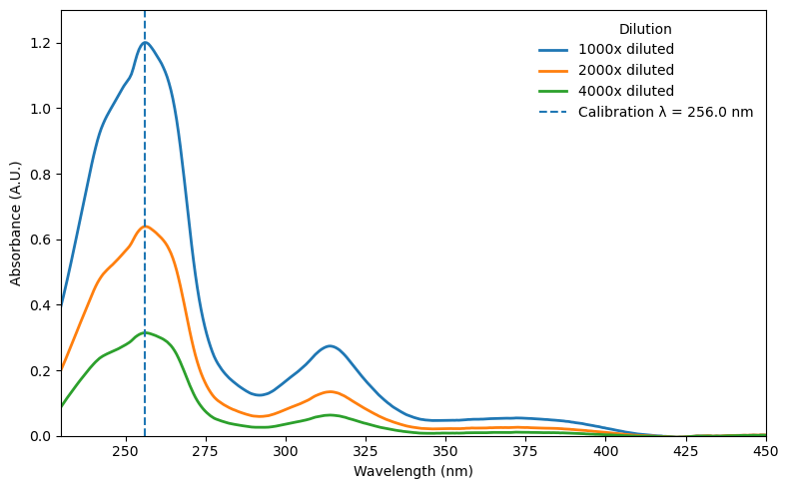}
  \caption{UV--Vis spectra of serial dilutions of the saturated \cmpd{bcc-BnPO3-5} supernatant in 0.5~M KOH. The dashed line indicates the calibration wavelength ($\lambda = 256$~nm).}
  \label{fig:solubility-013}
\end{figure}
\FloatBarrier

\begin{table}[ht]
  \centering
  \caption{UV--Vis solubility determination of \cmpd{bcc-BnPO3-5} in 0.5~M KOH.}
  \label{tab:solubility-uv-013}
  \begin{tabular}{l c c}
    \hline
    Dilution & Diluted conc.\ (mM) & Original conc.\ (mM) \\
    \hline
    1000$\times$ & $0.093$ & $93.45$ \\
    2000$\times$ & $0.048$ & $95.44$ \\
    4000$\times$ & $0.021$ & $85.25$ \\
    \hline
  \end{tabular}
\end{table}
\FloatBarrier

\begin{figure}[ht]
  \centering
  \includegraphics[width=0.8\textwidth]{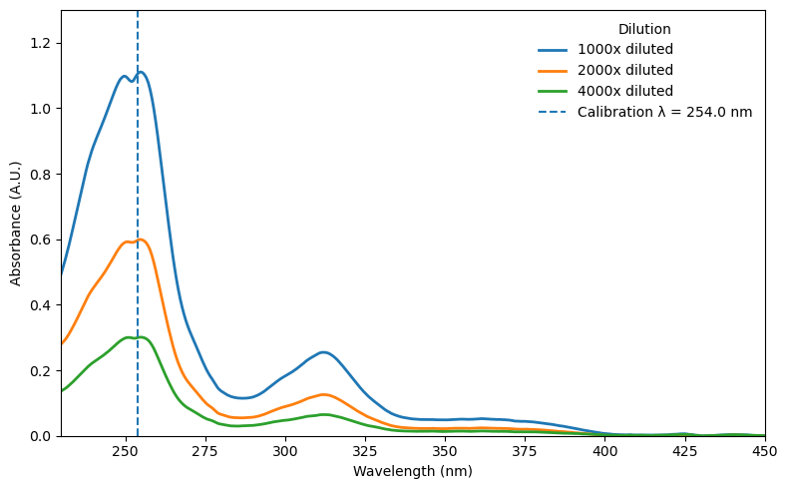}
  \caption{UV--Vis spectra of serial dilutions of the saturated \cmpd{bcc-BnSO3-5} supernatant in 0.5~M KOH. The dashed line indicates the calibration wavelength ($\lambda = 254$~nm).}
  \label{fig:solubility-014}
\end{figure}
\FloatBarrier

\begin{table}[ht]
  \centering
  \caption{UV--Vis solubility determination of \cmpd{bcc-BnSO3-5} in 0.5~M KOH.}
  \label{tab:solubility-uv-014}
  \begin{tabular}{l c c}
    \hline
    Dilution & Diluted conc.\ (mM) & Original conc.\ (mM) \\
    \hline
    1000$\times$ & $0.049$ & $48.75$ \\
    2000$\times$ & $0.025$ & $49.41$ \\
    4000$\times$ & $0.011$ & $42.27$ \\
    \hline
  \end{tabular}
\end{table}
\FloatBarrier

\FloatBarrier

\section{Synthetic procedures}\label{sec:si-synthesis}

\subsection{General information}
Reactions were performed under ambient atmosphere, unless otherwise noted. Yields refer to chromatographically and spectroscopically (\textsuperscript{1}H NMR) homogeneous materials, unless otherwise stated. Reagents were purchased from commercial vendors and used as received unless otherwise stated. All solvents were purchased as DriSolv grade and used as received without further drying. Column chromatography was performed on silica gel 60 (SiliCycle, 60--120 mesh). Thin-layer chromatography (TLC) utilized pre-coated plates (Sorbtech, silica gel 60 PF\textsubscript{254}, 0.25 mm) visualized with UV 254 nm, ninhydrin, or basic potassium permanganate stain. Analytical HPLC was performed on Waters HPLC systems using acetonitrile, methanol, and 0.1\% aq.\ formic acid as eluents. Mass spectra were recorded using a Micromass quattro ultima equipped with an ESI source. NMR spectra were recorded on a Bruker Avance (300 MHz) spectrometer.

\begin{figure}[h!]
  \centering
  \begin{overpic}[width=\textwidth]{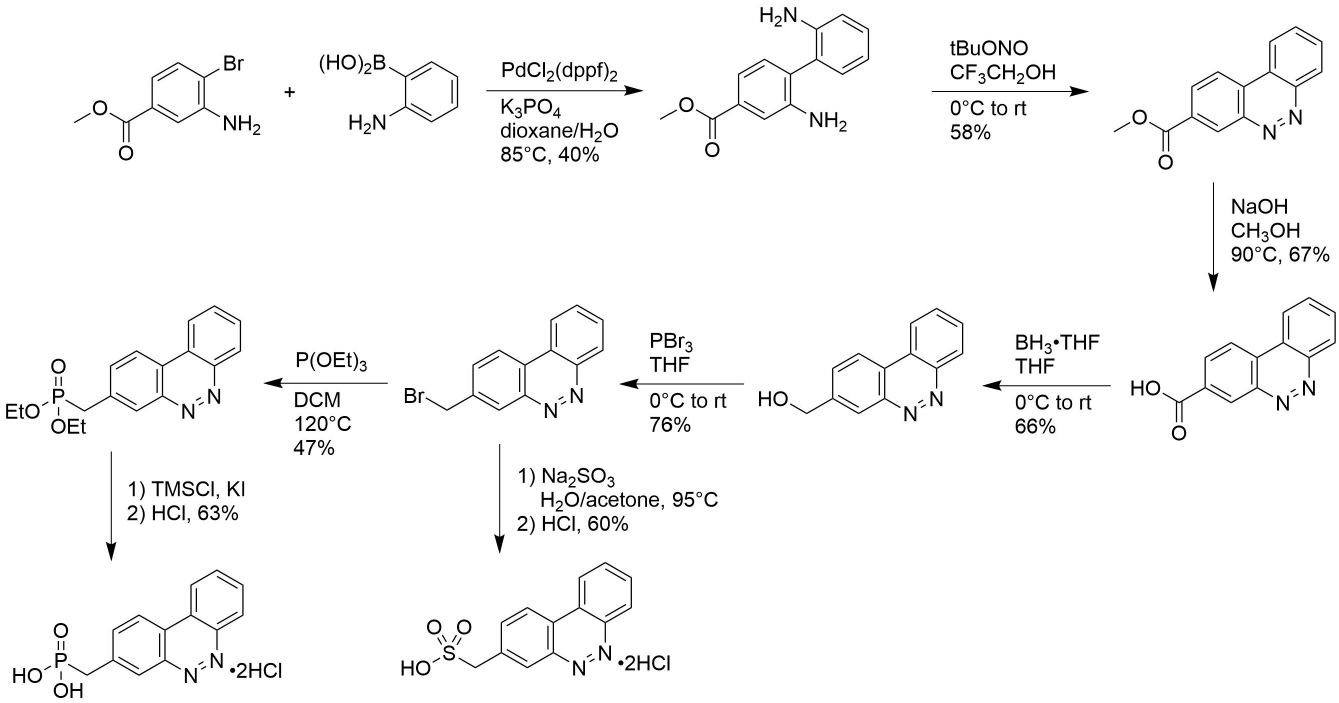}
    \put(58,40){\footnotesize\cmpd{si-diamino-biphenyl}}
    \put(93,40){\footnotesize\cmpd{si-ester}}
    \put(94,19){\footnotesize\cmpd{si-acid}}
    \put(65,19){\footnotesize\cmpd{si-alcohol}}
    \put(40,19){\footnotesize\cmpd{si-bromide}}
    \put(10,19){\footnotesize\cmpd{si-diethylphosphonate}}
    \put(10,-0.5){\footnotesize\cmpd{bcc-BnPO3-5}}
    \put(40,-0.5){\footnotesize\cmpd{bcc-BnSO3-5}}
  \end{overpic}
  \caption{Synthetic route to methanephosphonate (\cmpd{bcc-BnPO3-5}) and methanesulfonate (\cmpd{bcc-BnSO3-5}) benzo[c]cinnoline derivatives.}
  \label{fig:synthesis-scheme}
\end{figure}
\FloatBarrier

\subsection{Methyl 2,2\textquotesingle-diamino-4-biphenylcarboxylate (\cmpd{si-diamino-biphenyl})}
To a stirred solution of methyl 3-amino-4-bromobenzoate (5.0 g, 21.73 mmol) and 2-(dihydroxyboryl)phenylamine hydrochloride (4.145 g, 23.91 mmol) in 1,4-dioxane (75 mL)/water (25 mL), KOAc (13.84 g, 56.2 mmol) and Pd(dppf)\textsubscript{2}Cl\textsubscript{2} (0.794 g, 1.09 mmol) were added. The mixture was heated at 100~\textdegree C for 16 h. After cooling to room temperature, the mixture was diluted with EtOAc and filtered through a pad of Celite. The organic layer of the filtrate was separated, dried over Na\textsubscript{2}SO\textsubscript{4} and concentrated under reduced pressure. The dark red oil obtained was purified by flash column chromatography to afford the product as a reddish orange solid (2.10 g, 40\%).
\textsuperscript{1}H NMR (300 MHz, CDCl\textsubscript{3}) $\delta$ 7.55 (m, 2H), 7.26 (m, 2H), 7.17 (d, \textit{J} = 7.5 Hz, 1H), 6.90 (t, \textit{J} = 7.5 Hz, 1H), 6.85 (d, \textit{J} = 8.1 Hz, 1H), 3.97 (s, 3H), 3.94 (br s, 2H), 3.77 (br s, 2H); ES-MS: 243.7 [M+H]\textsuperscript{+}.

\subsection{Methyl 9,10-diaza-2-phenanthrenecarboxylate (\cmpd{si-ester})}
Methyl 2,2\textquotesingle-diamino-4-biphenylcarboxylate (2.10 g, 8.64 mmol, \cmpd{si-diamino-biphenyl}) was dissolved in 2,2,2-trifluoroethanol (30 mL) and cooled to 0~\textdegree C. \textit{tert}-Butyl nitrile (3.1 mL, 25.93 mmol) was added dropwise at 0~\textdegree C, and the mixture was stirred overnight in the same cooling bath. After completion, solvent was removed under reduced pressure, and the crude mixture was purified by flash column chromatography to afford the product as a brown solid (1.2 g, 58.3\%).
\textsuperscript{1}H NMR (300 MHz, CDCl\textsubscript{3}) $\delta$ 9.44 (s, 1H), 8.81 (m, 1H), 8.63 (m, 2H), 8.54 (m, 1H), 8.12 (m, 1H), 7.92 (m, 2H), 4.06 (s, 3H); ES-MS: 239.7 [M+H]\textsuperscript{+}.

\subsection{9,10-Diaza-2-phenanthrenecarboxylic acid (\cmpd{si-acid})}
A solution of methyl 9,10-diaza-2-phenanthrenecarboxylate (1.00 g, 4.197 mmol, \cmpd{si-ester}) in MeOH (60 mL) was treated with 1~M NaOH (6.30 mL, 6.29 mmol) and heated at 90~\textdegree C for 2 h. A solution of oxalic acid (2.0 equiv) in H\textsubscript{2}O (10 mL) was added dropwise, and the mixture was stirred for 5 min. After completion, methanol was removed under reduced pressure, and CH\textsubscript{2}Cl\textsubscript{2} was added. The solid obtained was filtered, washed with CH\textsubscript{2}Cl\textsubscript{2} and H\textsubscript{2}O, and dried on a high vacuum pump to afford the product as a brown solid (0.630 g, 67\%).
\textsuperscript{1}H NMR (300 MHz, CDCl\textsubscript{3}) $\delta$ 9.31 (s, 1H), 8.92 (m, 2H), 8.75 (m, 1H), 8.59 (m, 1H), 8.13 (m, 3H). ESI-MS: 225.7 [M+H]\textsuperscript{+}.

\subsection{(9,10-Diaza-2-phenanthryl)methanol (\cmpd{si-alcohol})}
9,10-Diaza-2-phenanthrenecarboxylic acid (0.630 g, 2.81 mmol, \cmpd{si-acid}) was suspended in anhydrous THF (20 mL) and cooled to 0~\textdegree C under an inert atmosphere. A solution of BH\textsubscript{3}$\cdot$THF complex (14 mL, 14.05 mmol) was added dropwise at 0~\textdegree C, and the mixture was stirred for 10 min at 0~\textdegree C, then allowed to warm to room temperature and stirred overnight. The reaction mixture was cooled back to 0~\textdegree C and carefully quenched by the slow addition of MeOH. After stirring for 10 min, the solvent was removed under reduced pressure. MeOH was added, and the mixture was concentrated again under reduced pressure to ensure complete removal of boron residues. The crude obtained was purified by flash column chromatography to afford pure product as a brown solid (0.390 g, 66\%).
\textsuperscript{1}H NMR (300 MHz, CDCl\textsubscript{3}) $\delta$ 8.67 (m, 2H), 8.49 (m, 2H), 7.91 (m, 3H), 5.02 (s, 2H).

\subsection{2-(Bromomethyl)-9,10-diazaphenanthrene (\cmpd{si-bromide})}
(9,10-Diaza-2-phenanthryl)methanol (0.354 g, 1.684 mmol, \cmpd{si-alcohol}) was dissolved in anhydrous THF (8 mL) under an inert atmosphere and cooled to 0~\textdegree C. A solution of PBr\textsubscript{3} (0.24 mL, 2.526 mmol) in CH\textsubscript{2}Cl\textsubscript{2} (2 mL) was added dropwise maintaining the temperature at 0 to 5~\textdegree C. The mixture was stirred at 0 to 5~\textdegree C for 30 min, allowed to warm up to room temperature and stirred for an additional 2 h. After completion, the mixture was diluted with CH\textsubscript{2}Cl\textsubscript{2} and carefully quenched with saturated aqueous NaHCO\textsubscript{3}. The organic layer was separated, dried over anhydrous Na\textsubscript{2}SO\textsubscript{4}, and concentrated under reduced pressure to afford the product as a yellow solid (0.350 g, 76\%) which was taken forward without purification.
\textsuperscript{1}H NMR (300 MHz, CDCl\textsubscript{3}) $\delta$ 8.75 (m, 2H), 8.56 (m, 2H), 7.96 (m, 3H), 4.76 (s, 2H).

\subsection{Diethyl [(9,10-diaza-2-phenanthryl)methyl]phosphonate (\cmpd{si-diethylphosphonate})}
2-(Bromomethyl)-9,10-diazaphenanthrene (0.160 g, 0.586 mmol, \cmpd{si-bromide}) and triethyl phosphite (1.1 mL, 5.84 mmol) were mixed in a pressure vessel and heated at 120~\textdegree C for 16 hours. After completion, the mixture was cooled to room temperature, concentrated under reduced pressure and purified by flash column chromatography to afford the product as a yellow solid (0.087 g, 45\%).
\textsuperscript{1}H NMR (300 MHz, CDCl\textsubscript{3}) $\delta$ 8.74 (m, 1H), 8.62 (br s, 1H), 8.56 (m, 2H), 7.92 (m, 3H), 4.10 (t, \textit{J} = 7.2 Hz, 4H), 3.52 (s, 1H), 3.44 (s, 1H), 1.29 (t, \textit{J} = 7.2 Hz, 6H). \textsuperscript{31}P NMR (121 MHz, CDCl\textsubscript{3}) $\delta$ 25.00.

\subsection{[(9,10-Diaza-2-phenanthryl)methyl]phosphonic acid hydrochloride (\cmpd{bcc-BnPO3-5})}
To a solution of diethyl [(9,10-diaza-2-phenanthryl)methyl]phosphonate (0.090 g, 0.272 mmol, \cmpd{si-diethylphosphonate}) in CH\textsubscript{3}CN (4 mL), KI (0.262 g, 1.36 mmol) was added, followed by the dropwise addition of TMSCl (0.175 mL, 1.362 mmol). The reaction mixture was stirred at room temperature for 20 minutes and then heated at 60~\textdegree C for 5 hours. After cooling to room temperature, the solvent was removed under reduced pressure. The crude obtained was washed with CH\textsubscript{2}Cl\textsubscript{2} ($3 \times 3$ mL) and then with MeOH ($2 \times 3$ mL). The residue was treated with concentrated HCl (2 mL) and stirred for 5 minutes. HCl solution was then evaporated under reduced pressure. The solid was washed with water and then with MeOH ($2 \times 2$ mL), to afford clean hydrochloride salt of the product as a yellow solid (60 mg, 63\%).
\textsuperscript{1}H NMR (300 MHz, DMSO-\textit{d}\textsubscript{6}) $\delta$ 8.86 (br s, 2H), 8.67 (br s, 1H), 8.56 (br s, 1H), 8.01 (br s, 3H), 3.15 (s, 2H). \textsuperscript{31}P NMR (121 MHz, DMSO-\textit{d}\textsubscript{6}) $\delta$ 19.9. ES-MS: 275.8 [M+H]\textsuperscript{+}. HPLC purity: 90.9\%.

\subsection{(9,10-Diaza-2-phenanthryl)methanesulfonic acid hydrochloride (\cmpd{bcc-BnSO3-5})}
2-(Bromomethyl)-9,10-diazaphenanthrene (0.150 g, 0.549 mmol, \cmpd{si-bromide}) was suspended in a mixture of H\textsubscript{2}O (5 mL) and acetone (2 mL). Na\textsubscript{2}SO\textsubscript{3} (0.070 g, 0.549 mmol) was added as a solid, and the mixture was stirred at room temperature for 20 min and then heated at 100~\textdegree C for 16 h. After cooling to room temperature, the solvent was removed under reduced pressure. The residue was washed with CH\textsubscript{2}Cl\textsubscript{2} ($3 \times 10$ mL), and the insoluble material was treated with concentrated HCl, resulting in the formation of a brown precipitate. The solid was collected by filtration and washed with CH\textsubscript{2}Cl\textsubscript{2} to afford the hydrochloride salt of the product as a brown solid (0.118 g, 60\%).
\textsuperscript{1}H NMR (300 MHz, CD\textsubscript{3}OD) $\delta$ 9.06 (t, \textit{J} = 9 Hz, 2H), 8.76 (br m, 2H), 8.45 (d, \textit{J} = 8.1 Hz, 1H), 8.36 (t, \textit{J} = 6.9 Hz, 1H), 8.28 (t, \textit{J} = 6.9 Hz, 1H), 4.51 (s, 2H). ESI-MS: 273.8 [M$-$H]\textsuperscript{$-$}. HPLC purity: 99.0\%.

\begin{figure}[h!]
  \centering
  \begin{overpic}[width=\textwidth]{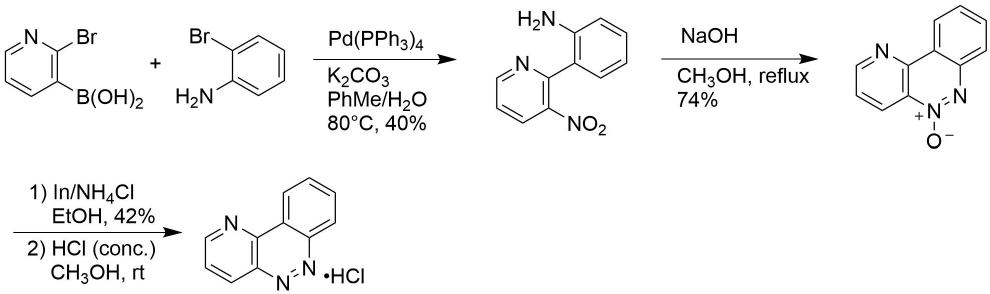}
    \put(55,12){\footnotesize\cmpd{si-nitropyr-amine}}
    \put(90,12){\footnotesize\cmpd{si-triaza-noxide}}
    \put(30,-2){\footnotesize\cmpd{exllm-aza-bcc}}
  \end{overpic}
  \vspace{8pt}
  \caption{Synthetic route to 4,9,10-triazaphenanthrene hydrochloride (\cmpd{exllm-aza-bcc}).}
  \label{fig:exllm-synthesis-1}
\end{figure}
\FloatBarrier

\subsection{2-(3-Nitropyridin-2-yl)benzen-1-amine (\cmpd{si-nitropyr-amine})}
To a stirred solution of 2-aminophenylboronic acid hydrochloride (5.00 g, 28.84 mmol) and 2-bromo-3-pyridylamine (5.85 g, 28.84 mmol) in a mixture of 1,4-dioxane (75 mL) and H\textsubscript{2}O (50 mL), K\textsubscript{3}PO\textsubscript{4} (18.36 g, 86.51 mmol) and Pd(dppf)\textsubscript{2}Cl\textsubscript{2} (1.05 g, 1.44 mmol) were added. The reaction mixture was heated at 85~\textdegree C for 48 h. After cooling to room temperature, 1~M HCl (200 mL) was added, and the mixture was stirred for 5 min. The suspension was filtered through a pad of Celite, and the filtrate was concentrated under reduced pressure to remove 1,4-dioxane. The remaining aqueous solution was washed with EtOAc ($3 \times 150$ mL). The aqueous layer was then carefully neutralized by dropwise addition of saturated aqueous Na\textsubscript{2}CO\textsubscript{3} and extracted with EtOAc ($3 \times 150$ mL). The combined organic extracts were dried over anhydrous Na\textsubscript{2}SO\textsubscript{4}, filtered, and concentrated under reduced pressure to afford a dark orange-red oil. Purification by silica gel chromatography (20--80\% EtOAc/hexanes) afforded the product as a reddish orange oil (40\%).
\textsuperscript{1}H NMR (300 MHz, CDCl\textsubscript{3}) $\delta$ 8.85 (dd, \textit{J} = 4.8, 1.2 Hz, 1H), 8.21 (dd, \textit{J} = 8.4 Hz, 1H), 7.45 (m, 1H), 7.26 (m, 2H), 7.07 (m, 1H), 6.82 (m, 2H), 7.45 (s, 2H). ESI-MS: 216.6 [M+H]\textsuperscript{+}.

\subsection{4,9,10-Triaza-9-phenanthrenium-9-olate (\cmpd{si-triaza-noxide})}
To a stirred solution of 2-(3-nitropyridin-2-yl)benzen-1-amine (\cmpd{si-nitropyr-amine}, 2.50 g, 11.62 mmol) in MeOH (50 mL), NaOH (2.33 g, 58.14 mmol) was added. The reaction mixture was heated at 80~\textdegree C for 2 h. After completion, the mixture was cooled to room temperature and concentrated \textit{in vacuo}. The residue was dissolved in EtOAc and washed with saturated aqueous NaHCO\textsubscript{3}. The layers were separated, and the aqueous layer was extracted with EtOAc ($3 \times 100$ mL). The combined organic extracts were dried over anhydrous Na\textsubscript{2}SO\textsubscript{4}, filtered, and concentrated under reduced pressure to afford the product (1.70 g, 74\%) as a brown solid.
\textsuperscript{1}H NMR (300 MHz, CDCl\textsubscript{3}) $\delta$ 9.21 (d, \textit{J} = 3.9 Hz, 1H), 9.08 (d, \textit{J} = 8.7 Hz, 1H), 8.91 (d, \textit{J} = 7.8 Hz, 1H), 8.04 (d, \textit{J} = 8.4 Hz, 1H), 7.89 (m, 1H), 7.79 (m, 2H).

\subsection{4,9,10-Triazaphenanthrene (\cmpd{si-triazaphen})}
To a stirred solution of 4,9,10-triaza-9-phenanthrenium-9-olate (\cmpd{si-triaza-noxide}, 1.70 g, 8.62 mmol) in EtOH (70 mL) and saturated aqueous NH\textsubscript{4}Cl (20 mL), indium powder (2.97 g, 25.9 mmol) was added. The reaction mixture was heated at 80~\textdegree C for 4 h and monitored by TLC. After completion, the mixture was concentrated under reduced pressure, diluted with EtOAc, and stirred for 5 min. The suspension was filtered through a pad of Celite, and the pad was washed with EtOAc ($3 \times 100$ mL). The combined filtrates were concentrated under reduced pressure and purified by flash column chromatography (5--20\% EtOAc/hexanes) to afford the product as a brown solid (42\%).
\textsuperscript{1}H NMR (300 MHz, CD\textsubscript{3}OD) $\delta$ 9.45 (m, 1H), 9.25 (m, 2H), 8.85 (m, 1H), 8.30 (m, 2H), 8.23 (m, 1H). ESI-MS: 182.7 [M+H]\textsuperscript{+}.

\subsection{4,9,10-Triazaphenanthrene hydrochloride (\cmpd{exllm-aza-bcc})}
To a solution of 4,9,10-triazaphenanthrene (\cmpd{si-triazaphen}, 900 mg, 4.97 mmol) in MeOH (20 mL), concentrated HCl (3 mL) was added dropwise at 0--5~\textdegree C. The mixture was stirred for 2 min at this temperature, then allowed to warm to room temperature and stirred for an additional 5 min. All volatiles were removed under reduced pressure. The resulting solid was washed with CH\textsubscript{2}Cl\textsubscript{2} ($3 \times 20$ mL) and dried under high vacuum overnight to afford the product (1.15 g, 80\%) as a light yellow solid.
\textsuperscript{1}H NMR (300 MHz, CD\textsubscript{3}OD) $\delta$ 9.42 (d, 1H), 9.25 (m, 1H), 9.20 (m, 1H), 8.22 (m, 2H), 8.15 (m, 1H). ESI-MS: 182.7 [M+H]\textsuperscript{+}. HPLC purity: 99\%.

\begin{figure}[h!]
  \centering
  \begin{overpic}[width=\textwidth]{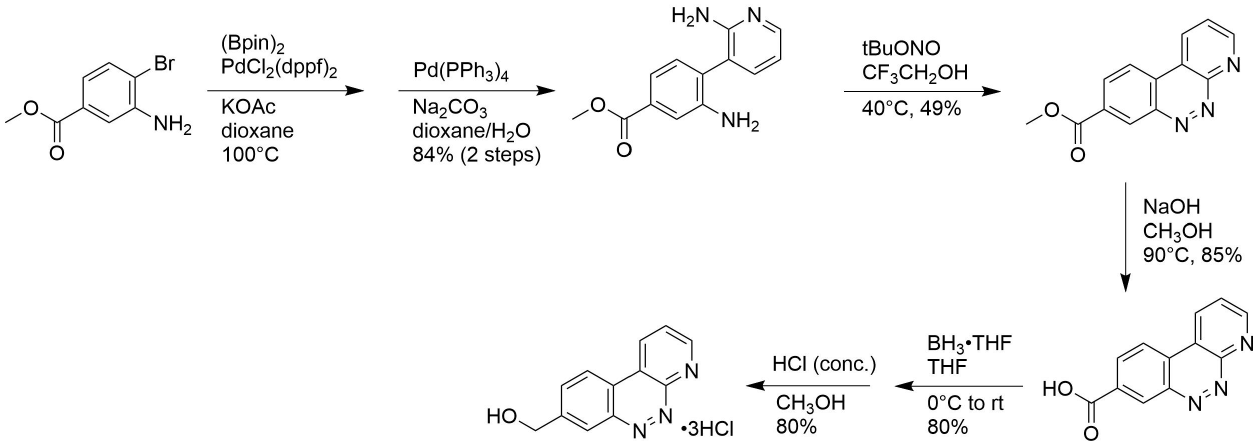}
    \put(55,22){\footnotesize\cmpd{si-amino-biphenyl-pyr}}
    \put(90,22){\footnotesize\cmpd{si-triaza-ester}}
    \put(90,-2){\footnotesize\cmpd{si-triaza-acid}}
    \put(50,-2){\footnotesize\cmpd{exllm-OH-aza-bcc}}
  \end{overpic}
  \vspace{8pt}
  \caption{Synthetic route to (1,9,10-triaza-7-phenanthryl)methanol hydrochloride (\cmpd{exllm-OH-aza-bcc}).}
  \label{fig:exllm-synthesis-2}
\end{figure}
\FloatBarrier

\subsection{Methyl 3-amino-4-(2-aminopyridin-3-yl)benzoate (\cmpd{si-amino-biphenyl-pyr})}
To a stirred solution of methyl 3-amino-4-bromobenzoate (6.0 g, 26.08 mmol) and 4,4,5,5-tetramethyl-2-(tetramethyl-1,3,2-dioxaborolan-2-yl)-1,3,2-dioxaborolane (6.622 g, 26.08 mmol) in 1,4-dioxane (100 mL), KOAc (5.138 g, 52.1 mmol) and Pd(dppf)\textsubscript{2}Cl\textsubscript{2} (0.953 g, 1.304 mmol) were added. The reaction mixture was heated at 100~\textdegree C for 16 h. After cooling to room temperature, 3-bromo-2-aminopyridine (3.74 g, 21.65 mmol) was added, followed by Pd(dppf)\textsubscript{2}Cl\textsubscript{2} (791 mg, 1.85 mmol), K\textsubscript{3}PO\textsubscript{4} (4.6 g, 43.3 mmol), and H\textsubscript{2}O (15 mL). The reaction mixture was heated at 80~\textdegree C for 16 h. After cooling to room temperature, 1~M HCl (200 mL) was added, and the mixture was stirred for 5 min. The mixture was passed through a pad of Celite, and the filtrate was concentrated under reduced pressure to remove 1,4-dioxane. The aqueous solution was washed with EtOAc ($3 \times 150$ mL). The aqueous layer was carefully neutralized with saturated aqueous Na\textsubscript{2}CO\textsubscript{3} (added dropwise) and extracted with EtOAc ($3 \times 150$ mL). The combined organic extracts were dried over Na\textsubscript{2}SO\textsubscript{4}, filtered, and concentrated under reduced pressure. The crude mixture was purified by flash column chromatography (EtOAc/hexanes, 20--80\%) to afford the product as a reddish orange solid (84\%).
\textsuperscript{1}H NMR (300 MHz, CDCl\textsubscript{3}) $\delta$ 8.15 (d, \textit{J} = 4.8 Hz, 1H), 7.51 (m, 2H), 7.41 (d, \textit{J} = 7.5 Hz, 1H), 7.26 (s, 1H), 7.18 (d, \textit{J} = 7.5 Hz, 1H), 6.8 (m, 1H), 4.50 (s, 2H), 3.92 (s, 3H), 3.83 (s, 2H). ESI-MS: 244.7 [M+H]\textsuperscript{+}.

\subsection{Methyl 1,9,10-triaza-7-phenanthrenecarboxylate (\cmpd{si-triaza-ester})}
Methyl 3-amino-4-(2-aminopyridin-3-yl)benzoate (\cmpd{si-amino-biphenyl-pyr}, 3.50 g, 14.40 mmol) was dissolved in 2,2,2-trifluoroethanol (50 mL) and cooled to 0--5~\textdegree C. \textit{tert}-Butyl nitrile (5.14 mL, 43.21 mmol) was added dropwise at 0--5~\textdegree C, and the reaction mixture was stirred in the same cooling bath for 16 h. The reaction was quenched with 1~M HCl (150 mL), and the solvent was removed under reduced pressure. The residue was dissolved in an immiscible mixture of EtOAc and 1~M HCl, and the aqueous layers were separated. The organic layer was washed again with 1~M HCl. The combined aqueous layers were carefully neutralized with solid Na\textsubscript{2}CO\textsubscript{3} (added portionwise), and the aqueous phase was extracted with EtOAc ($3 \times 100$ mL). The combined organic extracts were dried over Na\textsubscript{2}SO\textsubscript{4}, filtered, and concentrated under reduced pressure to afford a brown crude solid. The crude product was purified by flash column chromatography (EtOAc/hexanes, 20--80\%) to afford the product as a light yellow solid (50\%).
\textsuperscript{1}H NMR (300 MHz, CDCl\textsubscript{3}) $\delta$ 11.37 (s, 1H), 8.63 (m, 1H), 8.41 (d, \textit{J} = 7.5 Hz, 1H), 8.26 (s, 1H), 8.11 (d, \textit{J} = 7.5 Hz, 1H), 7.99 (d, \textit{J} = 10.2 Hz, 1H), 7.28 (m, 1H), 4.00 (s, 3H). ESI-MS: 240.6 [M+H]\textsuperscript{+}.

\subsection{1,9,10-Triaza-7-phenanthrenecarboxylic acid (\cmpd{si-triaza-acid})}
A solution of methyl 1,9,10-triaza-7-phenanthrenecarboxylate (\cmpd{si-triaza-ester}, 1.50 g, 62.76 mmol) in MeOH (59 mL) was treated with 1~M NaOH (12.6 mL, 125.52 mmol) and heated at 90~\textdegree C for 2 h. After completion, a solution of oxalic acid (2.0 equiv) in H\textsubscript{2}O (10 mL) was added dropwise, and the mixture was stirred for 5 min. The solvent was partially removed under reduced pressure, and CH\textsubscript{2}Cl\textsubscript{2} was added. The resulting solid was collected by filtration, washed with CH\textsubscript{2}Cl\textsubscript{2} and H\textsubscript{2}O, and dried on high vacuum to afford the product as a brown solid (85\%).
\textsuperscript{1}H NMR (300 MHz, DMSO-\textit{d}\textsubscript{6}) $\delta$ 11.5 (br s, 1H), 8.45 (m, 2H), 8.15 (m, 2H), 7.8 (m, 1H), 7.20 (br s, 1H).

\subsection{(1,9,10-Triaza-7-phenanthryl)methanol (\cmpd{si-triaza-alcohol})}
1,9,10-Triaza-7-phenanthrenecarboxylic acid (\cmpd{si-triaza-acid}, 1.2 g, 5.33 mmol) was suspended in anhydrous THF (20 mL) and cooled to 0--5~\textdegree C under an inert atmosphere. A solution of BH\textsubscript{3}$\cdot$THF complex (27 mL, 26.677 mmol) was added dropwise at 0--5~\textdegree C. The mixture was stirred for 10 min at this temperature, then allowed to warm to room temperature and stirred for 16 h. The reaction mixture was cooled to 0~\textdegree C and carefully quenched by the slow addition of MeOH. After stirring for 10 min, the solvent was removed under reduced pressure. MeOH was added, and the mixture was concentrated again under reduced pressure to ensure complete removal of boron residues. The crude product was purified by flash column chromatography (0--15\% MeOH in CH\textsubscript{2}Cl\textsubscript{2}) to afford the product as a light yellow solid (80\%).
\textsuperscript{1}H NMR (300 MHz, CD\textsubscript{3}OD) $\delta$ 8.65 (d, \textit{J} = 7.8 Hz, 1H), 8.39 (d, \textit{J} = 6.0 Hz, 1H), 8.15 (d, \textit{J} = 8.1 Hz, 1H), 7.72 (s, 1H), 7.38 (m, 1H), 4.82 (s, 2H).

\subsection{(1,9,10-Triaza-7-phenanthryl)methanol hydrochloride (\cmpd{exllm-OH-aza-bcc})}
(1,9,10-Triaza-7-phenanthryl)methanol (\cmpd{si-triaza-alcohol}, 840 mg, 3.98 mmol) was dissolved in MeOH (20 mL) and cooled to 0~\textdegree C. Concentrated HCl (3 mL) was added dropwise at 0--5~\textdegree C, and the reaction mixture was stirred for 2 min at this temperature, then allowed to warm up to room temperature and stirred for an additional 5 min. All volatiles were removed under reduced pressure. The resulting solid was washed with CH\textsubscript{2}Cl\textsubscript{2} ($3 \times 20$ mL) and dried under high vacuum overnight to afford the hydrochloride salt as a light yellow solid (80\%).
\textsuperscript{1}H NMR (300 MHz, CD\textsubscript{3}OD) $\delta$ 9.10 (d, \textit{J} = 7.5 Hz, 1H), 8.49 (d, \textit{J} = 6.0 Hz, 1H), 8.30 (d, \textit{J} = 8.4 Hz, 1H), 7.77 (s, 1H), 7.69 (m, 1H), 7.51 (d, \textit{J} = 8.1 Hz, 1H), 4.86 (s, 2H). HPLC purity: 99.7\%.

\begin{figure}[h!]
  \centering
  \begin{overpic}[width=\textwidth]{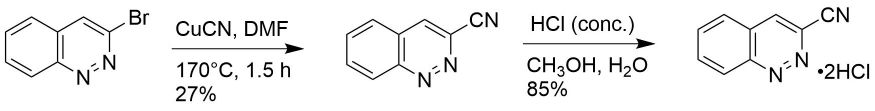}
    \put(45,-1){\footnotesize\cmpd{si-cinnoline-CN}}
    \put(89,-1){\footnotesize\cmpd{exllm-CN-bcc}}
  \end{overpic}
  \vspace{10pt}
  \caption{Synthetic route to 3-cinnolinecarbonitrile hydrochloride (\cmpd{exllm-CN-bcc}).}
  \label{fig:exllm-synthesis-3}
\end{figure}
\FloatBarrier

\subsection{3-Cinnolinecarbonitrile (\cmpd{si-cinnoline-CN})}
CuCN (430 mg, 4.798 mmol) was added to a solution of 3-bromocinnoline (500 mg, 2.399 mmol) in DMF (8 mL) in a sealed microwave vial. The reaction mixture was heated at 170~\textdegree C under microwave irradiation for 1.5 h. After cooling to room temperature, 2~N HCl was added, and the mixture was stirred for 10 min at room temperature. The solution was then neutralized with 1~N NaOH and extracted with EtOAc. The organic layer was separated, dried over Na\textsubscript{2}SO\textsubscript{4}, and concentrated \textit{in vacuo}. The crude mixture was purified by flash column chromatography (10--40\% EtOAc/hexanes) to afford the product as a brown solid (27\%).
\textsuperscript{1}H NMR (300 MHz, CD\textsubscript{3}OD) $\delta$ 8.72 (d, \textit{J} = 8.7 Hz, 1H), 8.34 (s, 1H), 8.10 (m, 1H), 7.95 (m, 2H).

\subsection{3-Cinnolinecarbonitrile hydrochloride (\cmpd{exllm-CN-bcc})}
3-Cinnolinecarbonitrile (\cmpd{si-cinnoline-CN}, 100 mg, 0.64 mmol) was dissolved in MeOH (3 mL) and cooled to 0~\textdegree C. Concentrated HCl (0.5 mL) was added dropwise at 0~\textdegree C, and the reaction mixture was stirred for 2 min at this temperature, then allowed to warm to room temperature and stirred for an additional 5 min. All volatiles were removed under reduced pressure. The resulting solid was washed with CH\textsubscript{2}Cl\textsubscript{2} ($3 \times 5$ mL) and dried under high vacuum overnight to afford the hydrochloride salt as a brown solid (85\%).
\textsuperscript{1}H NMR (300 MHz, CD\textsubscript{3}OD) $\delta$ 8.85 (s, 1H), 8.65 (d, \textit{J} = 8.7 Hz, 1H), 8.20 (m, 2H), 8.10 (m, 1H). ESI-MS: 156.5 [M+H]\textsuperscript{+}. HPLC purity: $>$~99.9\%.







